\documentclass[pdflatex,sn-basic]{sn-jnl}

\usepackage{graphicx}%
\usepackage{multirow}%
\usepackage{amsmath,amssymb,amsfonts}%
\usepackage{amsthm}%
\usepackage{mathrsfs}%
\usepackage[title]{appendix}%
\usepackage{xcolor}%
\usepackage{textcomp}%
\usepackage{manyfoot}%
\usepackage{booktabs}%
\usepackage{algorithm}%
\usepackage{algorithmicx}%
\usepackage{algpseudocode}%
\usepackage{listings}%
\usepackage{subcaption}
\usepackage{array}

\newlength{\alglabelwidth}
\algnewcommand{\Input}[1]{%
  \Statex
  \makebox[\alglabelwidth][l]{\textbf{Input}}\textbf{: }%
  \parbox[t]{\dimexpr\linewidth-\alglabelwidth-1em\relax}{#1}%
}

\algnewcommand{\Output}[1]{%
  \Statex
  \makebox[\alglabelwidth][l]{\textbf{Output}}\textbf{: }%
  \parbox[t]{\dimexpr\linewidth-\alglabelwidth-1em\relax}{#1}%
}

\newcommand{\dd}{\mathop{}\!\mathrm{d}} 

\theoremstyle{thmstyleone}%
\theoremstyle{thmstyletwo}%

\theoremstyle{thmstylethree}%

\begin{document}

\title[]{Spatial Principal Component Analysis and Moran's $I$ for Multivariate Functional Areal Data}

\author*[1,2,3]{\fnm{Dharini} \sur{Pathmanathan}}\email{dharini@um.edu.my}

\author[4,5]{\fnm{Issa-Mbenard} \sur{Dabo}}

\author[1,7]{\fnm{Tzung Hsuen} \sur{Khoo}}

\author[6]{\fnm{Alaa} \sur{Ali-Hassan}}

\author[7]{\fnm{Sophie} \sur{Dabo-Niang}}

\affil*[1]{\orgdiv{Institute of Mathematical Sciences, Faculty of Science},
\orgname{Universiti Malaya},
\orgaddress{\postcode{50603}, \city{Kuala Lumpur}, \country{Malaysia}}}

\affil[2]{\orgdiv{Universiti Malaya Centre for Data Analytics},
\orgname{Universiti Malaya},
\orgaddress{\postcode{50603}, \city{Kuala Lumpur}, \country{Malaysia}}}

\affil[3]{\orgdiv{Center of Research for Statistical Modelling and Methodology, Faculty of Science},
\orgname{Universiti Malaya},
\orgaddress{\postcode{50603}, \city{Kuala Lumpur}, \country{Malaysia}}}

\affil[4]{\orgdiv{Institut de Mathématiques de Bordeaux, UMR 5251},
\orgname{CNRS, Université de Bordeaux, Bordeaux INP},
\orgaddress{\city{Bordeaux}, \country{France}}}

\affil[5]{\orgdiv{Mathematics, Division of Science},
\orgname{New York University Abu Dhabi},
\orgaddress{\city{Abu Dhabi}, \country{United Arab Emirates}}}

\affil[6]{\orgname{Independent Researcher},
\orgaddress{\city{Québec City}, \country{Canada}}}

\affil[7]{\orgdiv{CNRS, UMR 8524--Laboratoire Paul Painlevé, Inria-Datavers},
\orgname{Université de Lille},
\orgaddress{\postcode{F-59000}, \city{Lille}, \country{France}}}









\abstract{The paper introduces a multivariate functional areal spatial principal component analysis (mfasPCA) framework, together with multivariate functional Moran’s $I$ statistics, to enable the assessment of spatial autocorrelation and dimension reduction for multivariate functional data observed over areal units. The proposed framework is spatial-functional in scope: the functional argument may represent time, age, wavelength, or another ordered continuum, while spatial dependence is introduced across areal units through a spatial weight matrix. The principal component method is defined through a Moran-type spatially weighted criterion. We propose eigenvalue-based permutation tests to assess the significance of spatially structured components. The testing framework includes omnibus tests, componentwise tests with Holm adjustment, and sequential rank-wise tests based on tail sums of eigenvalues. Simulation studies show that mfasPCA captures positive and negative spatial-functional structures and concentrates them in the leading components under the respective autocorrelation regimes. A real-data application illustrates how mfasPCA identifies spatially structured modes of multivariate functional variation.}



\maketitle


\section{Introduction}\label{intro}

Functional principal component analysis (FPCA) is a central tool for dimension reduction in functional data analysis \citep{ramsay2005functional}. It describes the dominant modes of variation in functional observations. Several extensions have been proposed for both univariate and multivariate functional data \citep{bali2014robust, berrendero2011principal, chiou2014linear, happ2018multivariate, hormann2015dynamic}. However, when functional observations are associated with spatial locations or areal units, dimension reduction should account not only for functional variation, but also for spatial relationships among the observational units.

Spatial functional data analysis integrates techniques from functional data analysis and spatial statistics to study collections of functions observed at different locations within a region. Such data are commonly referred to as spatially correlated functional data \citep{mateu2017advances}. \citet{delicado2010statistics} introduced a framework that combines classical spatial data structures, including geostatistical data, point patterns, and areal data, with functional data. In the areal setting considered in this paper, each region is represented by one or more functions observed over a domain such as time or age.

Principal component analysis has also been adapted in geospatial and spatially indexed functional data contexts \citep{kuenzer2021principal, li2014functional,liu2017functional}. For functional areal data, however, classical FPCA does not directly account for spatial autocorrelation among nearby areal units. Spatial principal component analysis (sPCA), introduced by \citet{jombart2008revealing}, was developed as a spatially explicit multivariate method that incorporates spatial information into PCA. In sPCA, the components are interpreted through global structures associated with positive spatial autocorrelation and local structures associated with negative spatial autocorrelation. In this paper, we introduce multivariate functional areal spatial principal component analysis (mfasPCA). The proposed mfasPCA extends the spatial PCA framework of \citet{jombart2008revealing} to multivariate functional areal data by formulating a Moran-type spatial criterion on the multivariate functional score matrix. We focus on the multivariate functional areal setting, while additional univariate fasPCA simulation results and a real-data illustration are provided in the Supplementary Material.

Spatio-temporal principal component analysis (STPCA), introduced by \citet{krzysko2024stpca}, is a Moran's $I$-based extension of principal component analysis applied to multivariate, temporally smoothed trajectories. The procedure begins with a temporal smoothing step employing a Fourier basis; the resulting Fourier coefficients for all variables are concatenated into a single coefficient matrix. Subsequently, the method conducts a multivariate analysis on these concatenated, low-dimensional Fourier-coefficient representations of the functional data. Formally, the associated eigenproblem is defined by $\mathbf{A}^\top \mathbf{W} \mathbf{A}$, where $\mathbf{A} \in \mathbb{R}^{n \times M}$ is the matrix that aggregates the Fourier basis coefficients in all multivariate temporal trajectories and $\mathbf{W}$ is a $n\times n$ spatial weight matrix that is invariant with respect to the temporal argument. Consequently, the approach consists of a temporal basis-expansion step followed by a spatially weighted eigendecomposition of the coefficient matrix. In this sense, the Moran-type criterion is applied after temporal smoothing rather than formulated as a joint criterion over the spatial and temporal domains.

In contrast, mfasPCA is designed for multivariate functional areal data. The multivariate functional observations are first represented through a functional basis expansion, which yields a functional score-coordinate matrix, and then a Moran-type spatially weighted criterion is imposed on this matrix. The corresponding principal components are subsequently mapped back to multivariate functional eigenfunctions, ensuring that the extracted directions preserve a coherent functional interpretation. The functional argument may represent time, age, wavelength, or another ordered continuum.

We further introduce multivariate functional Moran's $I$ statistics to quantify spatial autocorrelation in functional areal observations and to support the interpretation of the spatial structure captured by mfasPCA. Building on the univariate functional Moran's $I$ developments in \citet{hassan2021spatial} and \citet{romano2022spatial}, the present paper develops bivariate and multivariate functional Moran's $I$ statistics and integrates them into a spatial-functional dimension-reduction framework for functional areal data. The univariate functional Moran's $I$ developed in \citet{hassan2021spatial} was subsequently used by \citet{math11030674} to investigate spatial dependence in global stock indices during the 2015--2016 global market sell-off.

This work introduces a spatial-functional dimension-reduction framework for functional areal data, with main emphasis on the multivariate setting. The main contributions are as follows:
\begin{enumerate}
  \renewcommand{\labelenumi}{(\roman{enumi})}

  \item Functional Moran's $I$ statistics are developed from the univariate setting to the bivariate and multivariate settings to assess spatial autocorrelation in functional areal data.
  
  \item Spatial principal component analysis methods are introduced for functional areal data. In the multivariate case, mfasPCA is defined through a Moran-type spatially weighted criterion on the multivariate functional score matrix. The univariate case corresponds to $d=1$ and is reported in the Supplementary Material.

  \item Eigenvalue-based permutation tests are proposed for mfasPCA to assess positive/global and negative/local spatial structure, including omnibus, componentwise, and sequential tests.
\end{enumerate}

\noindent The proposed framework is evaluated through simulation studies and illustrated using a multivariate real-data application.

The remainder of the paper is organized as follows: Section~\ref{sec:method} develops the methodology for multivariate functional Moran's $I$ statistics and multivariate functional areal spatial PCA, including the associated spatial weight matrices and eigenvalue-based permutation tests. Section~\ref{sec:sim-study} presents the simulation study. Section~\ref{sec:real-data} applies the proposed framework to Polish regional socioeconomic data. Section~\ref{sec:discussion} discusses the main findings and limitations. Section~\ref{sec:conclusion} concludes. Additional univariate fasPCA simulation results and a real-data illustration are provided in the Supplementary Material.


\section{Methodology}\label{sec:method}

\subsection{Multivariate functional principal component analysis on areal data}

Consider $n$ areal units indexed by $i=1,\ldots,n$, with representative locations $\mathbf{s}_i\in\mathcal I$. For each areal unit $i$, we observe a $d$-dimensional functional measurement
\begin{equation}
    \mathbf{Y}_{i,\mathbf{x}} = (Y^1_{i,x_1}, \ldots, Y^d_{i,x_d})^\top, \qquad d\ge 1. 
    \nonumber
\end{equation}
Here, $\mathbf{x} = (x_1, \ldots, x_d)^\top \in \mathcal{X} = \mathcal{X}_1 \times \cdots \times \mathcal{X}_d$, $\mathcal{X}$ being the $d$-fold Cartesian product of the $\mathcal{X}_j$, $\mathcal{X}_j$ being a compact set of $\mathbb{R}^{d_j}$ ($d_j\in \mathbb{N}^*$). These data points $\mathbf{Y}_{i,\mathbf{x}}$ are assumed to be noisy observations of a smooth areal stochastic multivariate functional process $\{\mathbf{S}_{i} = (S^1_i, \ldots, S^d_i)^\top\}_{i \in \mathcal{I}}$:
\begin{equation}\label{yixt}
    \mathbf{Y}_{i,\mathbf{x}} 
    = \boldsymbol{\mu}(\mathbf{x}) + \mathbf{S}_{i}(\mathbf{x}) + \boldsymbol{\epsilon}_{i,\mathbf{x}} 
    = \mathbf{X}_{i}(\mathbf{x}) + \boldsymbol{\epsilon}_{i,\mathbf{x}}.
\end{equation}
Here, $\boldsymbol{\mu}(\cdot) = (\mu^1(\cdot), \ldots, \mu^d(\cdot))^\top$ is the mean function. The unobserved variables $\{\boldsymbol{\epsilon}_{i,\mathbf{x}},\, i = 1, \ldots, n\}$ are independent and identically distributed with zero mean Gaussian measurement errors of variance $\sigma^2$. Contrary to usual scalar approach, we consider a $d$-tuple of $d_1$, ..., $d_d$ dimensional vector, given by $\mathbf{x}= (x_1, x_2,\ldots, x_d) \in \mathcal{X}$, addressing then the problem of regression of multivariate spatial functional data in different domains.

The $n$ multivariate functions $\mathbf{S}_{i}(\cdot)$ are centered square-integrable functional random variables indexed by the spatial domain $\mathcal{I}$. For each component $j=1,\ldots,d$, let $\mathcal{X}_j$ have finite Lebesgue measure, and let $S_i^j:\mathcal{X}_j\longrightarrow\mathbb{R}$ be a real-valued square-integrable function. We write $S_i^j=\{S_i^j(x_j):\,x_j\in\mathcal X_j\}$ and assume that $S_i^j\in\mathcal{L}^2(\mathcal{X}_j)$. Note that the special case $d=1,d_1=1$ corresponds to the univariate spatial-functional case \citep{hassan2021spatial}.

So, $\mathbf{S}_i$ is a multivariate functional random variable function of $\mathbf{x}=(x_1,\cdots,x_d) \in \mathcal{X}$ and taking values in the $d$-fold Cartesian product space $\mathcal{H}:= \mathcal{L}^2(\mathcal{X}_1)  \times \cdots \times 
\mathcal{L}^2(\mathcal{X}_d)$. Let the inner product $\langle\langle \cdot, \cdot \rangle\rangle: \mathcal{H} \times \mathcal{H} \to \mathbb{R}$, for $\mathbf{f}, \mathbf{g} \in \mathcal{H}$:
\begin{equation}
  \langle\langle \mathbf{f}, \mathbf{g} \rangle\rangle
    := \sum_{j=1}^{d} \langle f_j, g_j \rangle
     = \sum_{j=1}^{d} \int_{\mathcal{X}_j} f_j(t_j)\,\overline{g_j(t_j)}\,\dd t_j.
     \nonumber
\end{equation}
Then, $\mathcal{H}$ is a Hilbert space with respect to the scalar product $\langle\langle \cdot, \cdot \rangle\rangle$ \citep{happ2018multivariate}.

The focus is on a multivariate functional PCA investigation, wherein the classical PCA is substituted with its spatial counterpart to consider spatial autocorrelation on the functional variable of interest at the sampling locations. This autocorrelation may be quantified by a weight matrix depending on the neighboring locations. Additional univariate fasPCA illustrations are provided in the Supplementary Material. In those univariate examples, the functional argument is denoted by $t$, but it should be understood as a generic ordered functional domain rather than being restricted to time.

We postulate in the following a Karhunen-Lo\`eve expansion \citep{ash1975topics}:
\begin{equation}\label{sit}
    \mathbf{S}_{i}(\mathbf{x}) = \sum_{k=1}^{\infty} \beta_{k,i}\boldsymbol{\phi}_{k}(\mathbf{x}),
\end{equation}
where $\boldsymbol{\phi}_{k}$'s are the orthonormal eigenfunctions (functional principal components, FPC) and $\beta_{k,i}$ are auto-correlated scores (see \citet{happ2018multivariate} in the geostatistical case). In practice, the sum is truncated to a finite integer $K$, to be chosen.

Direct estimation of $\boldsymbol{\phi}_k$ is infeasible due to its infinite-dimensional nature. We therefore represent the functional sample data $(\mathbf{S}_{i})_{i=1,\ldots,n}$ by means of a basis expansion. For each fixed $j$, suppose the functional variables $S_i^j$ admit the following basis representation:
\begin{equation}\label{sit2}
    S^j_{i}(x_j)
    = \sum_{m=1}^{\infty} c^j_{i,m} B^j_{m}(x_j) 
    \approx \sum_{m=1}^{p_j} c^j_{i,m} B^j_{m}(x_j), 
    \quad x_j \in \mathcal{X}_j,
\end{equation}
where $\{B^j_m(\cdot)\}_{m\geq 1}$ denotes a collection of univariate orthonormal basis functions in $\mathcal{L}^2(\mathcal{X}_j)$, and $c^j_{i,m} = \langle S_i^j, B^j_m \rangle_{\mathcal{L}^2(\mathcal{X}_j)}$ are centered basis coefficients. In practice, only the first $p_j$ basis functions are retained, where $p_j$ is chosen sufficiently large to achieve good approximation.

In what follows, we concentrate on the classical setting in which the functions $B_m^j$ are chosen to be independent functional principal component basis functions. Nonetheless, alternative families of basis functions, such as Fourier bases which are particularly well suited for modeling periodic phenomena, or B-spline bases which are typically preferable for non-periodic or irregularly spaced data, can also be employed without loss of generality. Denote by $\{\mathbf{B}(\mathbf{x})\}_{\mathbf{x} \in \mathcal{X}}$ the matrix of basis function vectors indexed by $\mathbf{x} \in \mathcal{X}$:
\begingroup
\small
\begin{equation}
    \mathbf{B}(\mathbf{x}) = \begin{pmatrix}
    B_{1}^{1}(x_1) & \ldots & B_{p_1}^{1}(x_1)&  0 & \ldots& 0 & \ldots & & 0 \\
    0 & \ldots &0 & B_{1}^{2}(x_2) & \ldots& B_{p_2}^{2}(x_2)& 0&\ldots & 0 \\  
     & \ldots & &   & \ldots&    \\ 
     0 & \ldots &0 &  & \ldots& 0 &  B_{1}^{d}(x_d) & \ldots& B_{p_d}^{d}(x_d)
    \end{pmatrix} \in \mathbb{R}^{d \times p},
    \label{base_f}
\end{equation}
\endgroup
where $p=\sum_{j=1}^d p_j$. Let $\mathbf{c}^j_{i}=(c^j_{i,1}, \ldots, c^j_{i,p_j})^\top$ and $\mathbf{C}_i=((\mathbf{c}^{1}_{i})^\top, \ldots, (\mathbf{c}^{d}_{i})^\top)^\top = (c^1_{i,1}, \ldots, c^1_{i,p_1} , c^2_{i,1}, \ldots, c^2_{i,p_2},\ldots, c^d_{i,1}, \ldots, c^d_{i,p_d})^\top$ be respectively the vectors of the univariate basis coefficient expansion for each element $S^j_i$ of $\mathbf{S}_i$ and the concatenation of $\mathbf{c}^j_{i}$, then we can express $\mathbf{S}_i(\cdot)=(S^1_i(\cdot),\ldots,S^d_i(\cdot))^\top$ as $\mathbf{S}_i(\cdot)\approx \mathbf{B}(\cdot)\mathbf{C}_i$.

Note that the univariate basis functions $B^j_m$ and the corresponding coefficients $c^j_{i,m}$ can be derived from the multivariate basis expansion $\boldsymbol{\psi}_m = (\psi_{1,m}, \ldots, \psi_{d,m})^\top$ of the observed vector-valued functions $(\mathbf{S}_{i})_{i=1,\ldots,n}$, since the multivariate basis representation is equivalent to $d$ separate univariate basis expansions. A detailed description of this procedure is provided in \cite{happ2018multivariate}.

In this work, functional Moran's $I$ statistics and mfasPCA are developed as complementary components of a Moran-based spatial-functional framework. The functional Moran's $I$ statistics quantify spatial autocorrelation in multivariate functional areal data, whereas mfasPCA uses a Moran-type spatially weighted variance criterion to extract spatially structured functional principal components from the functional score matrix. Thus, functional Moran's $I$ supports the motivation, diagnosis, and interpretation of the spatial structure captured by mfasPCA.


\subsubsection{From multivariate functional Moran's \texorpdfstring{$I$}{I} 
to multivariate functional principal components for areal data}\label{sec:multivariate_functional}

The univariate form of the functional Moran's $I$, as previously mentioned, has been extensively covered in \citet{hassan2021spatial} and \citet{romano2022spatial}, where detailed derivations are presented. The well-known Moran's $I$ statistic has been generalized to the multivariate functional context. This generalization takes into account spatial dependence in PCA to evaluate the degree of spatial autocorrelation among observations within the spatial domain $\mathcal{I}$ \citep{jombart2008revealing}. Let $w_{ij}$ denote the raw spatial weight between locations $i$ and $j$. Throughout the methodology, $\mathbf{W}=(W_{ij})$ denotes the spatial weight matrix used in the corresponding calculation after the appropriate transformation has been applied. For the functional Moran's $I$ statistics, $\mathbf{W}$ is row-standardized, so that each non-empty row sums to one. For the mfasPCA eigenproblem and the associated eigenvalue-based permutation tests, $\mathbf{W}$ is first symmetrized by averaging $w_{ij}$ and $w_{ji}$, and then globally rescaled so that $\sum_i\sum_j W_{ij}=1$. The specific transformation applied to $\mathbf{W}$ is stated explicitly where it enters the corresponding formula.

The functional Moran's index for the $n$-vector $\{\mathbf{S}_i(\mathbf{x})\}_{i=1,\ldots,n}$ is then introduced as follows:
\begin{equation}\label{it}
{I}_{n}(\mathbf{S}(\mathbf{x}))
  = \frac{\sum_{i=1}^n \sum_{j=1}^n W_{ij}\, \mathbf{S}_{i}^{\top}(\mathbf{x}) \mathbf{S}_{j}(\mathbf{x})}
         {\sum_{i=1}^n \mathbf{S}_{i}^{\top}(\mathbf{x}) \mathbf{S}_{i}(\mathbf{x})}
  = \frac{{C}_{n}(\mathbf{S}(\mathbf{x}))}{{\sigma}_{n}(\mathbf{S}(\mathbf{x}))},
\end{equation}
where $\mathbf{W}$ denotes the row-standardized spatial weight matrix used for the
functional Moran's $I$ calculation,

\begin{align}
{C}_{n}(\mathbf{S}(\mathbf{x}))
&= \frac{1}{n}\sum_{i=1}^n \sum_{j=1}^n W_{ij}\, \mathbf{S}_{i}^{\top}(\mathbf{x}) \mathbf{S}_{j}(\mathbf{x}) \approx \frac{1}{n}\sum_{i=1}^n\sum_{j=1}^n
W_{ij} \mathbf{C}_i^\top\mathbf{B}(\mathbf{x})^\top \mathbf{B}(\mathbf{x}) \mathbf{C}_j,
\end{align}

\begin{equation}\label{cchapotapproximation}
\begin{aligned}
{\sigma}_{n}(\mathbf{S}(\mathbf{x})) 
&= \frac{1}{n} \sum_{i=1}^n \mathbf{S}_i^\top(\mathbf{x})\mathbf{S}_i(\mathbf{x}) \approx
\frac{1}{n}\sum_{i=1}^n
\mathbf{C}_i^\top
\mathbf{B}(\mathbf{x})^\top \mathbf{B}(\mathbf{x})
\mathbf{C}_i.
\end{aligned}
\end{equation}

The trace functional Moran's index is then introduced as
\begin{equation}\label{traceit}
    {I}_{n}(\mathbf{S})=\int_\mathcal{X} {I}_{n}(\mathbf{S}(\mathbf{x}))\,\dd x.
\end{equation}

The purpose of the proposed mfasPCA method is to identify the estimated multivariate functional eigenfunctions $\widehat{\boldsymbol{\phi}}_k(\cdot)$ associated with unit-norm vectors $\mathbf{u} \in \mathbb{R}^{p}$, with $\|\mathbf{u}\|=1$, such that the projected functional score vector $\boldsymbol{\chi}=\mathbf{C}\mathbf{u}$, where $\mathbf{C}$ is the stacked $n\times p$ functional score-coordinate matrix whose $i$-th row is $\mathbf{C}_i^\top$, exhibits spatially structured variation. In other words, this aims to find the extreme values \citep{jombart2008revealing} of
\begin{equation}\label{eq:mfaspca_objective}
\mathcal{C}(\mathbf{u})
= \frac{1}{n}\mathbf{u}^{\top}\mathbf{C}^{\top}\mathbf{W}\mathbf{C}\mathbf{u},
\end{equation}
where  $\mathbf{W}$ denotes the symmetrized and globally rescaled spatial weight matrix used for the mfasPCA eigenproblem. 
Thus, mfasPCA is defined by optimizing  a Moran-type spatially weighted variance criterion for the projected functional scores $\boldsymbol{\chi}$.

The solutions $\mathbf{u}_k$ (see \citet{jombart2008revealing} in the multivariate case) are the eigenvectors  of
\begin{equation}
  \mathbf{Z} = \mathbf{C}^\top \mathbf{W} \mathbf{C}. 
  \nonumber
\end{equation}

The eigenvectors are associated with the largest and smallest eigenvalues $\alpha_k$. These eigenvalues quantify a spatially weighted variance criterion for the component score vectors $\boldsymbol{\chi}_k=\mathbf{C}\mathbf{u}_k$. Since spatial autocorrelation may be positive or negative, some eigenvalues $\alpha_k$ may also be negative.

Using orthonormal vectors $\mathbf{u}_k$ and their eigenvalues $\alpha_{k}$, the estimated functional loading (eigen-function), $\widehat{\boldsymbol{\phi}}_k(\mathbf{x})$ can be derived. In fact, approximating $\mathbf{C}$ by
\begin{equation}
{\mathbf{C}}\approx \widehat{\mathbf{C}}= \sum_{k=1}^K \boldsymbol{\chi}_{k} \mathbf{u}_{k}^\top,
\nonumber
\end{equation}
based on $K$ sufficiently large relevant score vectors $\boldsymbol{\chi}_{k}$ corresponding to the $K$ largest eigenvalues in absolute value, leads to
\begin{equation}
    \mathbf{S}(\mathbf{x})\approx \widehat{\mathbf{C}}\mathbf{B}(\mathbf{x})^\top 
    = \sum_{k=1}^K \boldsymbol{\chi}_{k} \mathbf{u}_{k}^\top \mathbf{B}(\mathbf{x})^\top.
    \nonumber
\end{equation}
The functional multivariate spatial PCA is then obtained by defining the estimated eigenfunctions as $\widehat{\boldsymbol{\phi}}_k(\mathbf{x})^\top=\mathbf{u}_{k}^\top \mathbf{B}(\mathbf{x})^\top$.

Using the mapped multivariate functional eigenfunctions, the mfasPCA approximation can be written as
\begin{equation}\label{sitapproximation}
	\mathbf{S}_{i}(\mathbf{x}) \approx \sum_{k=1}^K \hat\beta_{k,i}\widehat{\boldsymbol{\phi}}_k(\mathbf{x}),
\end{equation}
\begin{equation}\label{xitapproximate}
	\mathbf{X}_{i}(\mathbf{x}) \approx \widehat{\boldsymbol{\mu}}(\mathbf{x}) + \sum_{k=1}^K \hat\beta_{k,i}\widehat{\boldsymbol{\phi}}_k(\mathbf{x}),\; 
\end{equation}
where $\widehat{\boldsymbol{\mu}}(\mathbf{x})=\frac{1}{n}\sum_{i=1}^n \mathbf{X}_{i}(\mathbf{x})$ is the empirical mean with $\hat\beta_{k,i}= \langle \langle  \mathbf{S}_{i}, \widehat{\boldsymbol{\phi}}_k \rangle\rangle$.
Equivalently, $\hat\beta_{k,i}$ is the $i$-th entry of the mfasPCA score vector $\hat{\boldsymbol{\chi}}_k = \hat{\mathbf{C}}\mathbf{u}_k$.

The principal component (PC) scores derived from mfasPCA exhibit two distinct types of patterns, classified as global and local structures \citep{jombart2008revealing}. The global pattern distinguishes between two spatial groups or illustrates a cline (or any intermediate state), whereas the local pattern captures stronger differentiation among neighboring entities compared to random pairs \citep{jombart2008revealing}. The global pattern is indicative of positive spatial autocorrelation, while the local pattern signifies negative spatial autocorrelation \citep{jombart2008revealing}.

Note that ${I}_{n}(\mathbf{S}(\mathbf{x}))$ does not take into account the interrelation between the measurements of two distinct components of $\mathbf{S}_i$ at the same spatial location $\mathbf{s}_i$ \citep{eckardt2021partial}. To overcome this limitation, we extend the bivariate Moran's $I$ statistic of \citep{eckardt2021partial} to the functional case:
\begin{equation}\label{itp}
I_{kl}(\mathbf{S}(\mathbf{x}))
  = \frac{\sum_{i=1}^n \sum_{j=1}^n W_{ij}\, S_i^{k}(\mathbf{x})\, S_j^{l}(\mathbf{x})}
         {\sqrt{\sum_{i=1}^n S_i^{k}(\mathbf{x})^2}\, \sqrt{\sum_{i=1}^n S_i^{l}(\mathbf{x})^2}},
  \quad k,l = 1,\ldots,d,
\end{equation}
where $\mathbf{W}$ denotes the row-standardized spatial weight matrix used for the functional Moran's $I$ calculation.

Taken together, the multivariate functional Moran's indices and the mfasPCA decomposition provide a spatial-functional framework for multivariate areal data.


\subsection{Spatial weight matrices for mfasPCA and functional Moran's $I$}\label{sec:weights}

We evaluate functional Moran's $I$ and mfasPCA using two common classes of spatial weights: distance-based weights, which encode proximity between areal centroids, and contiguity-based weights, which encode shared boundaries between neighbouring regions. These two choices also allow direct comparison with the best-performing specifications reported by \citet{krzysko2024stpca} for STPCA. Let $d_{ij}$ be the centroid-to-centroid distance, and let $b_{ij}$ indicate whether regions $i$ and $j$ share a common boundary.

\paragraph{Row standardization, symmetrization, and global rescaling}
Starting from raw spatial weights $w_{ij}$, two transformed versions of the spatial weight matrix are used. For functional Moran's $I$, $\mathbf{W}$ denotes the row-standardized spatial weight matrix with entries
\begin{equation}
  W_{ij}
  =
  \begin{cases}
      \dfrac{w_{ij}}{\sum_k w_{ik}}, & \text{if } \sum_k w_{ik} > 0,\\[0.8ex]
      0,                             & \text{otherwise.}
    \end{cases}
    \nonumber
\end{equation}
Thus, each non-empty row of $\mathbf{W}$ sums to one.

For the mfasPCA eigenproblem and the eigenvalue-based permutation tests, $\mathbf{W}$ denotes the symmetrized and globally rescaled spatial weight matrix. This is obtained by first defining
\begin{equation}
  \widetilde w_{ij}
  =
  \frac{1}{2}(w_{ij}+w_{ji}),
  \nonumber
\end{equation}
and then setting
\begin{equation}
  W_{ij}
  =
  \frac{\widetilde w_{ij}}
       {\sum_i\sum_j \widetilde w_{ij}},
  \nonumber
\end{equation}
so that $\mathbf{W}$ is symmetric and $\sum_i\sum_j W_{ij}=1$. Throughout the formulas, the relevant transformation of $\mathbf{W}$ is stated according to whether functional Moran's $I$ or mfasPCA is being computed.

\paragraph{Radial distance (distance-based)}
Two regions are neighbors if their centroid-to-centroid distance is at or below a fixed radius $r$. We first define raw binary weights by setting $w_{ij}=1$ for neighboring regions and $w_{ij}=0$ otherwise. These raw weights are then transformed as described above: $\mathbf{W}$ is row-standardized for functional Moran's $I$, and $\mathbf{W}$ is symmetrized and globally rescaled for the mfasPCA eigenproblem and eigenvalue-based permutation tests.

\paragraph{Shared-boundary contiguity (contiguity-based)}
Two regions are treated as neighbours if they share a common boundary. We define the raw contiguity weights by
\begin{equation}
  w_{ij}
  =
  \begin{cases}
  1, & \text{if regions } i \text{ and } j \text{ share a boundary},\\
  0, & \text{otherwise},
  \end{cases}
  \qquad w_{ii}=0.
  \nonumber
\end{equation}
These raw binary contiguity weights are then transformed as described above: $\mathbf{W}$ is row-standardized for functional Moran's $I$, and $\mathbf{W}$ is symmetrized and globally rescaled for the mfasPCA eigenproblem and eigenvalue-based permutation tests.

To assess spatial autocorrelation in the patterns identified by mfasPCA, we use eigenvalue-based permutation tests adapted to the functional score-matrix setting.


\subsection{Eigenvalue-based permutation tests for mfasPCA}\label{sec:hyp-mfaspca}

To assess the spatial-functional principal components identified by mfasPCA, we use eigenvalue-based permutation tests. The tests are based on the eigenvalues of
\begin{equation}
  \mathbf{Z} = \mathbf{C}^\top \mathbf{W} \mathbf{C},
  \nonumber
\end{equation}
where $\mathbf{W}$ denotes the symmetrized and globally rescaled spatial weight matrix used for the mfasPCA eigenproblem. Since the spectrum of $\mathbf{Z}$ may contain both positive and negative eigenvalues, the positive and negative parts of the spectrum are treated separately.

Let
\begin{equation}
  \alpha_1^+ \ge \cdots \ge \alpha_{r^+}^+ > 0
  \nonumber
\end{equation}
denote the ordered positive spatial eigenvalues of $\mathbf{Z}$. For the negative spectrum, we order the magnitudes of the negative eigenvalues $\{|\alpha|:\alpha<0\}$ as
\begin{equation}
  \alpha_1^- \ge \cdots \ge \alpha_{r^-}^- > 0,
  \nonumber
\end{equation}
where each $\alpha_k^- := |\tilde{\alpha}_k|$ is the absolute value of a negative eigenvalue $\tilde{\alpha}_k < 0$. Let $A_k^+$ and $A_k^-$ denote the corresponding population spatial eigenvalues.

We consider three complementary families of tests: omnibus tests, componentwise per-eigen tests, and sequential tests. For the positive spectrum, the sequential hypotheses are
\begin{equation}
  H_{0i}^+:\ A_i^+ = A_{i+1}^+ = \cdots = A_{r^+}^+ = 0,
  \qquad i = 1,\ldots,r^+ .
  \nonumber
\end{equation}
Thus, $H_{0i}^+$ tests whether there is no remaining positive spatial signal from the $i$-th positive component onward. The corresponding tail-sum statistic is
\begin{equation}
  T_i^+ = \sum_{j=i}^{r^+} \alpha_j^+ .
  \nonumber
\end{equation}

A parallel family is defined for the negative spectrum:
\begin{equation}
  H_{0i}^-:\ A_i^- = A_{i+1}^- = \cdots = A_{r^-}^- = 0,
  \qquad i = 1,\ldots,r^- ,
  \nonumber
\end{equation}
with test statistic
\begin{equation}
  T_i^- = \sum_{j=i}^{r^-} \alpha_j^- .
  \nonumber
\end{equation}
Here, $H_{0i}^-$ tests whether there is no remaining negative, or local, spatial signal from the $i$-th negative component onward.

For the componentwise per-eigen tests, each eigencomponent is tested separately. For the positive spectrum, we test
\begin{equation}
  H_{0j}^+:\ A_j^+ = 0
  \nonumber
\end{equation}
using the observed positive eigenvalue $\alpha_j^+$ as the test statistic. For the negative spectrum, we test
\begin{equation}
  H_{0j}^-:\ A_j^- = 0
  \nonumber
\end{equation}
using the ordered magnitude $\alpha_j^-$ of the corresponding negative eigenvalue as the test statistic. Holm's step-down adjustment is applied separately within the positive and negative families of componentwise tests to control the family-wise error rate within each sign-specific family.

To avoid a full refitting of the smoothing/FPCA representation for each permuted dataset, the permutation test is implemented using the score-matrix formulation of the mfasPCA eigenproblem. After obtaining the fitted functional score-coordinate matrix $\mathbf{C}$ and the processed spatial weight matrix $\mathbf{W}$, permutation $p$-values are computed by randomly permuting the $n$ region labels consistently across all functional variables. Equivalently, the same permutation is applied to the rows of $\mathbf{C}$. For the $b$-th permutation, we compute
\begin{equation}
  \mathbf{Z}^{(b)}
  = (\mathbf{P}_b \mathbf{C})^\top \mathbf{W}(\mathbf{P}_b \mathbf{C}),
  \nonumber
\end{equation}
where $\mathbf{P}_b$ is the corresponding permutation matrix and $\mathbf{W}$, the symmetrized and globally rescaled spatial weight matrix, is kept fixed.

In this score-matrix implementation, the retained univariate FPC bases and the resulting score-coordinate matrix $\mathbf{C}$ are treated as fixed after fitting the observed data; the permutation step randomizes the assignment of the fitted score rows to regions while keeping the processed spatial weight matrix $\mathbf{W}$ fixed.

The full eigenvalue spectrum of $\mathbf{Z}^{(b)}$ is then obtained. The positive and negative spectra are then separated, and the corresponding statistics $T_i^+$, $T_i^-$, $\alpha_j^+$, and $\alpha_j^-$ are recomputed. Greater-tail permutation $p$-values are computed by comparing the observed statistics with their permutation distributions. For the negative spectrum, the comparison is based on the magnitudes of the negative eigenvalues.

Finally, omnibus tests are used to assess whether there is any spatial signal in the positive or negative subspace. The positive omnibus statistic is
\begin{equation}
  M^+ = T_1^+ = \sum_{j=1}^{r^+} \alpha_j^+,
  \nonumber
\end{equation}
and the negative omnibus statistic is
\begin{equation}
  M^- = T_1^- = \sum_{j=1}^{r^-} \alpha_j^- .
  \nonumber
\end{equation}
Permutation $p$-values for $M^+$ and $M^-$ are computed analogously.


\subsection{Implementation summary}\label{sec:impl-summary}

The following steps summarize how to run mfasPCA, compute functional Moran's $I$, and apply the eigenvalue-based permutation tests of Sec.~\ref{sec:hyp-mfaspca}, using the definitions in Secs.~\ref{sec:multivariate_functional}--\ref{sec:weights}.

\begin{algorithm}[!htbp]
\caption{mfasPCA: estimation and testing pipeline}\label{alg:mfaspca}
\begin{algorithmic}[1]
\vspace{0.2em}
\Input{The raw data $\{\mathbf{Y}_{i,\mathbf{x}}\}_{i=1}^n$ with $d$ variables; spatial weights $\mathbf{W}$.}

\vspace{0.3em}

\Output{Eigenfunctions $\{\widehat{\boldsymbol{\phi}}_k\}$, loadings $\{\mathbf{u}_k\}$, score vectors $\{\boldsymbol{\chi}_k\}$; sign-split eigenvalues $\{\alpha_k^+\}$ and $\{\alpha_k^-\}$; permutation $p$-values.}

\vspace{0.6em}

\State \textbf{Basis/smoothing.}
For each variable, perform a preliminary univariate FPCA/smoothing step and retain the resulting univariate FPC scores. Stack these retained score-coordinate vectors across variables to form $\mathbf{C}$.

\vspace{0.45em}

\State \textbf{Spatial weights.}
Construct $\mathbf{W}$ from the raw spatial weights according to the calculation being performed: row-standardization for functional Moran's $I$, and symmetrization followed by global rescaling for the mfasPCA eigenproblem and eigenvalue-based permutation tests, as described in Sec.~\ref{sec:weights}.

\vspace{0.45em}

\State \textbf{Core matrix and eigenpairs.}
Compute $\mathbf{Z}=\mathbf{C}^\top \mathbf{W}\mathbf{C}$, where $\mathbf{W}$ denotes the symmetrized and globally rescaled spatial weight matrix used for the mfasPCA eigenproblem, and obtain its eigenpairs $(\alpha_k,\mathbf{u}_k)$. Equivalently, the eigenvectors maximize or minimize $\mathcal{C}(\mathbf{u})=n^{-1}\mathbf{u}^\top\mathbf{Z}\mathbf{u}$, subject to $\|\mathbf{u}\|=1$.

\vspace{0.45em}

\State \textbf{Scores and functional loadings.}
Set score vectors $\boldsymbol{\chi}_k=\mathbf{C}\mathbf{u}_k$; define $\widehat{\boldsymbol{\phi}}_k$ and, if needed, $\hat\beta_{k,i}$ through Eqs.~\eqref{sitapproximation}--\eqref{xitapproximate}.

\vspace{0.45em}

\State \textbf{Sign split.}
Partition the eigenvalues into the ordered positive spectrum $\{\alpha_k^+\}$ and the ordered magnitudes of the negative spectrum $\{\alpha_k^-\}$, corresponding to global and local spatial structures, respectively.

\vspace{0.45em}

\State \textbf{Reporting and tests.}
Summarize sign-specific CPVE within each positive or negative eigenspace, map and interpret score signs, and apply the omnibus, componentwise per-eigen, and sequential eigenvalue-based permutation tests of Sec.~\ref{sec:hyp-mfaspca} to $\{\alpha_k^+\}$, $\{\alpha_k^-\}$, $T_i^+$, and $T_i^-$.

\vspace{0.2em}
\end{algorithmic}
\end{algorithm}

\noindent We quantify spatial autocorrelation of the smoothed curves using the functional Moran's indices in Eqs.~\eqref{it}--\eqref{traceit}; for pairs of variables we also report the bivariate version in Eq.~\eqref{itp}, using $\mathbf{W}$, where $\mathbf{W}$ is row-standardized for the functional Moran's indices, and assess significance using the permutation framework in Sec.~\ref{sec:hyp-mfaspca}.

\noindent\textbf{Implementation in R.}
In practice, we implement mfasPCA and the associated permutation tests in R using the packages \texttt{fda} \citep{ramsay2005functional,ramsay2025fda}, 
\texttt{adegenet} \citep{jombart2008adegenet},
\texttt{ade4} \citep{chessel2004ade4,dray2007ade4,dray2007ade4a,bougeard2018supervised},
\texttt{adespatial} \citep{dray2019adespatial},
and the spatial infrastructure provided by \texttt{spdep} and related packages
\citep{R-spdep-test2018,R-spdep-gean2022,R-asdar-2013,R-sf-book}.


\subsection{Evaluation framework}\label{sec:evaluation-framework}

We evaluate mfasPCA using both descriptive and inferential criteria. The descriptive assessment is based on sign-specific CPVE, defined as the cumulative percentage of spatially weighted variance explained within the positive or negative eigenspace. For the negative eigenspace, the percentages are computed using the magnitudes of the negative eigenvalues. Spatial signal assessment is based on the eigenvalue-based permutation tests described in Section~\ref{sec:hyp-mfaspca}. The simulation and real-data settings are summarized in Table~\ref{tab:eval_metrics}.

\begin{table}[!htbp]
\caption{Evaluation criteria and testing settings for the simulation study and real-data application}\label{tab:eval_metrics}
\begin{tabular*}{\textwidth}{@{\extracolsep{\fill}}
>{\raggedright\arraybackslash}p{0.22\textwidth}
>{\raggedright\arraybackslash}p{0.34\textwidth}
>{\raggedright\arraybackslash}p{0.34\textwidth}
@{}}
\toprule
& Simulation study & Real-data application \\
\midrule
Sign-specific CPVE targets / reporting
& Targets: sign-specific CPVE $(+)$ = 90\%, $(-)$ = 70\%. The sign-specific dimensions $(K^+,K^-)$ are selected from the mfasPCA spectrum once per $\mathbf{W}$ and then used for both mfasPCA and STPCA sign-specific CPVE summaries, giving a matched-dimension descriptive comparison across $\rho$.
& Report sign-specific CPVE (\%) for the first two $(+)$ PCs and the first two $(-)$ PCs. \\
\addlinespace[1.5em]
& Positive/negative through eigenvalue sums $M^+=T_1^+$, $M^-=T_1^-$; permutation $p$-values (999/run).
& Positive/negative through eigenvalue sums; permutation $p$-values (9999/method). \\
\addlinespace[1.5em]
Componentwise (per-eigen)
& Holm FWER within sign over Top-20.\footnotemark[1]
& Holm FWER within-sign over all retained PCs. \\
\addlinespace[1.5em]
Sequential test
& Tail-sum $T_i^\pm$ with Holm step-down within sign (Top-20 cap).
& Tail-sum $T_i^{\pm}$ with Holm step-down within sign over all retained PCs. \\
\botrule
\end{tabular*}
\footnotetext{Note: Holm FWER refers to Holm family-wise error rate.}
\footnotetext[1]{``Top-20'' = largest $|\alpha|$ by sign; if fewer exist, all are used.}
\end{table}


\section{Simulation study}\label{sec:sim-study}

We compare the proposed mfasPCA with STPCA as a related spatially weighted reference approach on simulated multivariate functional areal data, using the metrics and testing settings in Table~\ref{tab:eval_metrics}. For comparability with STPCA, the functional argument is denoted by $t$, and the simulated curves are treated as temporally indexed trajectories; within mfasPCA, $t$ plays the role of the functional argument. The simulations are run on the French departmental lattice under two regimes: positive spatial autocorrelation (radial distance weights, $\rho>0$) and negative spatial autocorrelation (shared-boundary weights, $\rho<0$). These choices are aligned with the simulation settings reported by \citet{krzysko2024stpca}, who found distance-based weights to be favourable for positive spatial autocorrelation and contiguity-based weights to be favourable for negative spatial autocorrelation in their STPCA framework. 

In our study, mfasPCA is evaluated under all nine spatial weight matrices considered by \citet{krzysko2024stpca}. The radial and shared-boundary cases are reported here as representative examples. For each regime, we compare mfasPCA and STPCA descriptively in terms of (i) sign-specific CPVE for their leading components, (ii) the performance of eigenvalue-based permutation tests in detecting spatial signal in the positive or negative subspace, and (iii) the behavior of functional Moran's $I$ curves over the functional domain. The purpose is to examine how the two spatially weighted criteria behave under controlled spatial autocorrelation settings, rather than to claim a universal ranking between the methods.

Multivariate FPCA, as developed by \citet{happ2018multivariate}, is not included as a direct simulation benchmark because it addresses a non-spatial dimension-reduction objective, whereas mfasPCA targets components whose functional scores are spatially structured, as formalized by the Moran-type spatially weighted criterion in Eq.~\eqref{eq:mfaspca_objective}. For this reason, the simulation comparison is restricted to mfasPCA and STPCA as two spatially weighted dimension-reduction approaches, while recognizing that they are formulated through different representations.


\subsection{Multivariate areal data-generating process}

We consider a lattice of areal units corresponding to the second-level administrative divisions of mainland France (Figure~\ref{french_map}). The spatial data were obtained through \texttt{geodata::gadm} \citep{R-geodata}, processed using the \texttt{raster} package in R \citep{R-raster}, and further processed with \texttt{sf} \citep{R-sf-article,R-sf-book} and \texttt{terra} \citep{R-terra}.

We analyze 94 mainland departments in metropolitan France, excluding Corsica and the five overseas departments: Guadeloupe, Martinique, French Guiana, Réunion, and Mayotte. The polygons are reprojected to Lambert-93 (EPSG:2154) using \texttt{sf} \citep{R-sf-article} and \texttt{terra} \citep{R-terra}. For distance-based connectivity, we use radial weights with a 120\,km cutoff based on Lambert-93 centroid-to-centroid distances. Centroids are computed with \texttt{sf::st\_centroid}, and centroid-to-centroid planar distances are computed in meters and reported in kilometers with \texttt{units} \citep{R-units}. Distance-based weights use these distances, and contiguity (rook/shared-boundary) weights are derived from polygon boundaries through \texttt{spdep} \citep{R-asdar-2013,R-spdep-test2018,R-spdep-gean2022}.

Subsequently, the data are generated according to the following model:
\begin{equation}
X_{i}^j(t)=t\alpha_{i}^j+u_{i}^j(t),\qquad t\in \mathcal{X}_j=[0,1],
\nonumber
\end{equation}
where
\begin{equation}
\alpha_{i}^j\sim\mathcal{U}(-3,3),\qquad j=1,\ldots,d.
\nonumber
\end{equation}
Here, $\{u_{i}^j(t)\}$ is a Gaussian process with exponential covariance, where $i\in\{1,\ldots,n\}$ indexes spatial locations $\mathbf{s}_i$ on the French departmental lattice. The curves $X_i^j$ are observed at $T=101$ evenly spaced time points on $[0,1]$.

For each variable $j$, spatial dependence is introduced by applying a spatial autoregressive transformation to the latent functional signal:
\begin{equation}\label{eqrho2}
    \widetilde{\mathbf{Y}}^{j}(t)
    = (\mathbf{I}_n-\rho \mathbf{W})^{-1}\mathbf{X}^{j}(t),
    \qquad t\in[0,1],\quad j=1,\ldots,d,
\end{equation}
where $\mathbf{X}^{j}(t)=(X_1^j(t),\ldots,X_n^j(t))^\top$, $\widetilde{\mathbf{Y}}^{j}(t)=(\widetilde Y_1^j(t),\ldots,\widetilde Y_n^j(t))^\top$, and $\mathbf{W}$ is the row-standardized spatial weight matrix used in the data-generating process. The observed curves are then obtained by adding independent measurement error:
\begin{equation}\label{eqrho2_obs}
    \mathbf{Y}^{j}(t)
    = \widetilde{\mathbf{Y}}^{j}(t)+\boldsymbol{\varepsilon}^{j}(t),
    \qquad t\in[0,1],\quad j=1,\ldots,d.
\end{equation}
Here, $\mathbf{Y}^{j}(t)=(Y_1^j(t),\ldots,Y_n^j(t))^\top$ and $\boldsymbol{\varepsilon}^{j}(t)=(\varepsilon_1^j(t),\ldots,\varepsilon_n^j(t))^\top$. The $\varepsilon_i^j$ are independent centered Gaussian error processes with
\begin{equation}
\operatorname{var}\{\varepsilon_i^j(t)\}=\sigma^2,\qquad
\operatorname{cov}\{\varepsilon_i^j(t),\varepsilon_i^j(u)\}=0,\quad t\neq u.
\nonumber
\end{equation}
The raw spatial weight matrix follows the nine constructions of \citet{krzysko2024stpca} and is row-standardized before being used in the data-generating process. In our simulations, we set $d=10$ variables and observe $T=101$ time points.


\subsection{Simulation results}\label{sec:sim-results}

For comparability with \citet{krzysko2024stpca}, we focus on global (positive) components with distance-based weights and local (negative) components with shared-boundary weights. Accordingly, we report results for global components under radial distance weights with $\rho \in \{0.3, 0.5, 0.7, 0.9\}$ and for local components under shared-boundary weights with $\rho \in \{-0.3, -0.5, -0.7, -0.9\}$.

The sign-specific component counts $K^+$ and $K^-$ were calibrated separately for the positive and negative regimes. Under radial-distance weights, $K^+$ was selected from a single calibration replicate at $\rho = 0.9$ as the smallest number of positive components whose cumulative sum of positive Moran-type eigenvalues reached 90\% of the total positive eigenvalue mass, yielding $K^+ = 5$. Under shared-boundary weights, $K^-$ was selected analogously from a single calibration replicate at $\rho = -0.9$ using a 70\% threshold on the magnitudes of the negative Moran-type eigenvalues, yielding $K^- = 6$. Within each weight regime, the calibrated value of $K$ was then held fixed across all $\rho$ values. Since the calibration is performed on the mfasPCA spectrum, $\mathbf{C}^\top \mathbf{W}\mathbf{C}$, the matched-dimension comparison is anchored on the mfasPCA sign-specific CPVE scale; an analogous STPCA-anchored calibration based on the STPCA coefficient-based eigenspectrum need not yield the same $K^+$ and $K^-$ values. The lower threshold for the negative subspace, 70\% rather than 90\%, was used to provide a more parsimonious summary of the empirically more diffuse local spatial structure; increasing the threshold retained additional weak components without materially changing the interpretation. 

\begin{figure}[ht!]
  \centering

  \begin{subfigure}{0.23\textwidth}
    \includegraphics[width=\textwidth,height=5cm]{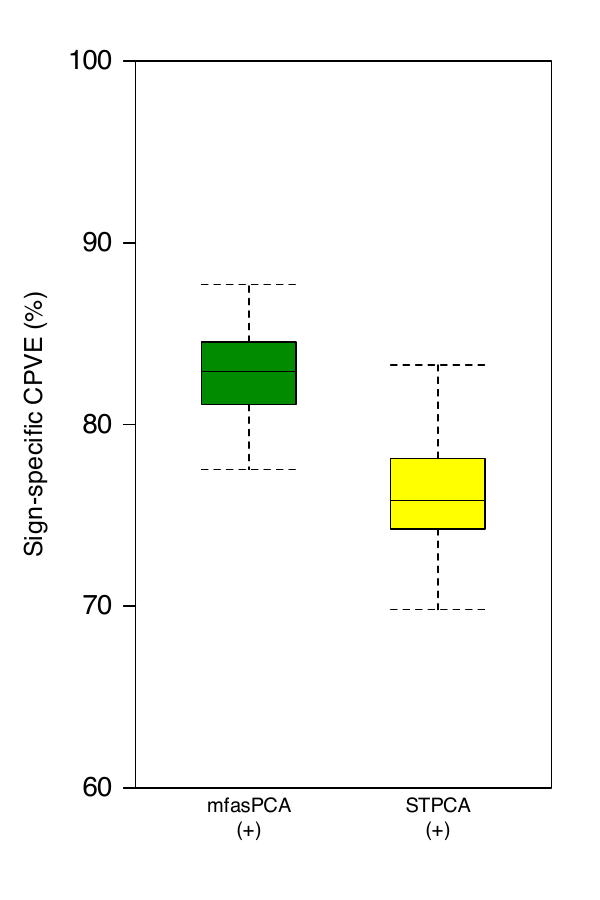}
    \caption{$\rho=0.3$}
  \end{subfigure}
  \hfill
  \begin{subfigure}{0.23\textwidth}
    \includegraphics[width=\textwidth,height=5cm]{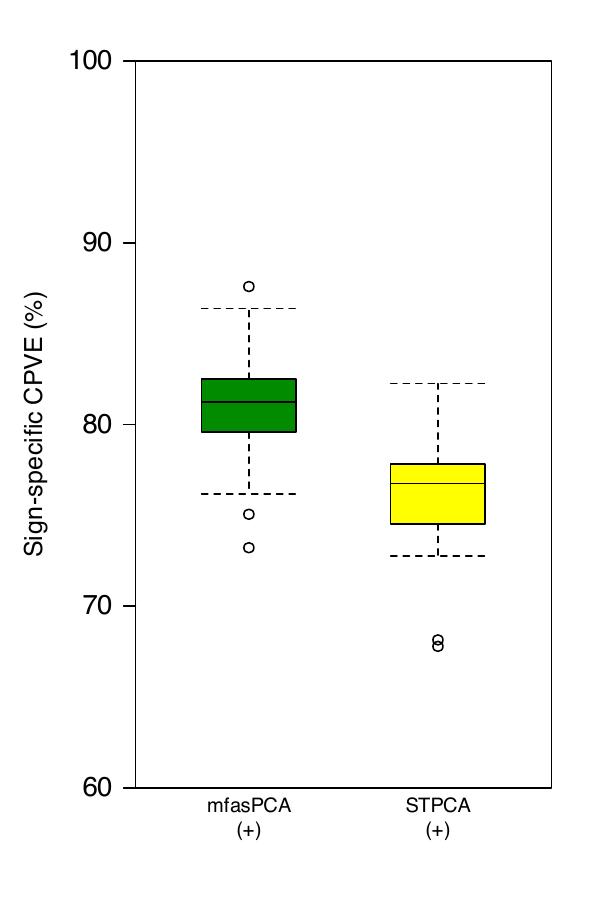}
    \caption{$\rho=0.5$}
  \end{subfigure}
  \hfill
  \begin{subfigure}{0.23\textwidth}
    \includegraphics[width=\textwidth,height=5cm]{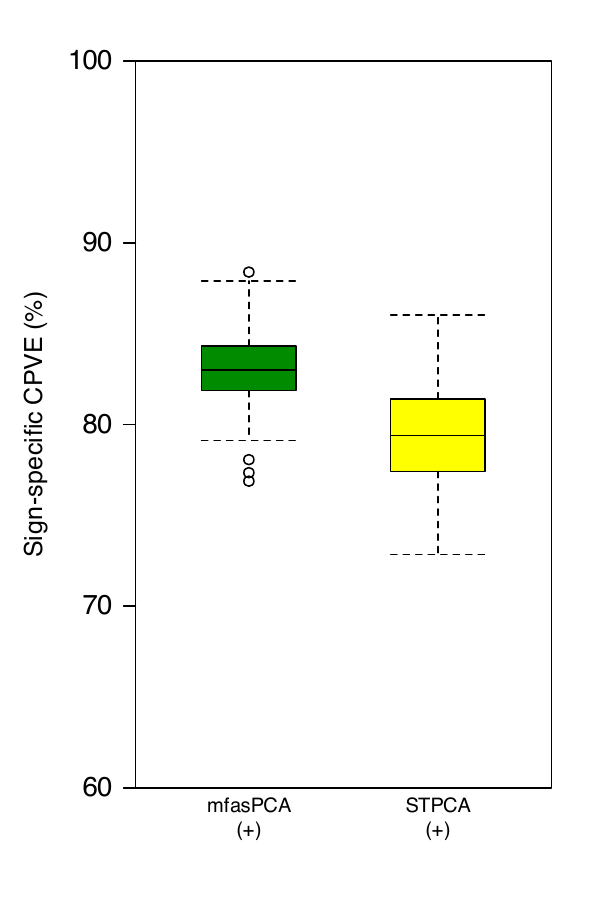}
    \caption{$\rho=0.7$}
  \end{subfigure}
  \hfill
  \begin{subfigure}{0.23\textwidth}
    \includegraphics[width=\textwidth,height=5cm]{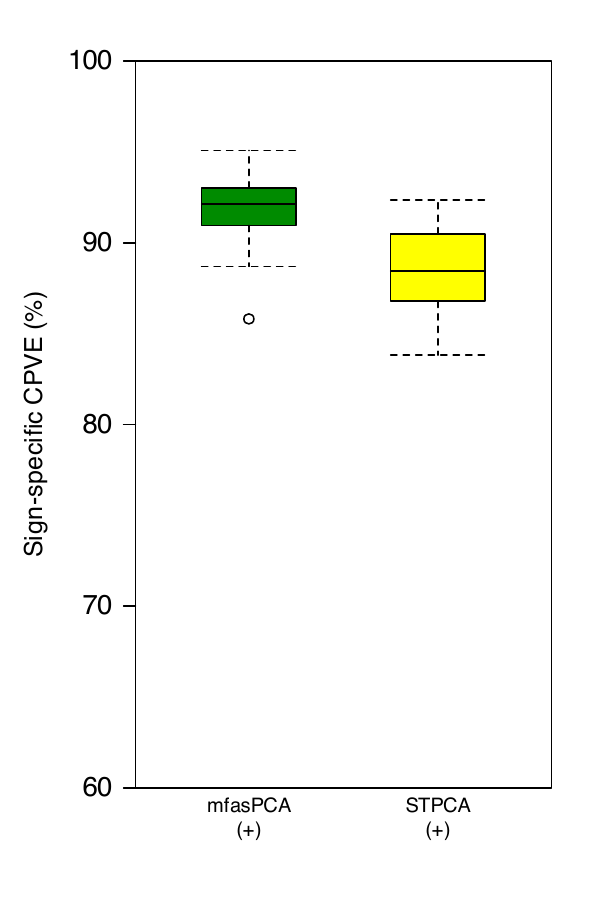}
    \caption{$\rho=0.9$}
  \end{subfigure}

\caption[Sign-specific CPVE under radial weights]{%
   {Sign-specific CPVE for mfasPCA and STPCA across 50 simulated datasets for components associated with positive eigenvalues, using radial distance weights ($r=120$ km; top five components).}%
  }
  \label{simu3a}
\end{figure}

\begin{figure}[ht!]
  \centering

  \begin{subfigure}{0.23\textwidth}
    \includegraphics[width=\textwidth,height=5cm]{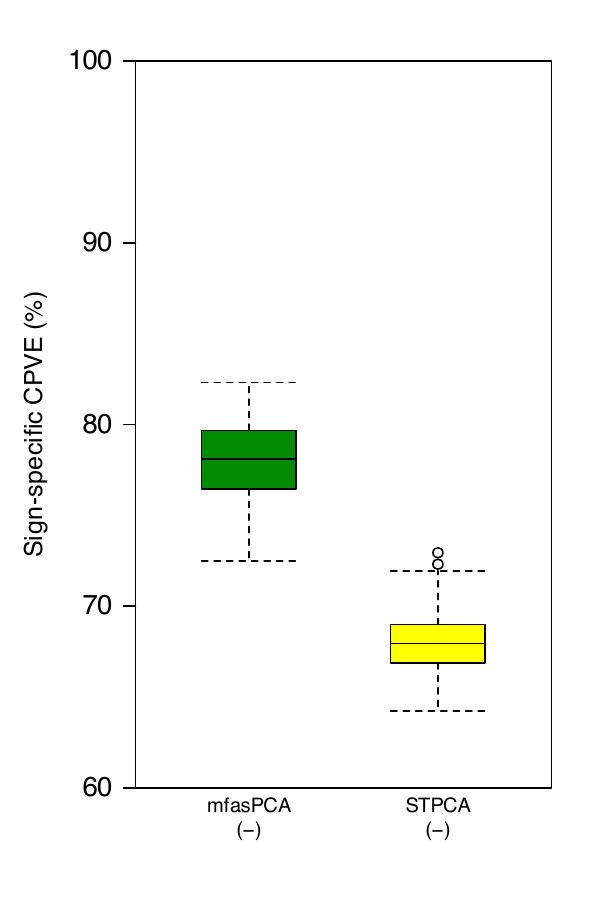}
    \caption{$\rho = -0.3$}
  \end{subfigure}
  \hfill
  \begin{subfigure}{0.23\textwidth}
    \includegraphics[width=\textwidth,height=5cm]{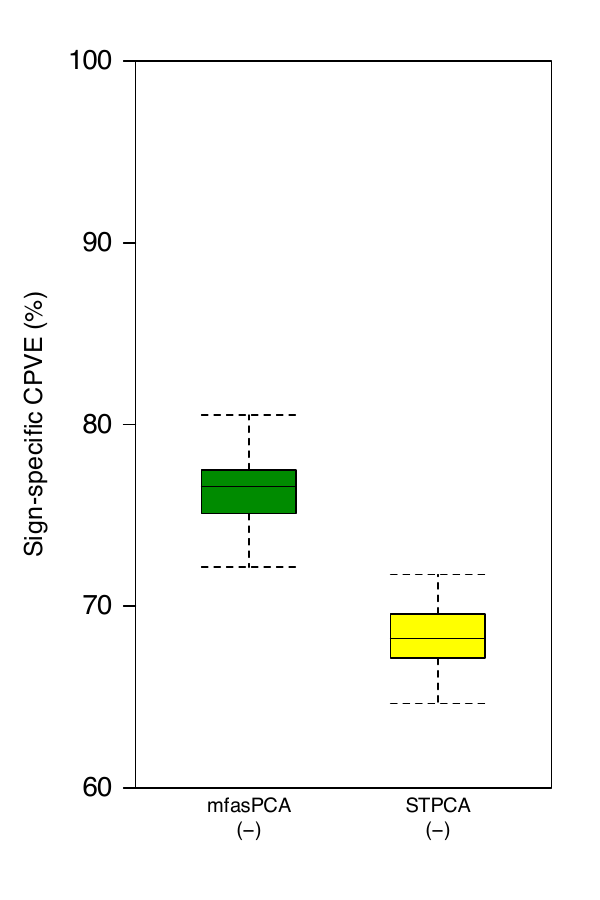}
    \caption{$\rho = -0.5$}
  \end{subfigure}
  \hfill
  \begin{subfigure}{0.23\textwidth}
    \includegraphics[width=\textwidth,height=5cm]{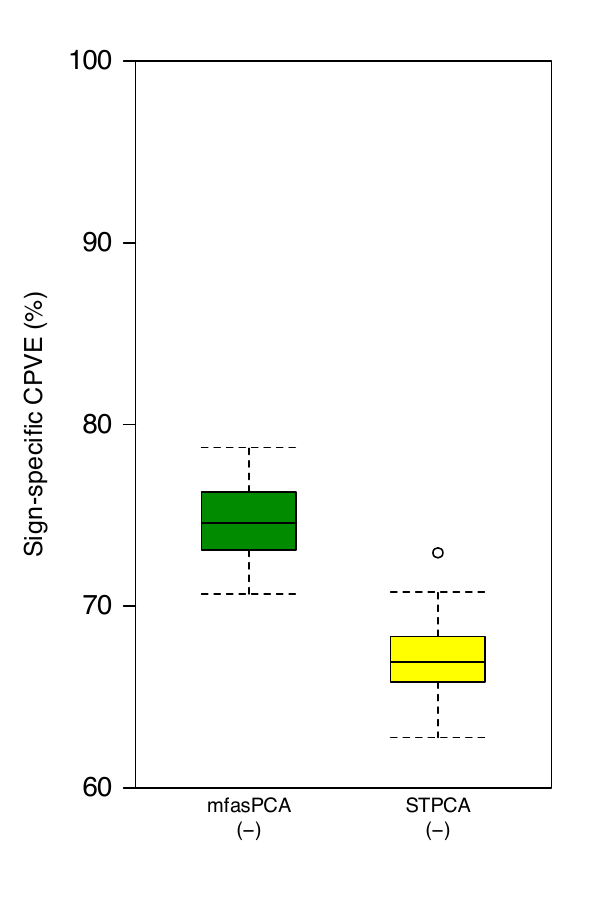}
    \caption{$\rho = -0.7$}
  \end{subfigure}
  \hfill
  \begin{subfigure}{0.23\textwidth}
    \includegraphics[width=\textwidth,height=5cm]{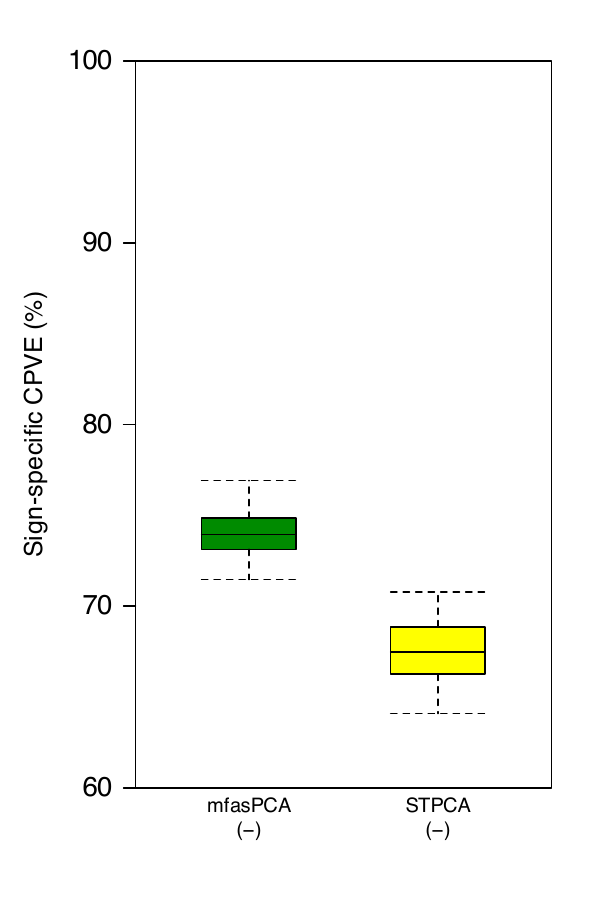}
    \caption{$\rho = -0.9$}
  \end{subfigure}

\caption[Sign-specific CPVE under shared-boundary weights]{%
    {Sign-specific CPVE for mfasPCA and STPCA across 50 simulated datasets for components associated with negative eigenvalues, using shared-boundary weights (top six components).}
  }
  \label{simu3b}
\end{figure}

For the STPCA comparison, the sign-specific CPVE summaries are computed using the same sign-specific component counts, giving a matched-dimension descriptive comparison rather than a separately optimized retention rule for each method. Across 50 simulations, the retained mfasPCA components achieve higher sign-specific CPVE than the corresponding STPCA components in the reported settings: the positive subspace under radial-distance weights (Figure~\ref{simu3a}) and the negative subspace under shared-boundary weights (Figure~\ref{simu3b}).

We compare the two regimes at matched absolute values of the spatial-dependence parameter, using $\rho=0.7$ and $\rho=-0.7$, representing mid-to-strong spatial dependence within the simulation grid. For STPCA, the analogous eigenvalue-based tests were applied to its coefficient-based eigenspectrum, while the mfasPCA tests were applied to the mfasPCA score-matrix eigenspectrum. Both testing procedures were implemented with 999 permutations. 

Both methods show significant omnibus spatial dependence in the relevant subspace (positive at $\rho=0.7$ and negative at $\rho=-0.7$; $p=0.001$ in both cases; see Table~\ref{tab:perm_rho07}). Sequential tests (Holm, top-20 per sign) give similar conclusions for mfasPCA and STPCA, detecting positive components and no negative components when $\rho = 0.7$ under radial distance weights, and detecting negative components and no positive components when $\rho = -0.7$ under shared-boundary weights. Thus, the sequential detections are confined to the signal-bearing side in each regime. The componentwise Holm tests also reject, on average, all 20 inspected components in the corresponding signal-bearing subspace for both methods. 

The top-20 cap used in the eigenvalue-based tests is distinct from the sign-specific CPVE dimensions $K^+=5$ and $K^-=6$, which are used only for the matched-dimension descriptive sign-specific CPVE summaries. Taken together, the omnibus, componentwise, and sequential results indicate that both mfasPCA and STPCA detect the designed sign-specific spatial signal in these controlled simulation regimes. The distinction between the methods is therefore not in whether the signal is detected by the permutation tests, but in how the spatially structured variation is represented and summarized, as reflected by the sign-specific CPVE comparisons.

\begin{figure}
  \centering
  \begin{subfigure}[b]{0.48\textwidth}
    \includegraphics[width=\textwidth]{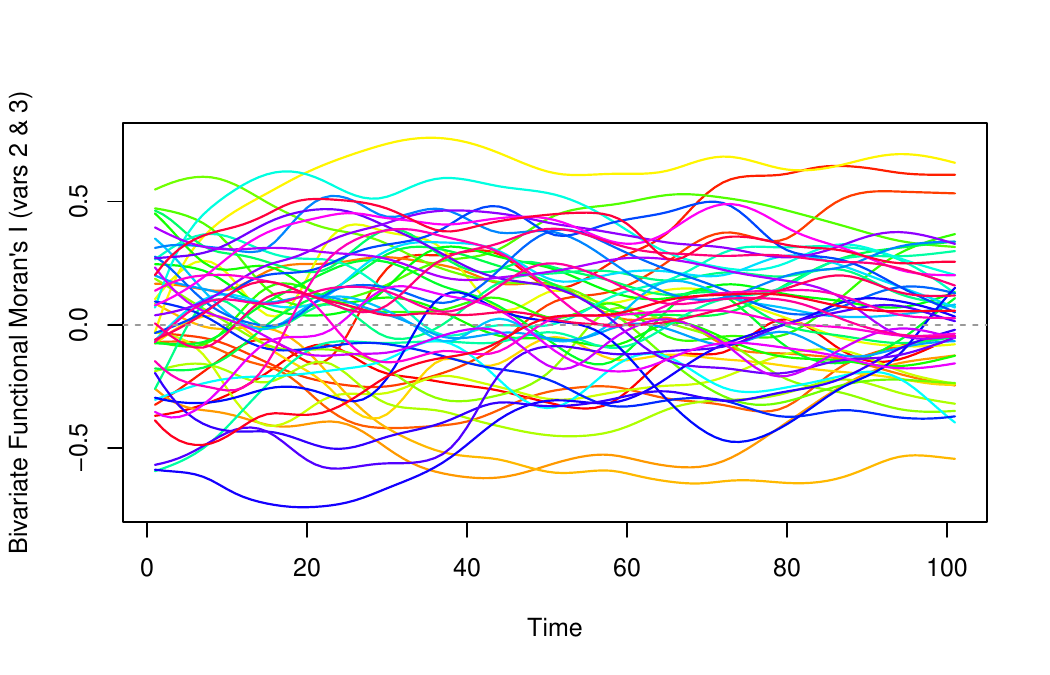}
    \caption{}
    \label{fig:figure1}
  \end{subfigure}
  \hfill
  \begin{subfigure}[b]{0.48\textwidth}
    \includegraphics[width=\textwidth]{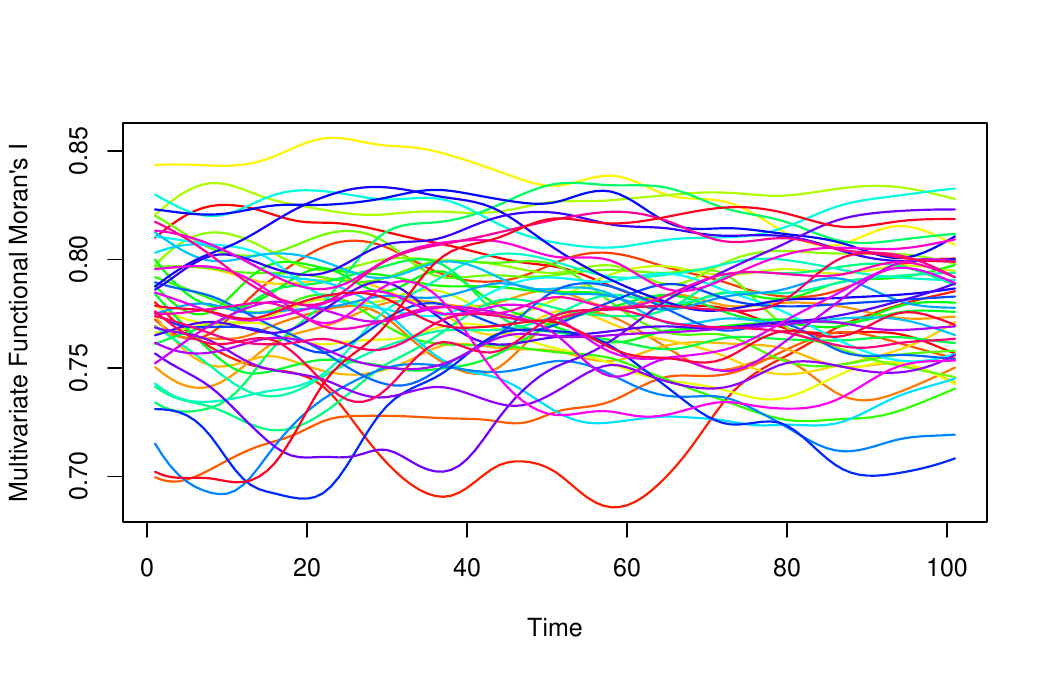}
    \caption{}
    \label{fig:figure2}
  \end{subfigure}
 \caption{Simulated (a) bivariate and (b) multivariate functional Moran's $I$ indices for radial distance weights ($r=120\,\mathrm{km}$; $\rho=0.9$). {\textit{Note: Each colored curve corresponds to one of 50 simulation runs}}}
  \label{sim_fun_moran}
\end{figure}

Figure~\ref{sim_fun_moran} shows the 50 simulated curves of the bivariate and multivariate functional Moran's $I$ under radial-distance weights at $\rho = 0.9$. The two panels measure different forms of spatial association and should be read accordingly. The bivariate functional Moran's $I$ for variables 2 and 3 (Figure~\ref{sim_fun_moran}a) measures cross-variable spatial association between one variable in a region and the other variable in neighbouring regions. In the present data-generating process, however, the spatial autoregressive transformation is applied separately to each variable, with variable-specific latent processes and measurement errors generated independently across variables. Thus, positive spatial autocorrelation is induced within each variable, but no systematic cross-variable spatial coupling is introduced between variables 2 and 3. Consequently, the bivariate functional Moran's $I$ fluctuates around zero, with both positive and negative values across the functional domain and across simulation runs.

By contrast, the multivariate functional Moran's $I$ (Figure~\ref{sim_fun_moran}b) is based on the multivariate inner product between neighbouring functional observations and therefore aggregates within-variable spatial agreement across all $d = 10$ simulated variables. Since the radial-distance simulation with $\rho = 0.9$ induces positive within-variable spatial autocorrelation in each variable, these contributions accumulate in the multivariate statistic. As a result, the multivariate functional Moran's $I$ remains strongly positive and concentrated in a narrow band across the functional domain, whereas the bivariate statistic for variables 2 and 3 remains centred near zero.

This distinction is also relevant for mfasPCA, because the mfasPCA eigenproblem is based on a multivariate Moran-type criterion on the stacked functional score matrix, not on a pairwise bivariate Moran's $I$ statistic. Thus, the multivariate Moran's $I$ diagnostic is more closely aligned with the type of joint spatial structure targeted by mfasPCA, while the bivariate Moran's $I$ remains useful as a complementary diagnostic for pairwise cross-variable spatial association.

\begin{figure}
  \centering
  \begin{subfigure}[b]{0.48\textwidth}
    \includegraphics[width=\textwidth]{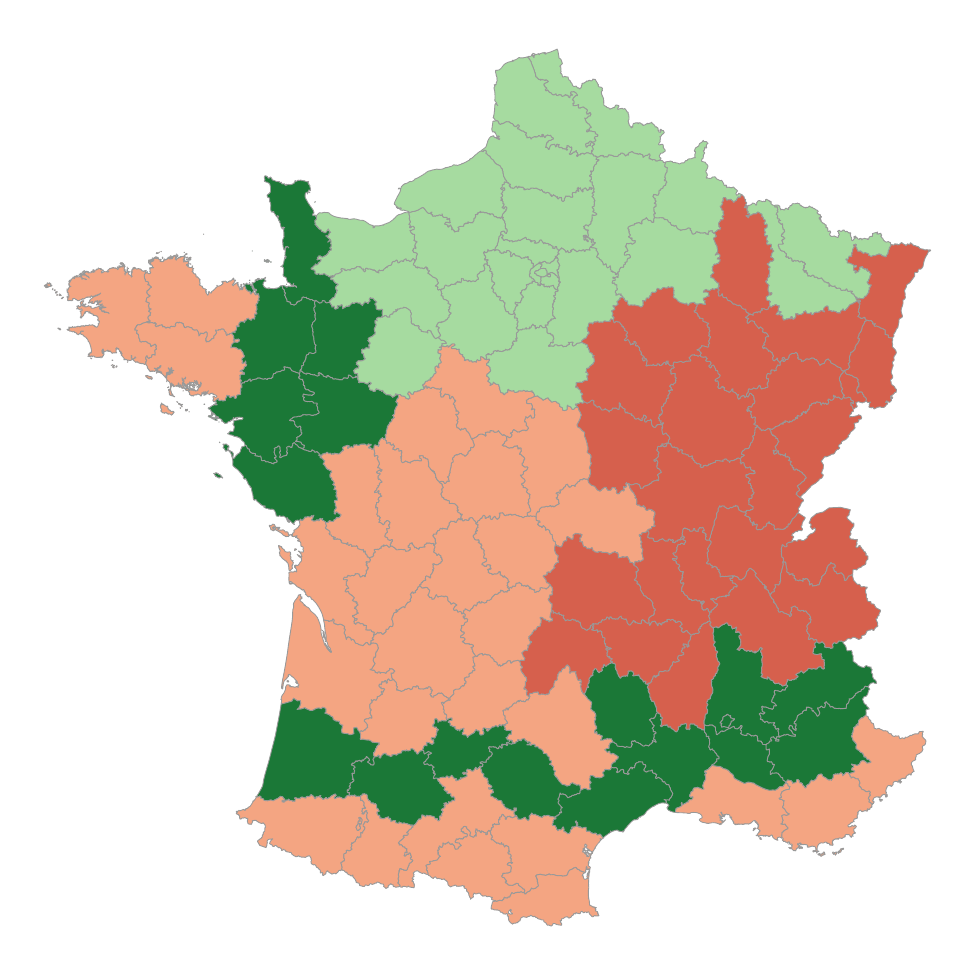}
    \caption{}
    \label{french_pos}
  \end{subfigure}
  \hfill
  \begin{subfigure}[b]{0.48\textwidth}
    \includegraphics[width=\textwidth]{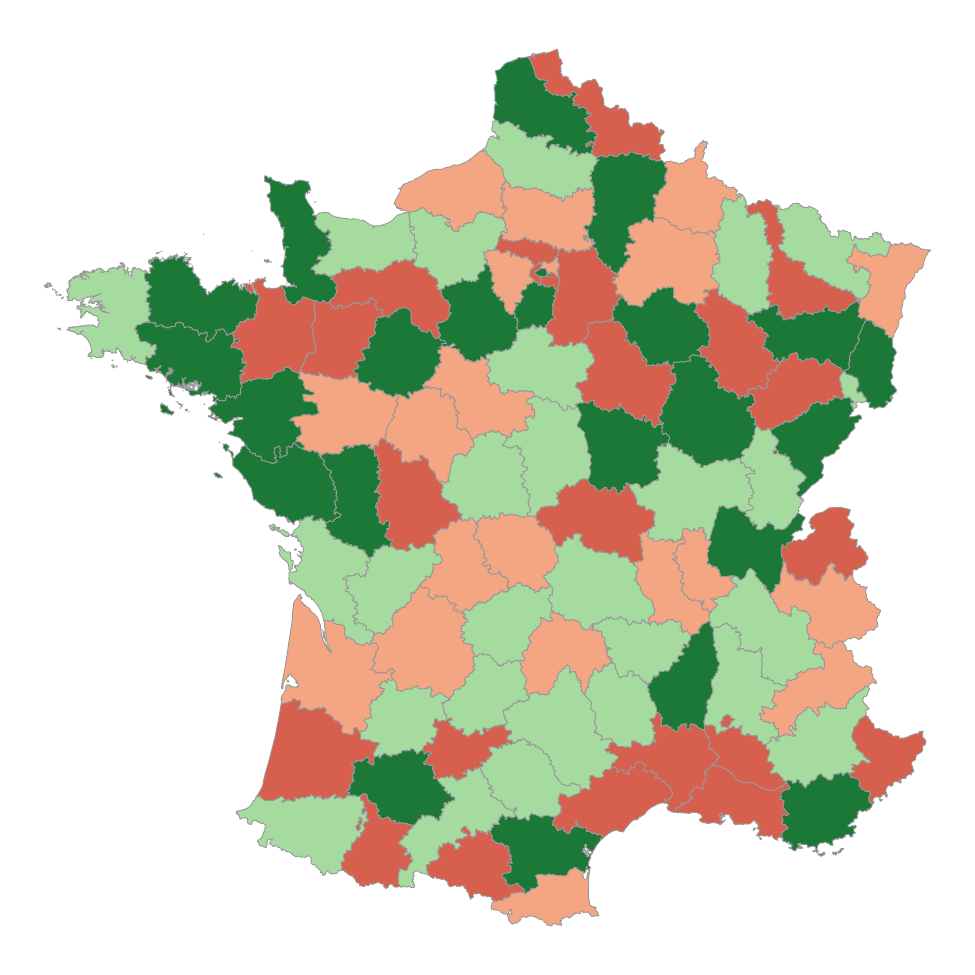}
    \caption{}
    \label{french_neg}
  \end{subfigure}
\caption{Second-level administrative divisions in mainland France generated by the model in Section~\ref{sec:sim-results}: (a) positive spatial autocorrelation using radial distance weights ($r=120\,\mathrm{km}$; $\rho=0.9$); (b) negative spatial autocorrelation using shared-boundary weights ($\rho=-0.9$). Colors encode the sign pair of the two mapped component scores: dark green = \texttt{+ +}, light green = \texttt{+ -}, light orange = \texttt{- +}, and dark orange = \texttt{- -}.}
  \label{french_map}
\end{figure}

Figure~\ref{french_map} visualizes the spatial clustering patterns generated by the model under these two regimes. With positive autocorrelation (radial distance weights, $\rho=0.9$; Figure~\ref{french_map}a), neighboring regions share similar component-score signs, forming broad homogeneous clusters. With negative autocorrelation (shared-boundary weights, $\rho=-0.9$; Figure~\ref{french_map}b), adjacent regions alternate in sign, producing a checkerboard pattern consistent with spatial repulsion.

Overall, the simulation results show broadly similar qualitative patterns across the nine spatial weight matrices considered by \citet{krzysko2024stpca}.


\section{Application to real data}\label{sec:real-data}

We illustrate our approach with an application that considers calendar year as the functional domain in a multivariate setting. Specifically, we analyze 16 regions with 12 variables (proxies for the socioeconomic development of Polish regions, Table~\ref{tab:table_variables}), measured annually from 2002 to 2018, with calendar year 2002--2018 as the functional domain \citep{krzysko2024stpca}. The dataset was introduced by \citet{krzysko2024stpca}, who also reported STPCA results; it is used here to provide a reference comparison with their reported STPCA analysis, while the simulation study in Section~\ref{sec:sim-study} provides the primary controlled evaluation of method behaviour.

\begin{figure}
  \centering
  \begin{subfigure}[b]{0.48\textwidth}
    \includegraphics[width=\textwidth]{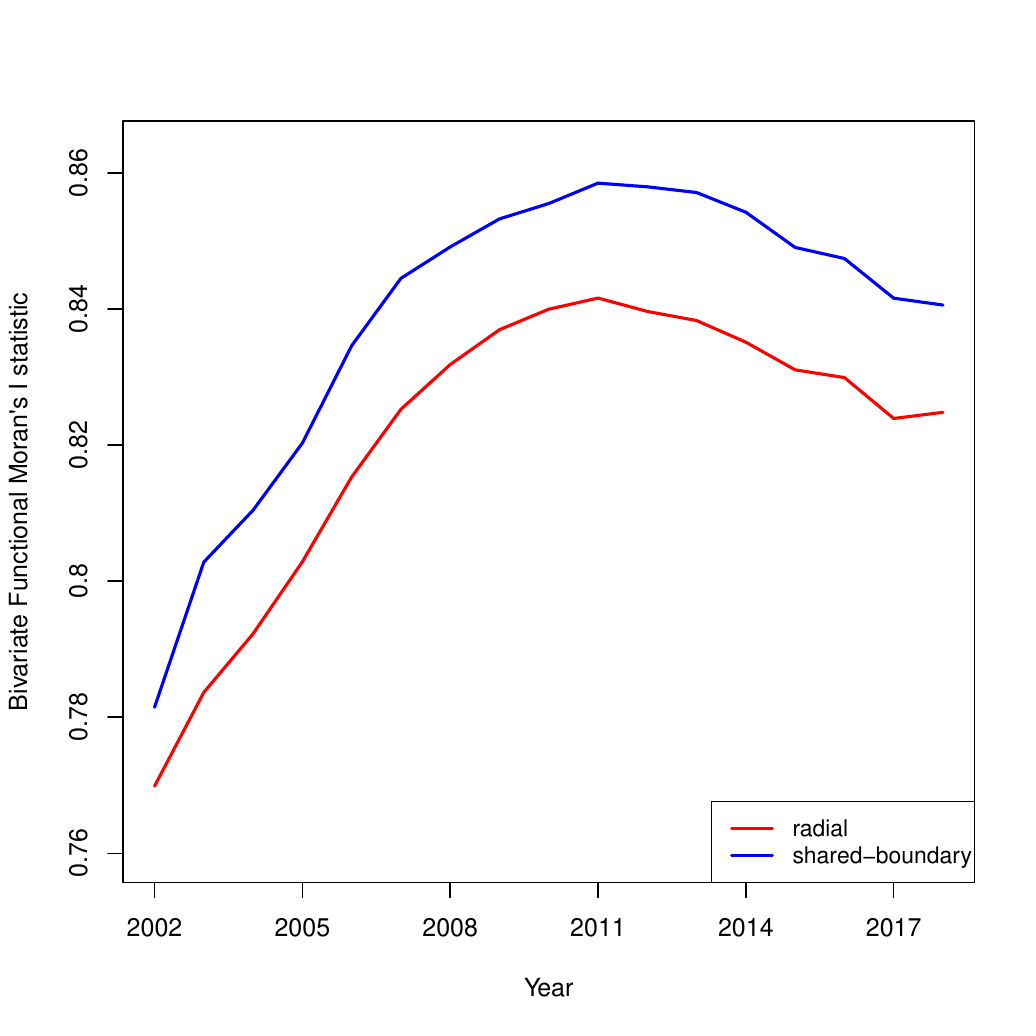}
    \caption{}
    \label{bi_functional_moran}
  \end{subfigure}
  \hfill
  \begin{subfigure}[b]{0.48\textwidth}
    \includegraphics[width=\textwidth]{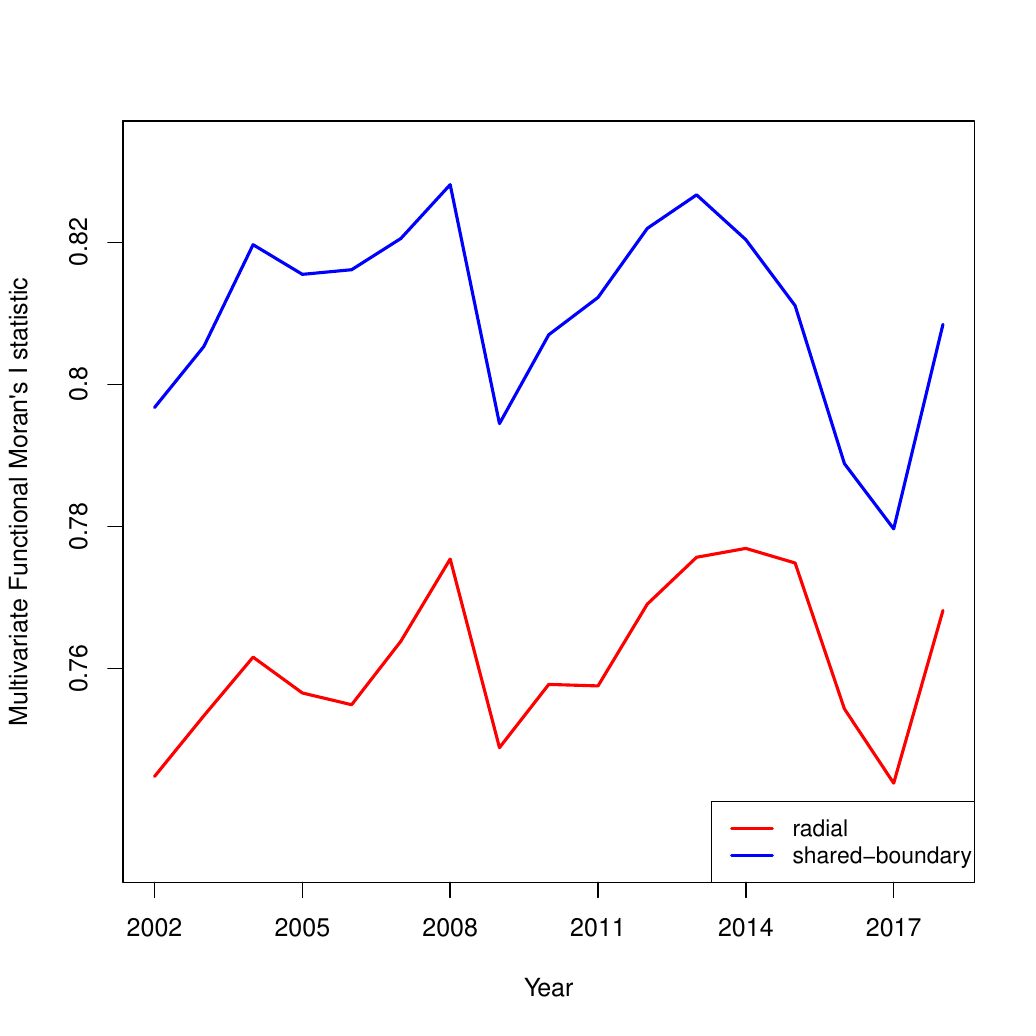}
    \caption{}
    \label{mul_functional_moran}
  \end{subfigure}
  \caption{Functional Moran's $I$ curves over the functional domain: (a) Bivariate case: variables 2 and 3 from Table~\ref{tab:table_variables} (Polish socioeconomic indicators), using the radial distance and shared-boundary weight matrices; (b) Multivariate case: 12 socioeconomic variables (2002--2018) for Polish regions, using the radial distance and shared-boundary weight matrices. Cases (a) and (b) are based on the dataset in \citet{krzysko2024stpca}.}
  \label{Poldata_fun_moran}
\end{figure}

We compute bivariate and multivariate functional Moran's $I$ statistics (Figure~\ref{Poldata_fun_moran}) using both radial-distance and shared-boundary weight matrices. For the Polish application, the radial cutoff is set to the mean plus one standard deviation of the off-diagonal centroid-to-centroid distances among the regions, providing a data-driven neighbourhood scale. This differs from the fixed $120\,\mathrm{km}$ cutoff used in the simulation study and reflects the different geographic scales of the two settings. In the bivariate case, variables 2 and 3 from Table~\ref{tab:table_variables} were selected as a representative pair of socioeconomic indicators to illustrate cross-variable functional spatial association; the multivariate Moran's $I$ analysis then extends this assessment by jointly considering all 12 indicators.

Both the bivariate and multivariate functional Moran's $I$ curves remain positive throughout the study period and vary within relatively narrow ranges. For the bivariate statistic (Figure~\ref{Poldata_fun_moran}a), Moran's $I$ increases during the early years and then stabilizes before declining slightly towards the end of the period. For the multivariate statistic (Figure~\ref{Poldata_fun_moran}b), Moran's $I$ shows more year-to-year fluctuation but remains consistently positive. In both panels, the shared-boundary weights generally yield larger Moran's $I$ values than the radial-distance weights. Overall, these curves provide descriptive evidence of positive functional spatial association among Polish regions under both neighbourhood definitions.

\begin{figure}
  \centering
  \begin{subfigure}[b]{0.48\textwidth}
    \includegraphics[width=\textwidth]{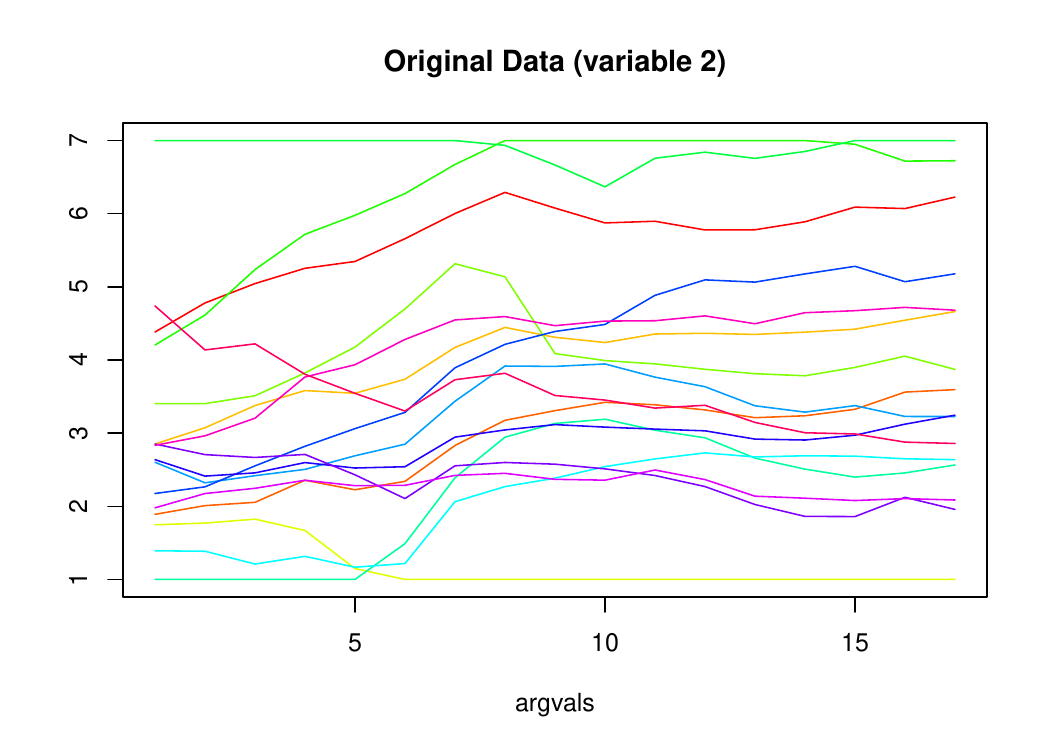}
    \caption{}
    \label{var2_original}
  \end{subfigure}
  \hfill
  \begin{subfigure}[b]{0.48\textwidth}
    \includegraphics[width=\textwidth]{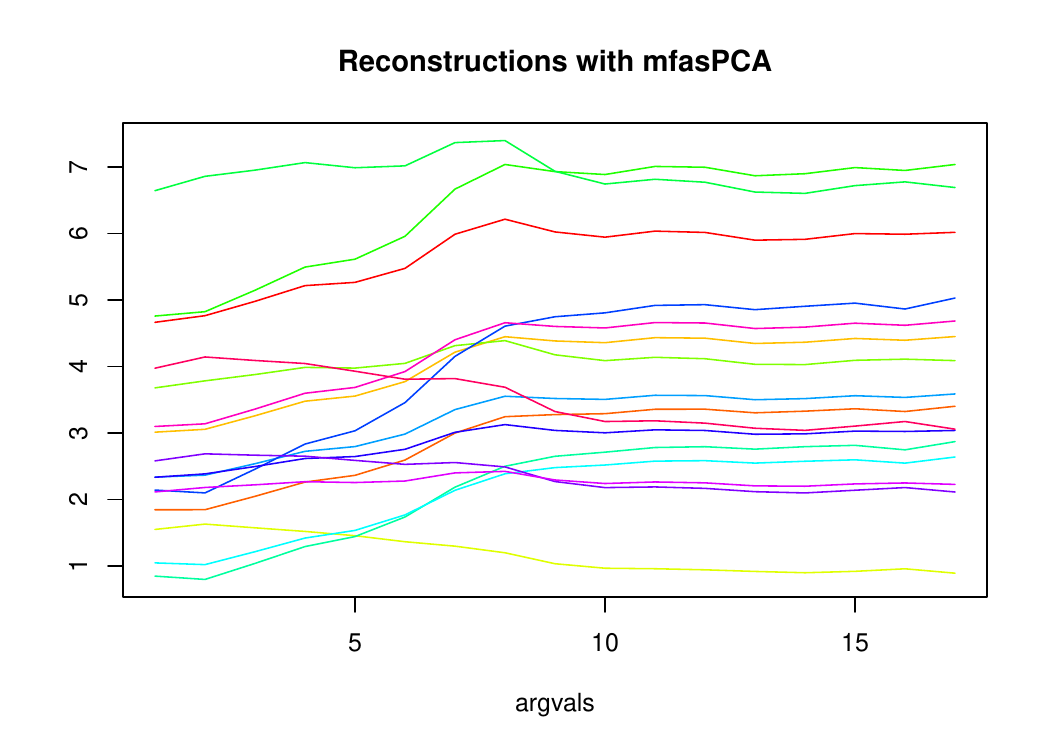}
    \caption{}
    \label{var2_recon_mfasPCA}
  \end{subfigure}
  \caption{Polish data (variable 2 in Table~\ref{tab:table_variables}): (a) Original data and (b) reconstruction with mfasPCA (radial weights). \textit{Note: Each colored line corresponds to one of the 16 Polish regions}}
  \label{Poldata_reconstruction_Var2}
\end{figure}

We compare mfasPCA with STPCA using radial-distance and shared-boundary weight matrices. In the simulation study (Figures~\ref{simu3a}--\ref{simu3b}), mfasPCA achieves higher sign-specific CPVE than STPCA in the reported positive and negative regimes. In the Polish application, Table~\ref{tab:var_explained_only} shows a different pattern: mfasPCA has higher positive-subspace sign-specific CPVE, whereas STPCA has higher negative-subspace sign-specific CPVE, especially under shared-boundary weights. Thus, the empirical results do not indicate a uniform dominance of either method; rather, the two methods emphasize different aspects of the spatial-functional structure. As a qualitative reconstruction diagnostic, Figure~\ref{Poldata_reconstruction_Var2} compares the original curves for variable 2 with their mfasPCA reconstruction under radial-distance weights, illustrating how the retained components summarize the dominant functional pattern in the Polish data.

\begin{figure}[!ht]
  \centering

\captionsetup{
  labelfont={bf},
  textfont={}
}
 \begin{subfigure}[b]{0.28\linewidth}
    \includegraphics[width=\textwidth]{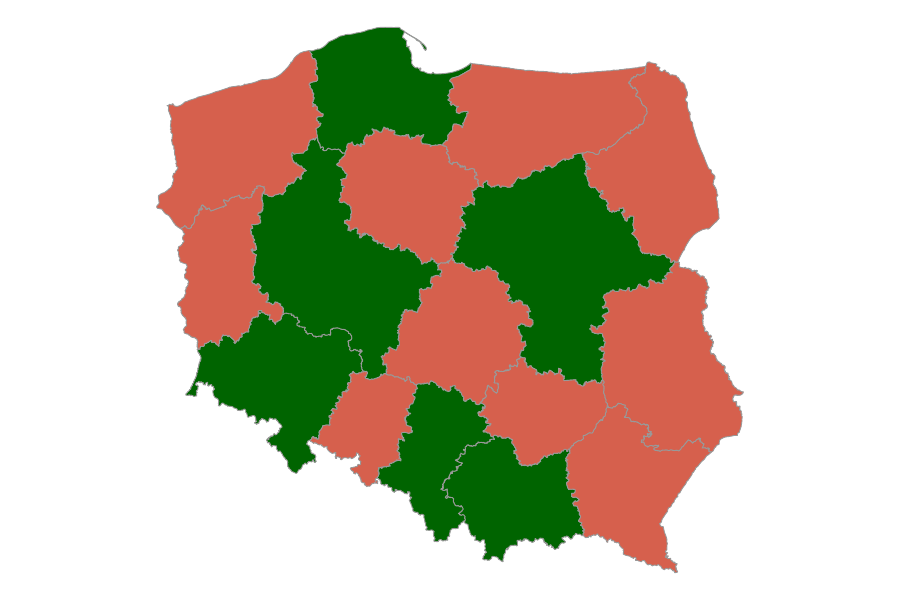}
    \caption{}
    \label{fig:mfaspca_shb_a}
  \end{subfigure}
  \hfill
  \begin{subfigure}[b]{0.28\linewidth}
    \includegraphics[width=\textwidth]{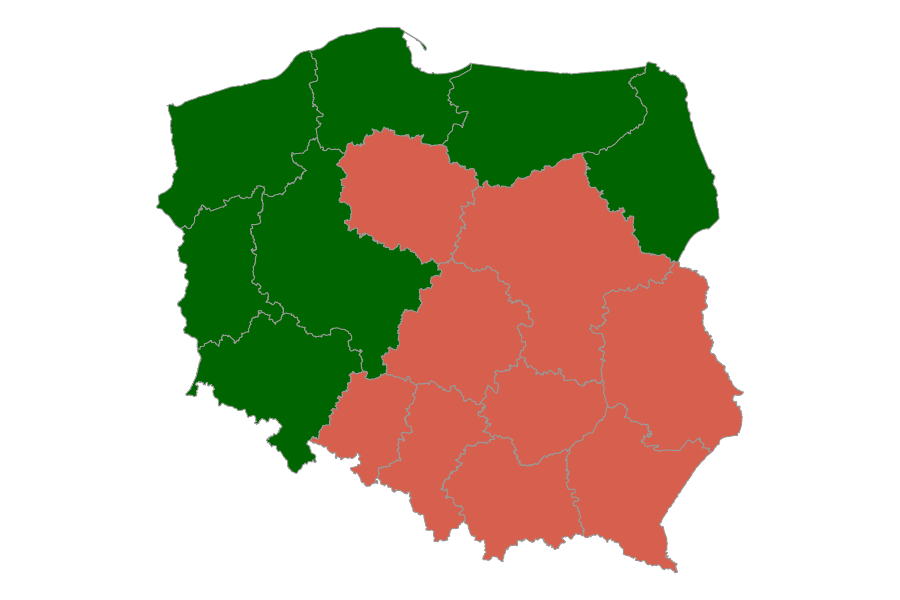}
    \caption{}
    \label{fig:mfaspca_shb_b}
  \end{subfigure}
  \hfill
  \begin{subfigure}[b]{0.28\linewidth}
    \includegraphics[width=\textwidth]{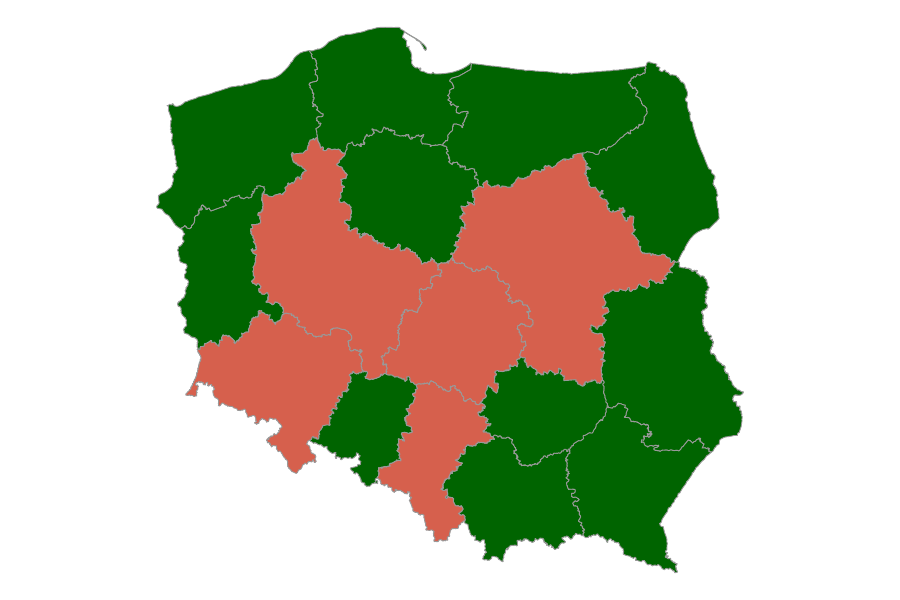}
    \caption{}
    \label{fig:mfaspca_shb_c}
  \end{subfigure}
  \hfill
  \begin{subfigure}[b]{0.28\linewidth}
    \includegraphics[width=\textwidth]{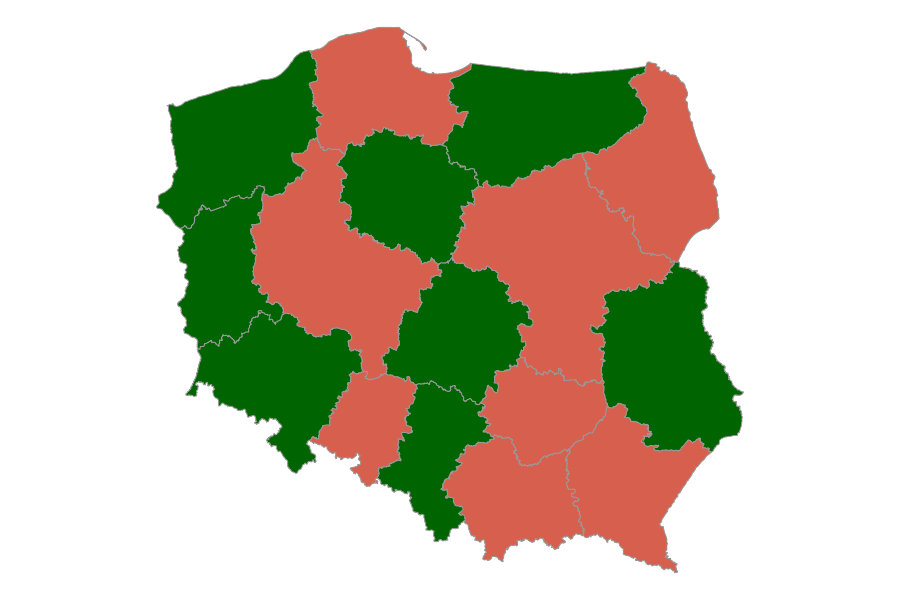}
    \caption{}
    \label{fig:mfaspca_shb_d}
  \end{subfigure}

\caption{mfasPCA sign maps under the shared-boundary weights. Color key: green = positive score sign; red = negative score sign. Panels (a) and (b) correspond to the first and second positive mfasPCA components, respectively; panels (c) and (d) correspond to the first and second negative mfasPCA components, respectively. Color flips across adjacent regions highlight transitions consistent with positive versus negative spatial dependence.}
  \label{fig:mfaspca_shb_signmaps}
\end{figure}

The spatial patterns of the mfasPCA component scores under shared-boundary weights (Figure~\ref{fig:mfaspca_shb_signmaps}) further clarify the sign-specific structure identified by mfasPCA in the Polish application. In the positive subspace, the second positive mfasPCA component (Figure~\ref{fig:mfaspca_shb_signmaps}b) displays clearer spatial clustering than the first positive component (Figure~\ref{fig:mfaspca_shb_signmaps}a), indicating a clearer contrast between groups of neighbouring regions. In the negative subspace, the first two negative mfasPCA components (Figures~\ref{fig:mfaspca_shb_signmaps}c--\ref{fig:mfaspca_shb_signmaps}d) show coherent spatial contrasts.

Table~\ref{tab:randtests_Polish} shows no omnibus significance under radial distance for either method. Under shared-boundary weights, the positive subspace is significant for mfasPCA ($p = 0.022$), whereas the negative subspace is significant for STPCA ($p = 0.013$). The componentwise Holm tests further indicate a single significant positive mfasPCA eigencomponent (PC2), with no per-component rejections in the negative subspace and none for STPCA under either sign. The sequential checks lead to the same sign-specific interpretation. This eigencomponent corresponds to the second positive mfasPCA component in Figure~\ref{fig:mfaspca_shb_signmaps}b, whose spatial pattern visually reflects the shared-boundary clustering among Polish regions. We interpret this component primarily as a spatially structured score contrast, and do not assign a specific socioeconomic meaning to it without further inspection of the variable-specific functional loadings.


\section{Discussion}\label{sec:discussion}

We synthesize the findings across the simulation study and the multivariate real-data application. The proposed mfasPCA framework extends earlier functional areal spatial PCA and functional Moran's $I$ work, including the univariate spatial-functional framework \citep{hassan2021spatial}, to the multivariate functional areal setting. In the present formulation, mfasPCA uses the multivariate functional score matrix and a Moran-type spatially weighted objective function to extract spatially structured functional components. Bivariate and multivariate functional Moran's $I$ statistics complement this decomposition by quantifying spatial autocorrelation in functional areal data. This formulation differs from STPCA, which is constructed through a coefficient representation of multivariate regional time series \citep{krzysko2024stpca}. Thus, although both methods use spatially weighted dimension-reduction criteria for regional multivariate data, they differ in representation and in the objective criterion used to define the components.

These functional Moran's $I$ diagnostics also behave coherently across the simulation and real-data settings. In the simulation, where the data-generating process induces within-variable spatial autocorrelation but does not introduce explicit cross-variable spatial coupling, the bivariate statistic for variables 2 and 3 fluctuates around zero, whereas the multivariate statistic remains strongly positive. In the Polish application, both diagnostics remain positive over the study period, providing descriptive evidence of functional spatial association among the regional socioeconomic indicators. These results illustrate the complementary roles of the two diagnostics: the bivariate functional Moran's $I$ isolates pairwise cross-variable spatial association, while the multivariate statistic aggregates joint spatial agreement across variables.

Beyond the main sign-specific CPVE and omnibus-test summaries, the per-eigen Holm and sequential tests in the Supplementary Material show that, for the Polish data, mfasPCA identifies the same second positive component as carrying spatial signal across most weight matrices, whereas the STPCA detections tend to occur in different negative components depending on the weight matrix. This suggests that, in this application, the mfasPCA signal is concentrated in a recurring positive component, while STPCA highlights negative-subspace structure that varies with the weighting scheme. The simulation results in Table~\ref{tab:perm_rho07} provide a controlled comparison: in the designed positive and negative regimes, both mfasPCA and STPCA detect the corresponding signal-bearing sign through the omnibus, componentwise Holm, and sequential tests, while the non-signal-bearing sign is not selected by the sequential tests. This comparison is useful because it shows that the mfasPCA score-matrix permutation procedure can detect the designed sign-specific spatial signal comparably to STPCA, while preserving the functional score-space formulation of the proposed method.

In the real-data application, permutation-based eigenvalue tests further clarify these contrasts. Under radial distance weights, neither method shows an omnibus signal. Under shared-boundary weights, mfasPCA shows an omnibus signal in the positive subspace and STPCA shows an omnibus signal in the negative subspace, with a single significant positive mfasPCA eigencomponent (PC2) and no component-level rejections in the negative subspace or for STPCA. Taken together, these results suggest that mfasPCA extracts Moran-type spatially structured functional components in this application, while STPCA is used here as a coefficient-based spatially weighted reference construction for temporally indexed regional trajectories, rather than as a direct equivalent to the proposed functional areal framework.


\section{Conclusion}\label{sec:conclusion}

This study shows that multivariate functional areal spatial PCA can be defined through a Moran-type spatial criterion on the functional score matrix. This formulation provides a coherent link between the proposed functional Moran's $I$ statistics and the construction of mfasPCA, rather than treating spatial autocorrelation assessment and dimension reduction as separate steps. The bivariate and multivariate functional Moran's $I$ statistics introduced here are also usable independently of mfasPCA as diagnostic tools for spatial autocorrelation in multivariate functional areal data.

The sign-split eigenspace formulation allows positive and negative spatial-functional structures to be examined separately. Together with the eigenvalue-based permutation tests, this provides a practical framework for identifying whether the extracted components carry statistically meaningful spatial signal.

The comparison with STPCA indicates that different spatially weighted criteria and representations can emphasize different sign-specific eigenspaces of the same multivariate regional dataset. This reflects the STPCA construction, which applies spatial weighting to basis-coefficient representations obtained after temporal smoothing, rather than defining a joint spatio-temporal functional criterion. In contrast, mfasPCA is explicitly spatial-functional rather than spatio-temporal: the functional argument may be time, age, wavelength, or another ordered continuum, while spatial dependence is introduced across areal units through the spatial weight matrix. The univariate illustration in the Supplementary Material, which uses age as the functional domain, further shows that the spatial-functional framework is not restricted to time-indexed data.

Methodologically, this contributes to the dimension-reduction toolkit for spatially indexed multivariate functional data: the leading components are, by construction, those that maximize or minimize the Moran-type spatially weighted criterion, so that positive components emphasize similarity among connected regions, whereas negative components emphasize contrasts under the chosen spatial weight matrix. The real-data application illustrates the practical relevance of this framework for applied researchers working with regional multivariate data, including in spatial economics and regional science. Future work may extend this framework to spatio-temporal functional data settings in which the temporal and spatial domains are modelled jointly, and further study the theoretical properties of the proposed score-space eigenspaces under suitable areal asymptotic regimes.


\clearpage 

\section*{Tables}

\begin{table}[!htbp]
\caption{Permutation tests on eigencomponents under radial-distance and shared-boundary weights ($\rho = 0.7$ and $\rho = -0.7$); averages over 50 simulation runs}\label{tab:perm_rho07}
\setlength{\tabcolsep}{3pt}
\renewcommand{\arraystretch}{1.15}
\begin{tabular*}{\textwidth}{@{\extracolsep{\fill}}
>{\raggedright\arraybackslash}p{0.19\textwidth}
>{\raggedright\arraybackslash}p{0.31\textwidth}
cccc
@{}}
\toprule
& &
\multicolumn{2}{c}{\shortstack[c]{$\rho=0.7$\\(radial distance)}} &
\multicolumn{2}{c}{\shortstack[c]{$\rho=-0.7$\\(shared-boundary)}} \\
\cmidrule{3-4}\cmidrule{5-6}
Test Block & Metric &
mfasPCA & STPCA &
mfasPCA & STPCA \\
\midrule
Omnibus $(+)$ & $p$-value & 0.001 & 0.001 & 1.000 & 1.000 \\
Omnibus $(-)$ & $p$-value & 1.000 & 1.000 & 0.001 & 0.001 \\
\midrule
Componentwise (Holm FWER) &
Mean no.\ significant components in signal-bearing sign\footnotemark[1] &
20 & 20 & 20 & 20 \\
\midrule
Sequential test $(+)$ &
No.\ of significant components\footnotemark[1] (average per run) &
20 & 20 & 0 & 0 \\
Sequential test $(-)$ &
No.\ of significant components\footnotemark[1] (average per run) &
0 & 0 & 20 & 20 \\
\botrule
\end{tabular*}
\footnotetext{Note: All $p$-values are based on 999 permutations per run. Omnibus $(+)$ tests the positive eigenvalue sum. Omnibus $(-)$ tests the negative subspace using the magnitudes of the negative eigenvalues. The reported numbers of significant components are averages over 50 simulation runs after Holm-Bonferroni adjustment. For the componentwise Holm FWER row, the reported count refers to the signal-bearing subspace: the positive subspace for $\rho=0.7$ under radial-distance weights and the negative subspace for $\rho=-0.7$ under shared-boundary weights.}
\footnotetext[1]{Counts are computed over the top 20 components in the relevant sign-specific spectrum, ordered by $|\alpha|$.}
\end{table}

\begin{table}[!htbp]
\caption{List of variables characterizing different spheres of economy and natural environment of the regions in Poland}\label{tab:table_variables}
\begin{tabular*}{\textwidth}{@{\extracolsep\fill}lp{0.85\textwidth}@{}}
\toprule
Number & Variable \\
\midrule
1  & Population per $\mathrm{km}^2$ \\
2  & Students per 10,000 inhabitants \\
3  & Libraries per 1,000 inhabitants \\
4  & Production sales of industry, total per capita \\
5  & Retail sales of goods per capita \\
6  & Forestry and logging \\
7  & Regional income \\
8  & Regional expenditure \\
9  & Targeted grants received from the state budget for indigenous tasks per $\mathrm{km}^2$ \\
10 & Fees and impact on the fund for environmental protection and water management \\
11 & Communal waste-water treatment plants \\
12 & Devastated and degraded land, remediated and developed \\
\botrule
\end{tabular*}
\end{table}

\begin{table}[!htbp]
\caption{Sign-specific CPVE for the first two positive and negative PCs under radial and shared-boundary weights}\label{tab:var_explained_only}
\begin{tabular}{@{}llccc@{}}
\toprule
& & & \multicolumn{2}{c}{Sign-specific CPVE (\%)} \\
\cmidrule{4-5}
Component & Weight matrix & Rank & mfasPCA & STPCA \\
\midrule
\multicolumn{5}{@{}l}{\textbf{Positive PCs}}\\
& Radial          & 1 & 99.55 & 81.59 \\
&                 & 2 &  0.40 & 17.63 \\
& Shared-boundary & 1 & 95.82 & 63.13 \\
&                 & 2 &  2.77 & 22.22 \\
\midrule
\multicolumn{5}{@{}l}{\textbf{Negative PCs}}\\
& Radial          & 1 & 37.07 & 48.69 \\
&                 & 2 & 25.59 & 20.17 \\
& Shared-boundary & 1 & 43.02 & 75.79 \\
&                 & 2 & 19.98 &  8.05 \\
\botrule
\end{tabular}
\end{table}

\begin{table}[!htbp]
\caption{Permutation tests on eigencomponents for radial distance weights and shared-boundary weights for Polish data}\label{tab:randtests_Polish}
\setlength{\tabcolsep}{3pt}
\renewcommand{\arraystretch}{1.15}
\begin{tabular*}{\textwidth}{@{\extracolsep{\fill}}
>{\raggedright\arraybackslash}p{0.19\textwidth}
>{\raggedright\arraybackslash}p{0.30\textwidth}
cccc
@{}}
\toprule
& & \multicolumn{2}{c}{Radial distance} & \multicolumn{2}{c}{Shared-boundary} \\
\cmidrule{3-4}\cmidrule{5-6}
Test Block & Metric & mfasPCA & STPCA & mfasPCA & STPCA \\
\midrule
Omnibus $(+)$ & $p$-value & 0.160 & 0.123 & \textbf{0.022} & 0.126 \\
Omnibus $(-)$ & $p$-value & 0.829 & 0.721 & 0.618 & \textbf{0.013} \\
\midrule
Componentwise (Holm FWER) & No.\ of significant components & 0 & 0 & 1 & 0 \\
\midrule
Sequential test $(+)$ & No.\ of significant components & 0 & 0 & 1 & 0 \\
Sequential test $(-)$ & No.\ of significant components & 0 & 0 & 0 & 0 \\
\botrule
\end{tabular*}
\footnotetext{Note: All $p$-values are based on 9999 permutations. Omnibus $(+)$ tests the positive eigenvalue sum. Omnibus $(-)$ tests the negative subspace using the magnitudes of the negative eigenvalues. ``No.\ of significant components'' denotes the number of eigencomponents rejected after Holm FWER adjustment in the corresponding outputs.}
\end{table}


\clearpage 

\section*{Figures}


\clearpage 

\backmatter

\bmhead{Supplementary information}
The Supplementary Material is provided as a separate file and contains additional univariate fasPCA results, simulation diagnostics, and supporting empirical results for the multivariate mfasPCA and STPCA analyses.


\bmhead{Use of AI-assisted tools}
The authors used AI-assisted tools, including ChatGPT and Claude, only for language polishing, readability improvement, and checking the clarity and consistency of wording. The research questions, methodology, analyses, code, results, interpretations, and conclusions are entirely the authors' own. The authors reviewed and edited all AI-assisted suggestions and take full responsibility for the final manuscript.


\bibliography{sn-bibliography}

\end{document}


\maketitle



\section{Univariate fasPCA}

This supplement presents the univariate case to illustrate how the proposed spatial-functional PCA framework behaves when \(d=1\). Throughout the univariate examples, the functional argument is denoted by \(t\). Although \(t\) may be interpreted as time in these numerical examples, the fasPCA framework is not restricted to temporal data; the functional argument may represent any ordered continuum, such as age, wavelength, or another measurement domain. Classical FPCA is used as a non-spatial reference method, not as a direct competitor, since FPCA and fasPCA optimize different criteria. Classical FPCA summarizes dominant functional variation, whereas fasPCA extracts components that are explicitly informed by spatial autocorrelation. The comparison is therefore intended to show the effect of incorporating spatial structure into the dimension-reduction step.


\subsection{Simulation studies using Models 1--3 for univariate fasPCA}

We conducted simulation studies to evaluate the proposed fasPCA in the univariate setting, using classical FPCA as a non-spatial reference. We considered spatial autocorrelation parameters \(\rho = 0.3, 0.5, 0.7, 0.9\) (see Eq.~\eqref{eqrho1}), where larger values of \(\rho\) correspond to stronger spatial dependence among functional observations. The number of components was determined by examining all values of $\rho$ and selecting the point at which one method first achieved the 95\% variance explained threshold. This led to the choice of four components for the univariate case, which were then applied consistently across all values of \(\rho\) to enable fair comparison. Performance is assessed for each value of $\rho$ by comparing the variance explained, where higher variance explained reflects a more effective representation of the underlying spatial-functional structure. Classical FPCA does not produce positive and negative spatial eigenspaces because its eigenproblem is based on the covariance structure rather than a Moran-type spatial criterion.

We consider a grid consisting of $50 \times 50$ locations with random allocation of $N=100,250,$ and $500$ spatial units. The data are generated according to the following spatial autoregressive model:
\begin{align}
    Y_{i}(t)=\rho \sum_{j=1}^{n} \omega_{i,j}Y_{j}(t)+X_{i}(t)+ \varepsilon_{i}(t),\;  t\in[0, 1] .\label{eqrho1}
\end{align}
Here, $Y_i(t)$ denotes the spatial functional observation at location $i$, $\varepsilon_i(t)$ denotes a Gaussian white noise process with variance $\sigma^2=0.1$, and $X_i(t)$ is a location-specific process. We consider three different models to generate $X_i(t)$. The descriptions of the three models are as follows:
\\~\\
\noindent \textbf{Model 1}:
\[X_{i}(t)= t\alpha_{i}+\epsilon_{i}(t),\; \alpha_{i}\backsim\mathcal{U}(-3, 3),\]\\
$\{\epsilon_{i}(t)\}$ is a Gaussian process with exponential covariance, where $i\in[1,N]$, denotes a spatial location $\mathbf{s}=(s_1,s_2)$, sampled on the grid. The curve $X_{i}$ is observed at 101 evenly spaced times of $[0,1]$.
\\~\\
\noindent \textbf{Model 2}:
\begin{equation}\label{mo2and3}
X_i(t)=u_i h_1(t)+(1-u_i)h_1(t+4)+f_i,
\qquad u_i\sim\mathcal U(0,1),
\end{equation}
where $f_{i}\backsim\mathcal{N}(0,1)$ are independent and identically distributed, 
\[h_1(t)=\max\{6-|t-11|,0\}.\]

\noindent \textbf{Model 3}: 
For a set of 50 neighbouring locations, we define
\[X_i(t)=u_i h_1(t)+(1-u_i)h_1(t-4)+f_i.\]
For the remaining locations, \(X_i(t)\) is generated according to Eq.~\eqref{mo2and3}. Model 3 differs from Model 2 by forcing the formation of a cluster of curves.

\begin{figure}[!htbp]
    \centering
    \renewcommand{\arraystretch}{1.5} 
    \begin{tabular}{cccc}
        \includegraphics[width=0.2\textwidth]{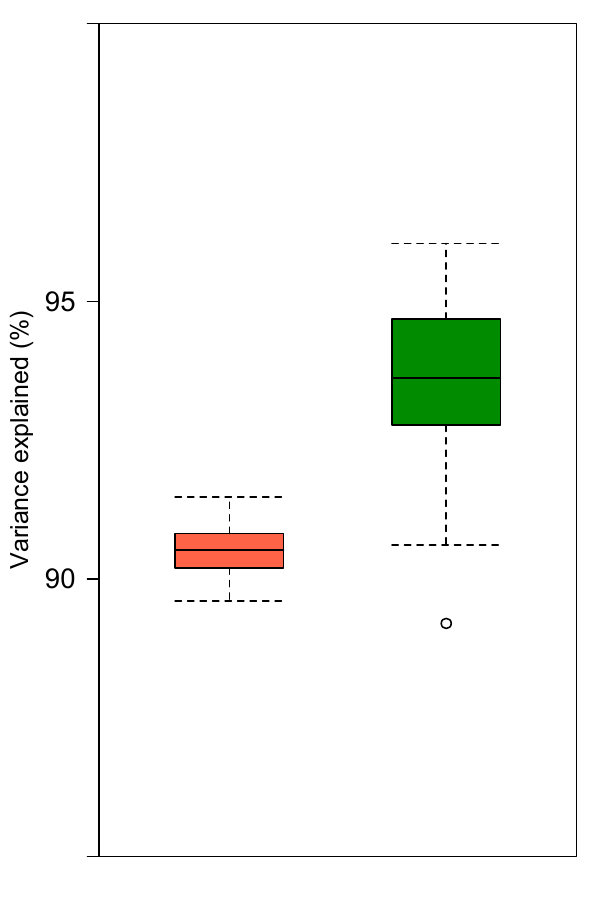} & \includegraphics[width=0.2\textwidth]{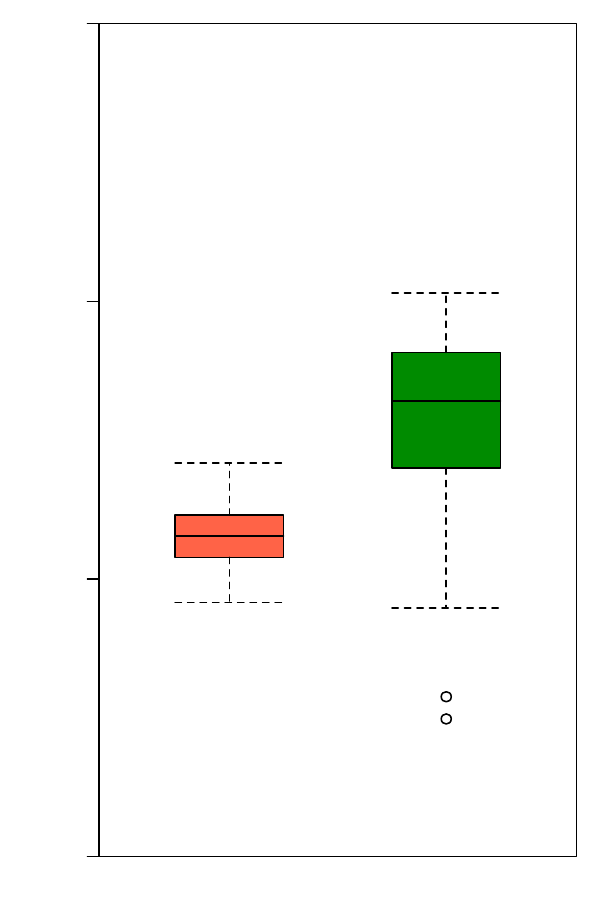} & \includegraphics[width=0.2\textwidth]{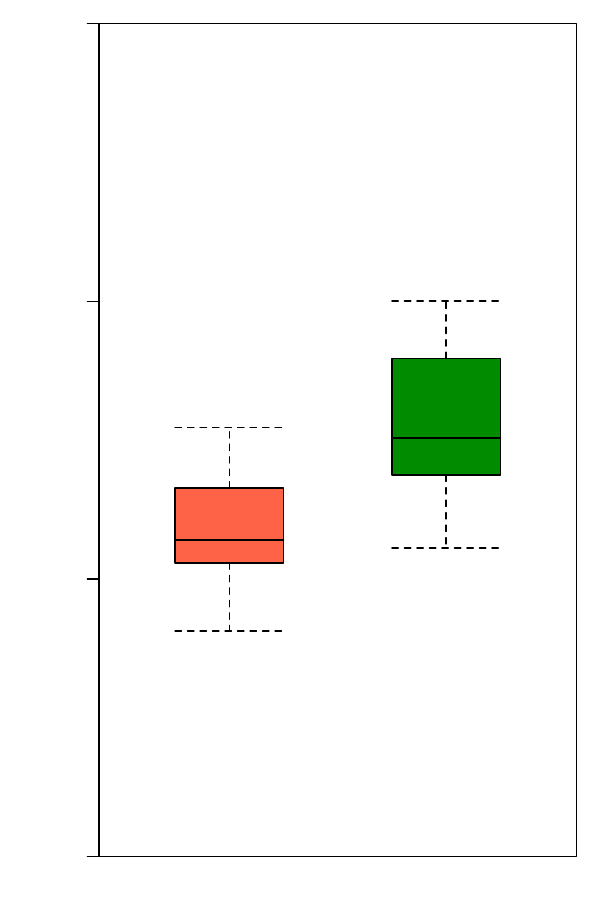} & \includegraphics[width=0.2\textwidth]{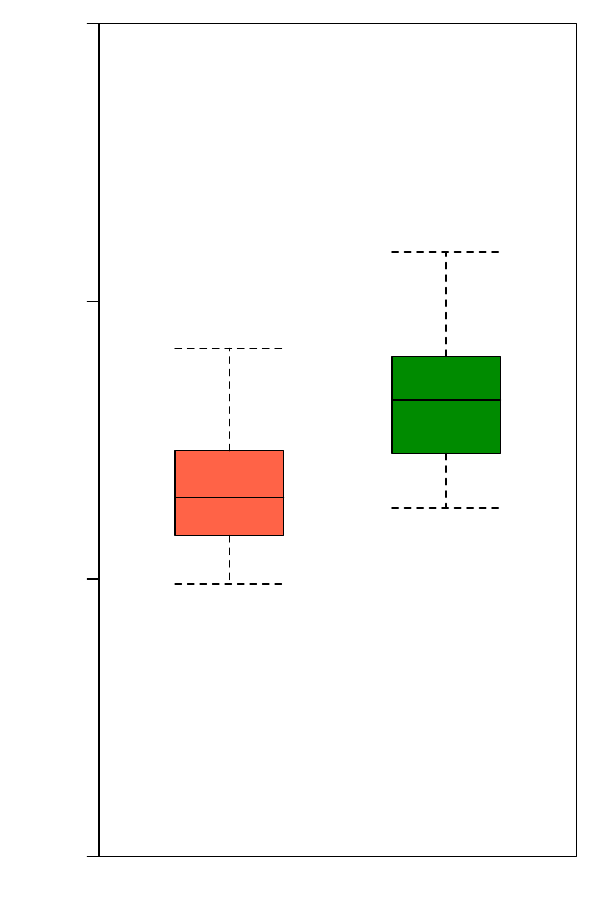} \\
     $\rho=0.3$ & $\rho=0.5$ & $\rho=0.7$ & $\rho=0.9$ \\
    \end{tabular}
    \caption{Variance explained by the top four principal components in 50 simulated data sets with \(N=500\) from Model 1, using classical FPCA and fasPCA (positive scores)}
    \label{simu1}
\end{figure}

\begin{figure}[!htbp]
    \centering
    \renewcommand{\arraystretch}{1.5} 
    \begin{tabular}{cccc}
        \includegraphics[width=0.2\textwidth]{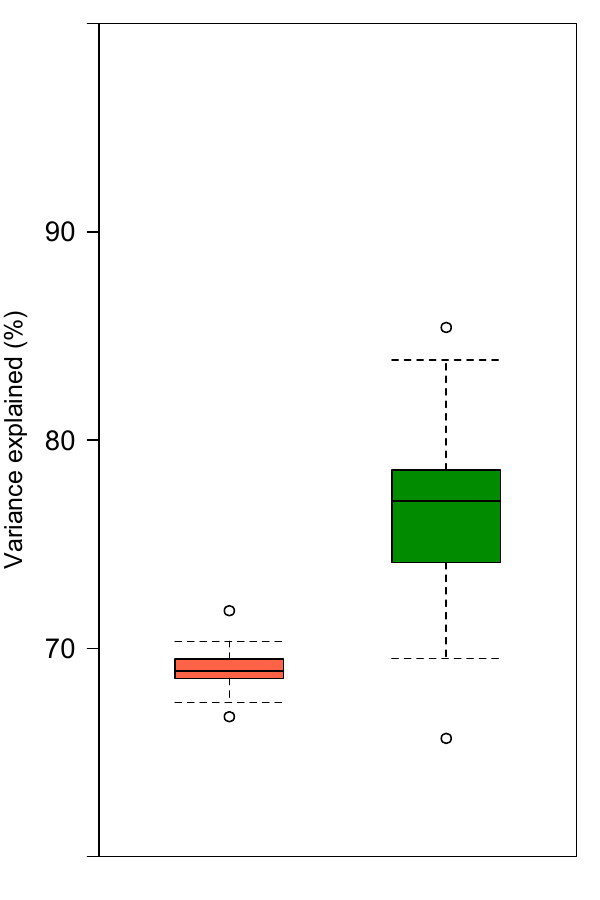} & \includegraphics[width=0.2\textwidth]{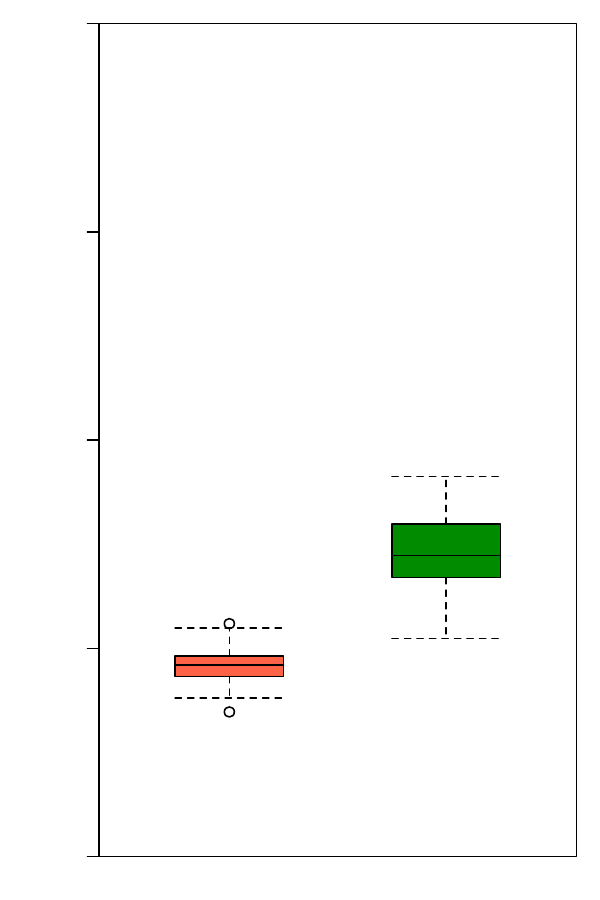} & \includegraphics[width=0.2\textwidth]{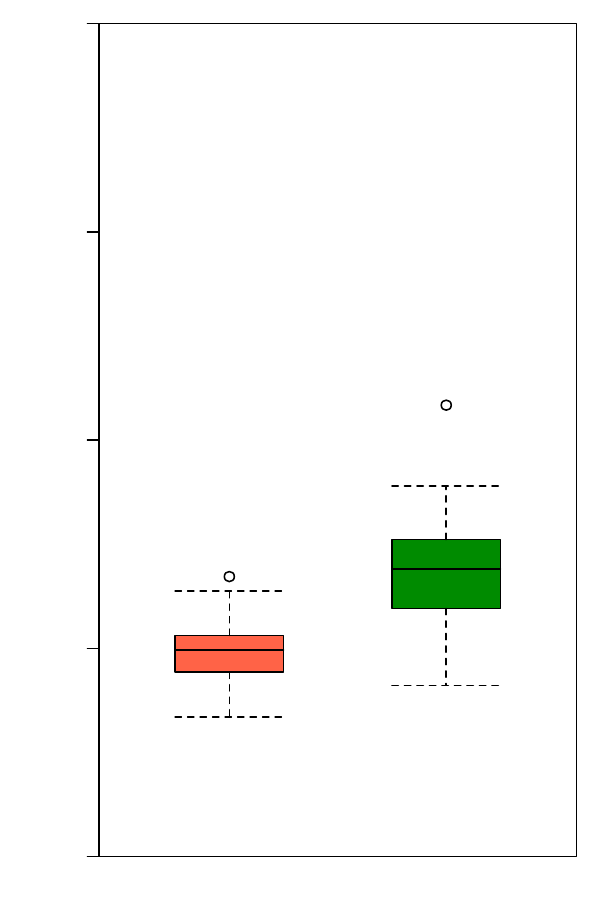} & \includegraphics[width=0.2\textwidth]{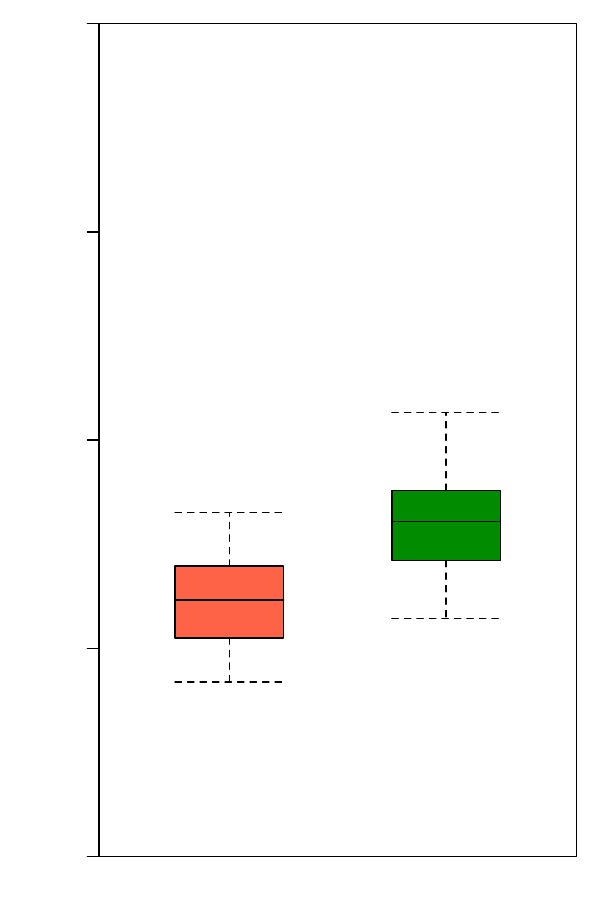} \\
     $\rho=0.3$ & $\rho=0.5$ & $\rho=0.7$ & $\rho=0.9$ \\
    \end{tabular}
    \caption{Variance explained by the top four principal components in 50 simulated data sets with \(N=500\) from Model 2, using classical FPCA and fasPCA (positive scores)}
    \label{simu2}
\end{figure}

\begin{figure}[!htbp]
    \centering
    \renewcommand{\arraystretch}{1.5} 
    \begin{tabular}{cccc}
        \includegraphics[width=0.2\textwidth]{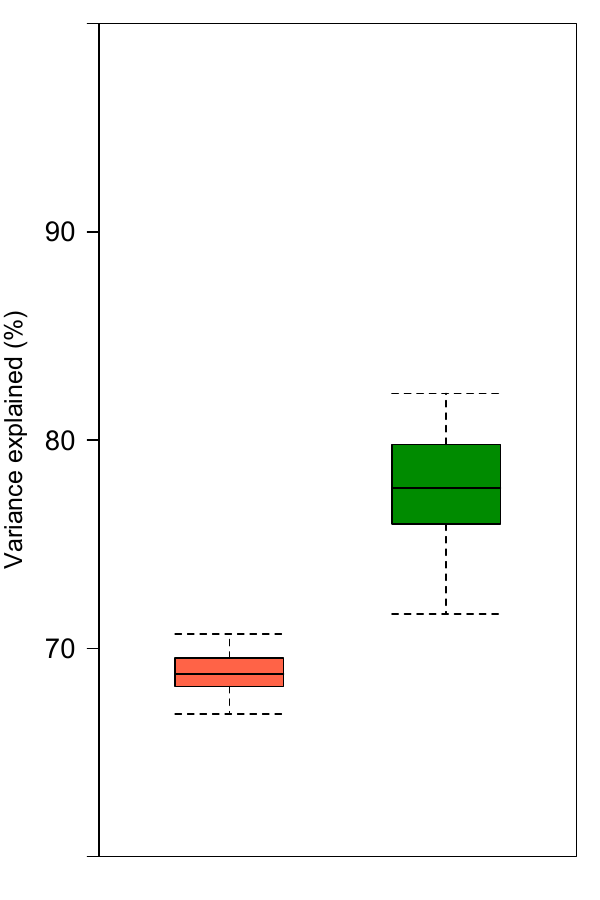} & \includegraphics[width=0.2\textwidth]{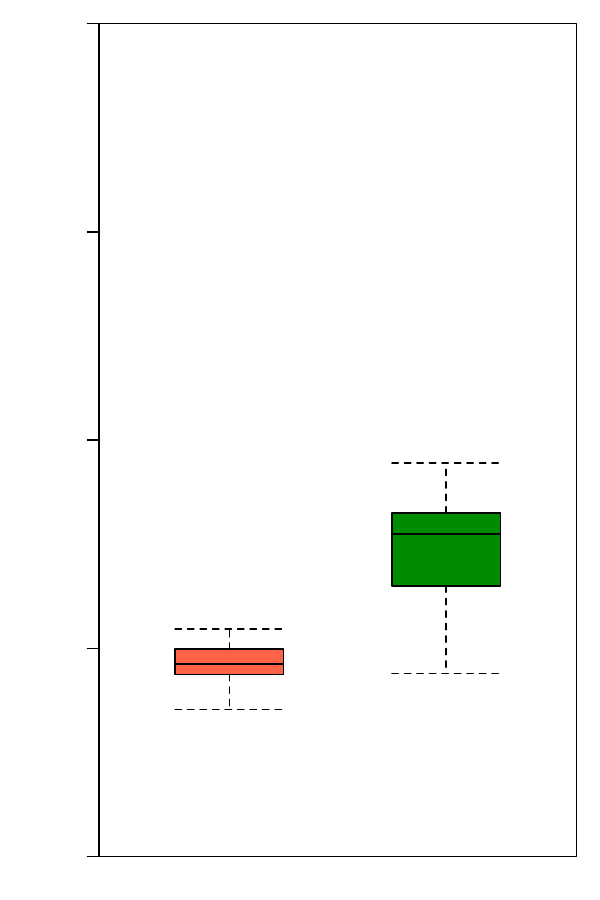} & \includegraphics[width=0.2\textwidth]{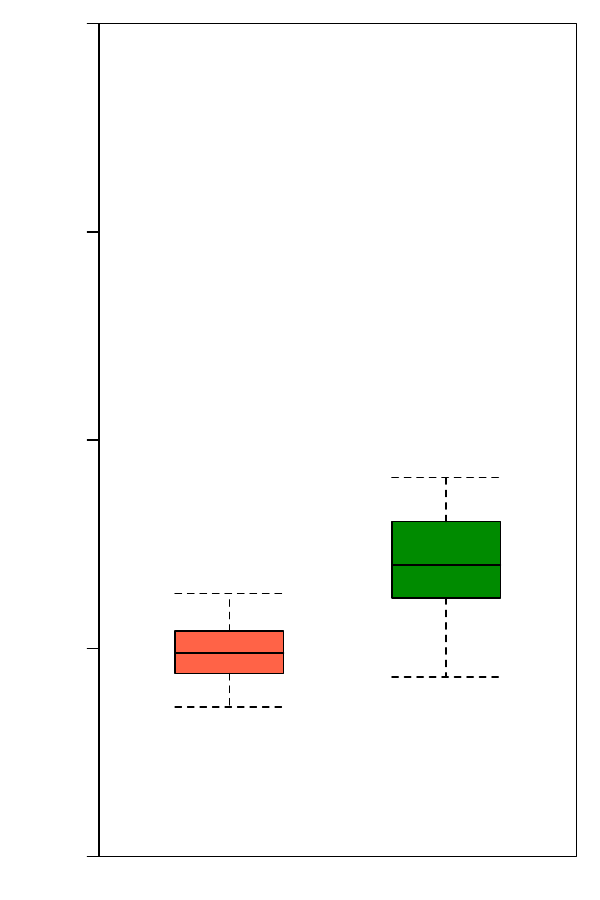} & \includegraphics[width=0.2\textwidth]{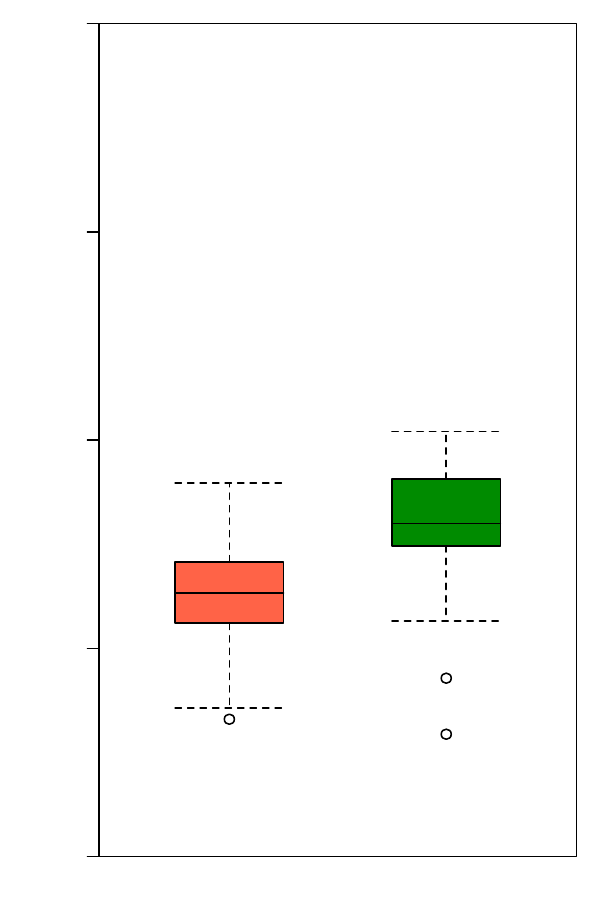} \\
     $\rho=0.3$ & $\rho=0.5$ & $\rho=0.7$ & $\rho=0.9$ \\
    \end{tabular}
    \caption{Variance explained by the top four principal components in 50 simulated data sets with \(N=500\) from Model 3, using classical FPCA and fasPCA (positive scores)}
    \label{simu3}
\end{figure}

The results of the simulation study with three different models are presented in Figures \ref{simu1}, \ref{simu2}, and \ref{simu3} which depict the variance explained (\%) for the top four principal components in both classical FPCA and fasPCA based on the kNN weight matrix. The eigenvalues from each simulation run are stored in increasing order. The variance explained by fasPCA is less dispersed as the spatial autocorrelation $\rho$ increases. In general, the results illustrate how incorporating spatial structure through fasPCA changes the component representation relative to classical FPCA, especially as spatial autocorrelation increases. Similar patterns were also observed under contiguity weights.

\clearpage

\subsection{Application to real data from HMD for univariate fasPCA}

\begin{figure}[!htbp]
	\centering
    \includegraphics[scale= 0.5]{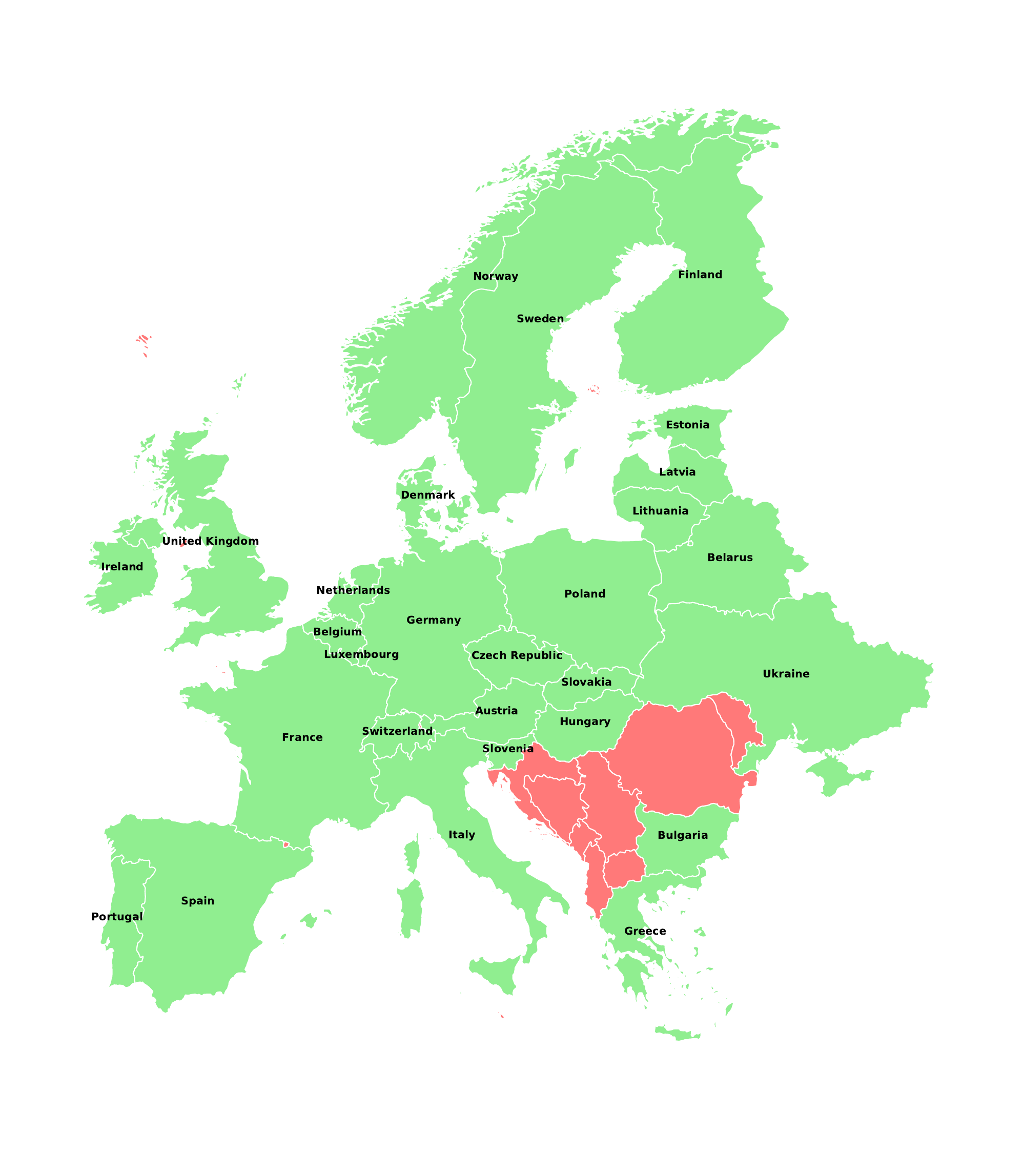}
	\caption{The map of European countries with mortality data from HMD. \textit{Note: Data for the red shaded region is not available on HMD}}
	\label{map}
\end{figure}

\begin{figure}[!htbp]
    \centering
    \includegraphics[scale= 0.35]{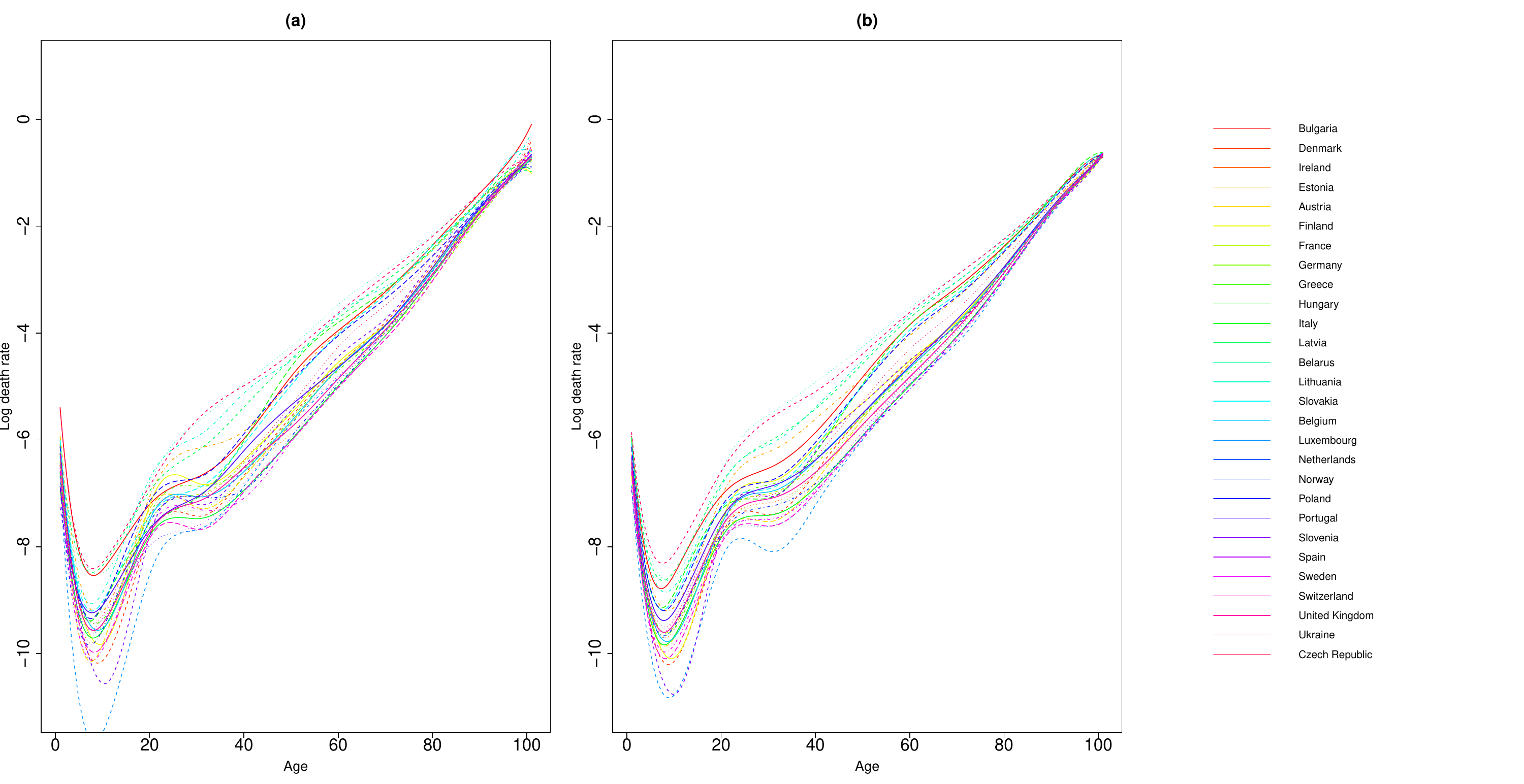}
    \caption{Male log mortality rates for 28 European countries in 2010: (a) smoothed functional curves over the age domain (0--100+ years), representing the functional domain; (b) reconstruction of the curves using the retained fasPCA components with KNN weights}
    \label{male_curves_2010}
\end{figure}

In this section, we analyze 2010 male age-specific mortality data for European countries (see Figure \ref{map}) from the Human Mortality Database \citep{HMD} using the proposed fasPCA. The discussion here focuses on KNN weights. To avoid undefined logarithms at zero, a constant equal to the minimum strictly positive observed rate was added to all death rates before log transformation. Figure \ref{male_curves_2010}(a) shows the smoothed log death rates data using B-splines for the male population in year 2010. The presence of spatial autocorrelation among the data is then investigated.

We first compute classical FPCA, then apply fasPCA to incorporate spatial structure via a weight matrix following \cite{jombart2008revealing}. The log death rates data for males in the 28 countries do not satisfy the normality assumption. The Shapiro-Wilk test, conducted using the \texttt{mvnormtest} R package \citep{mvnormtest}, was employed to assess multivariate normality of the log death rates data. The results indicated that this data violated the normality assumption. Hence, permutation tests for the Moran's $I$ statistics are calculated for these data using 999 random permutations of the log death rates for all cases studied based on the KNN weights. Table \ref{tablemoran3years} shows the presence of significant spatial autocorrelation for male log death rates for 2010 across the 28 countries. In this table, a classical Moran's $I$ index is calculated for each year using the raw data matrix treated as a panel data set. These statistics are aggregations of the functional index (equation (4) in the manuscript) as the functional trace index defined in (equation (7) in the manuscript).
    
\begin{figure}[!htbp]
    \centering
    \includegraphics[scale= 0.7]{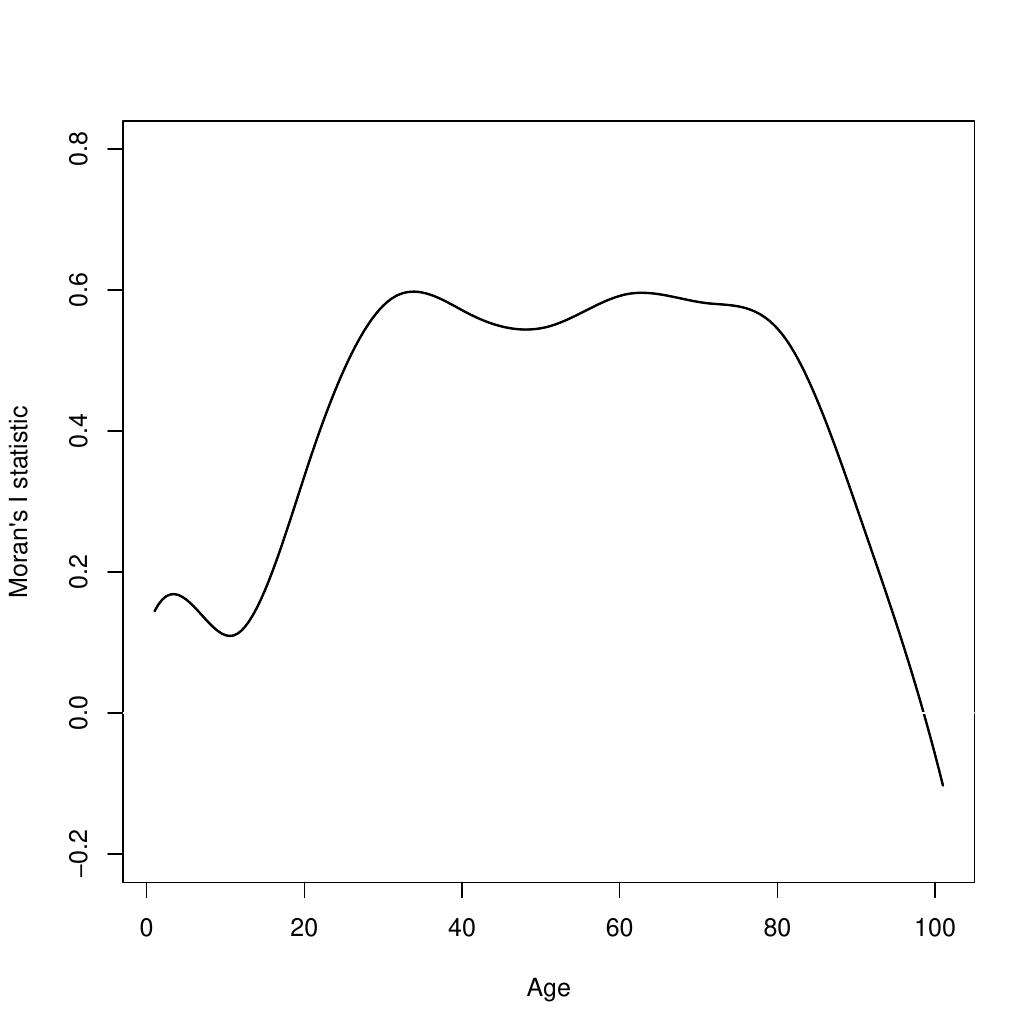}
    \caption{Functional Moran's $I$ for male log mortality rates over ages 0--100+ in 2010, using the KNN weight matrix}
    \label{functional_moran}
\end{figure}

The Moran's $I$ statistics reported values close to +1, suggesting that neighboring locations exhibit strong positive autocorrelations in male mortality rates for the year 2010. Consistently, Figure \ref{functional_moran} shows the functional Moran's $I$ over the age domain, with spatial dependence strengthening between ages 20–80 and tapering at older ages.

Under classical FPCA, the first PC explains about 99.67\% of variance  (Table \ref{pcaspca2010}) but does not account for spatial structure. We perform fasPCA on the basis functions of the male log death rates data using KNN weights. For fasPCA, we retain the top three positive and top two negative components (\ref{pcaspca2010}); their functional Moran’s $I$ values are significant, and together they explain 97.48\% of variability while explicitly capturing spatial autocorrelation.

Figure~\ref{male_curves_2010}(b) shows the reconstruction of the curves using the retained components. These PCs were mapped onto geographic spaces (representing 28 European countries generated using the \texttt{maps} \citep{maps} package  and the \texttt{rgdal} \citep{bivand} package where the black and white squares of the variable size represent positive and negative scores of the PCs respectively. The large black squares are well differentiated from the large white squares, while the small squares are less differentiated. The area of the square is proportional to the absolute value of the score. These graphical representations are applied to the significant PCs.

\begin{figure}[!htbp]
    \centering
    \includegraphics[scale= 0.4]{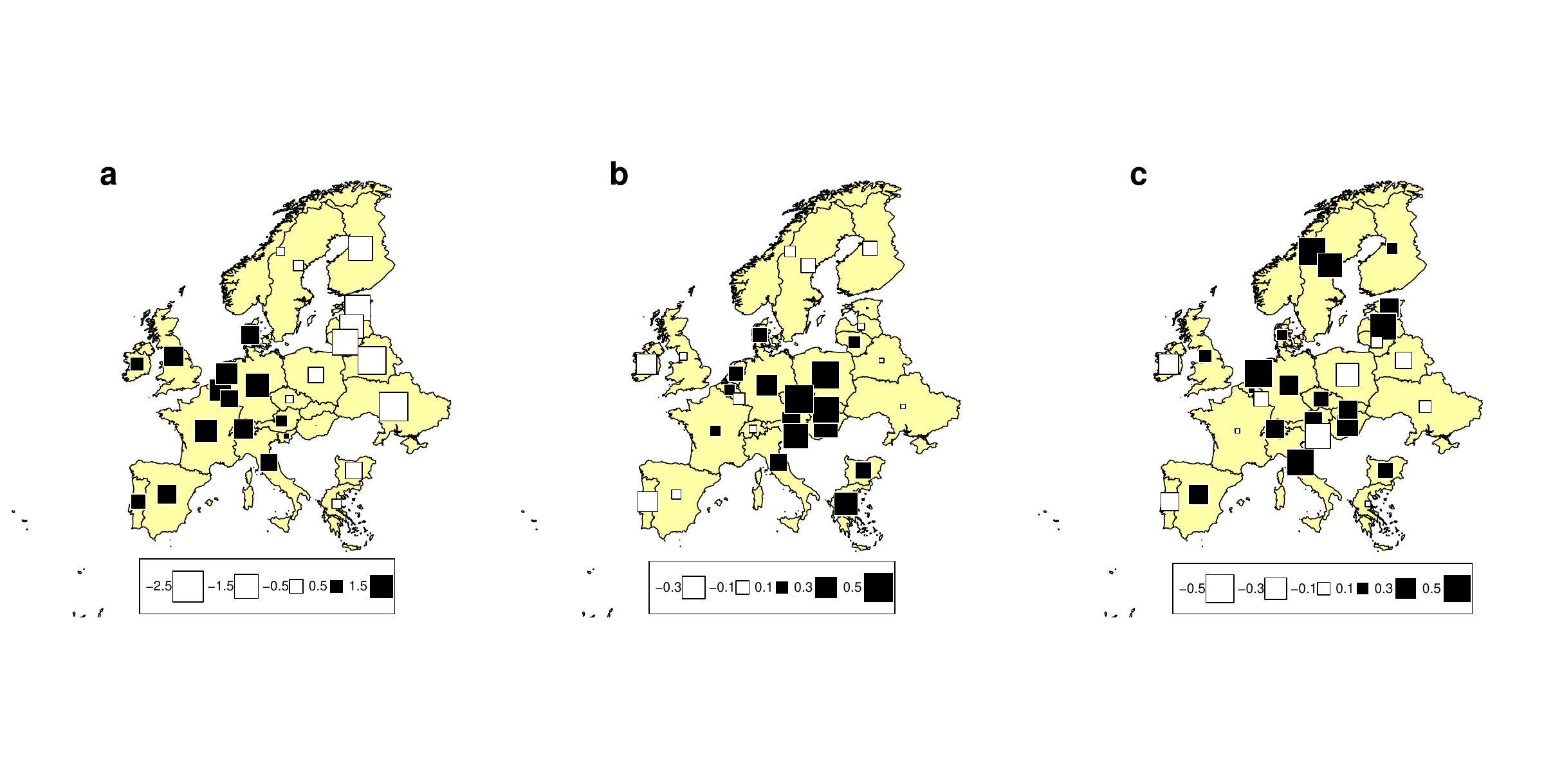}
    \caption{fasPCA spatial score maps for male log mortality rates in 2010 across 28 European countries: (a) first positive component, (b) second positive component, and (c) first negative component, based on KNN weights. \textit{Note: Square size is proportional to the absolute score value, and colour (white to black) indicates direction and magnitude, as shown in the legend bars}}
    \label{male_mapsknn_2010}
\end{figure}

The first positive PC (Figure \ref{male_mapsknn_2010}(a)) shows spatial connectivity between the countries, which are split into two clusters: one in the west and one in the east. The projection for the second positive PC (Figure \ref{male_mapsknn_2010}(b)) also shows spatial connectivity, with two clusters formed in the northern and southern regions. The ﬁrst negative PC (Figure \ref{male_mapsknn_2010}(c)) does not seem to display a particular spatial pattern. This outcome is anticipated because the negative principal components are associated with local structures that highlight dissimilarities on the geographical map at neighbouring locations. In general, spatial patterns are noticeable for the ﬁrst and second PCs.

Together, Figures \ref{male_curves_2010}, \ref{male_mapsknn_2010} and \ref{functional_moran}  illustrate how temporal variation in mortality curves is linked to spatial dependence. The mortality curves define the functional domain, the functional Moran’s $I$ identifies the ages at which spatial autocorrelation is strongest, and the score maps show where these patterns emerge geographically. This joint interpretation demonstrates that fasPCA connects temporal structure with spatial clustering, a feature not explicitly targeted by classical FPCA.

\begin{table} [!htbp]
    \centering
    \caption{Moran's test for spatial autocorrelation, based on the log of death rates for males in 28 European countries using KNN weights}
    \begin{tabular*}{\textwidth}{@{\extracolsep\fill} l c  @{\extracolsep\fill}}
        \toprule
        & Moran’s $I$ (Year 2010) \\
        \midrule
        KNN & $0.9831^{***}$ \\
        \bottomrule
    \end{tabular*}
    \begin{tablenotes}[flushleft]
        \item[] Note: "***" p < 0.001.
    \end{tablenotes}
    \label{tablemoran3years}
\end{table}

\begin{table}[!htbp]
    \centering
    \caption{Moran's $I$ values and variance explained by principal component scores using classical FPCA and fasPCA, based on KNN weights for male log mortality rates in 28 European countries in 2010}
    \label{pcaspca2010}
    \begin{threeparttable}
        \begin{tabular*}{\textwidth}{@{\extracolsep\fill} lcc @{\extracolsep\fill}}
            \toprule
            \textbf{Method/component} & \textbf{Moran's $I$} & \textbf{Variance explained (\%)} \\
            \midrule
            \multicolumn{3}{@{}l}{\textbf{Classical FPCA}} \\
            PC1 & \(0.5482^{***}\)  & 99.67 \\
            PC2 & \(-0.0873\)       & 0.14  \\
            PC3 & \(0.2687^{**}\)   & 0.08  \\
            PC4 & \(-0.0548\)       & 0.04  \\
            \textbf{Total} & -- & \textbf{99.93} \\
            \midrule
            \multicolumn{3}{@{}l}{\textbf{fasPCA, KNN \((3,2)\)}} \\
            PC1 positive & \(0.6021^{***}\)       & 85.14 \\
            PC2 positive & \(0.3239^{**}\)        & 2.54  \\
            PC3 positive & \(0.1821^{*}\)         & 0.86  \\
            PC1 negative & \(-0.1516^{\dagger}\)  & 6.65  \\
            PC2 negative & \(-0.1575^{\dagger}\)  & 2.29  \\
            \textbf{Total} & -- & \textbf{97.48} \\
            \bottomrule
        \end{tabular*}
        \begin{tablenotes}[flushleft]
            \footnotesize
            \item[] Note: \(^{\dagger}p<0.1\); \(^{*}p<0.05\); \(^{**}p<0.01\); \(^{***}p<0.001\).
            \item[] Classical FPCA is included as a non-spatial reference; fasPCA components are obtained from the spatial Moran-type criterion.
        \end{tablenotes}
    \end{threeparttable}
\end{table}

\FloatBarrier


\section{Multivariate fasPCA}

We report additional results for the multivariate functional areal spatial PCA (mfasPCA) and STPCA analyses. The supplementary results include additional simulation diagnostics across the spatial weight matrices considered in the study, as well as supporting results for the Polish regional socioeconomic data application. Across the nine spatial weight matrices proposed by \citet{krzysko2024stpca}, the simulations yield qualitatively similar conclusions for both positive and negative eigenspace diagnostics. To avoid redundancy, the main manuscript focuses on two representative weight structures: radial distance weights for positive spatial autocorrelation and shared-boundary weights for negative spatial autocorrelation.


\subsection{Additional simulation diagnostics for representative weight matrices}

{\small \textit{The cumulative proportion of variance explained (CPVE) summaries are reported using the retained positive and negative components specified in each figure caption.}}

\begin{figure}[!htbp]
    \centering
    \begin{tabular}{cccc}
        \multicolumn{1}{c}{(a) $\rho=0.3$} & \multicolumn{1}{c}{(b) $\rho=0.5$} & \multicolumn{1}{c}{(c) $\rho=0.7$} & \multicolumn{1}{c}{(d) $\rho=0.9$} \\
        \begin{subfigure}[t]{4cm}
            \includegraphics[width=\textwidth,height=6cm]{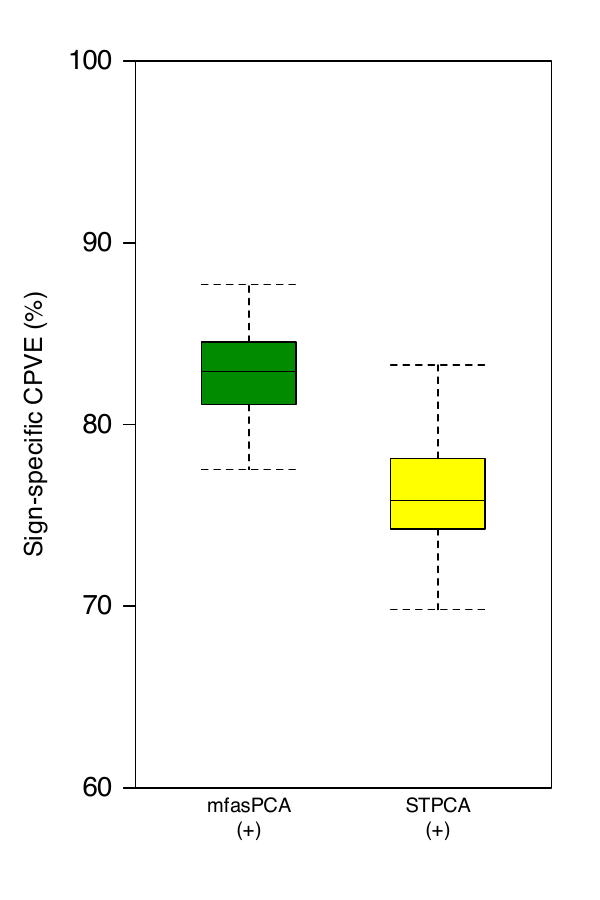}
        \end{subfigure} &
        \begin{subfigure}[t]{4cm}
            \includegraphics[width=\textwidth,height=6cm]{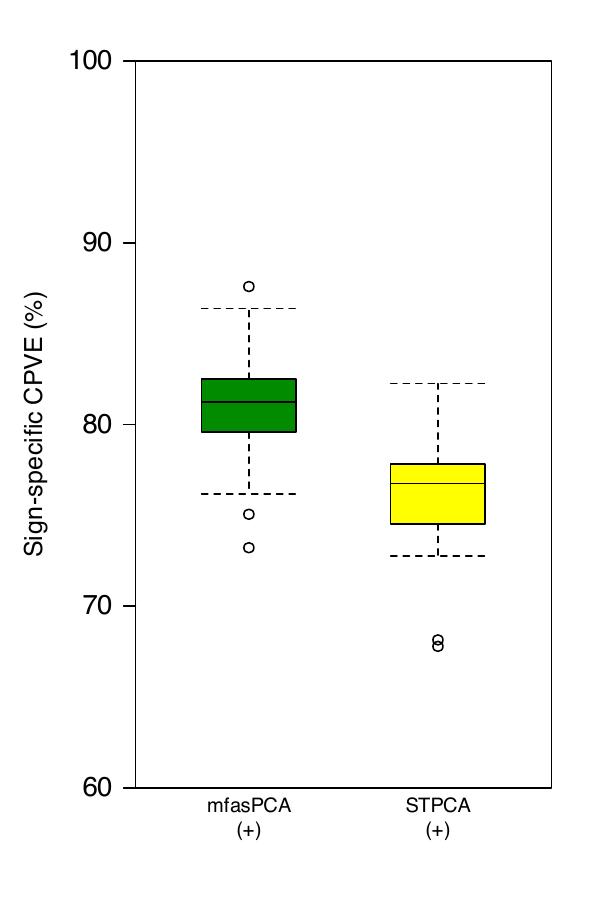}
        \end{subfigure} &
        \begin{subfigure}[t]{4cm}
            \includegraphics[width=\textwidth,height=6cm]{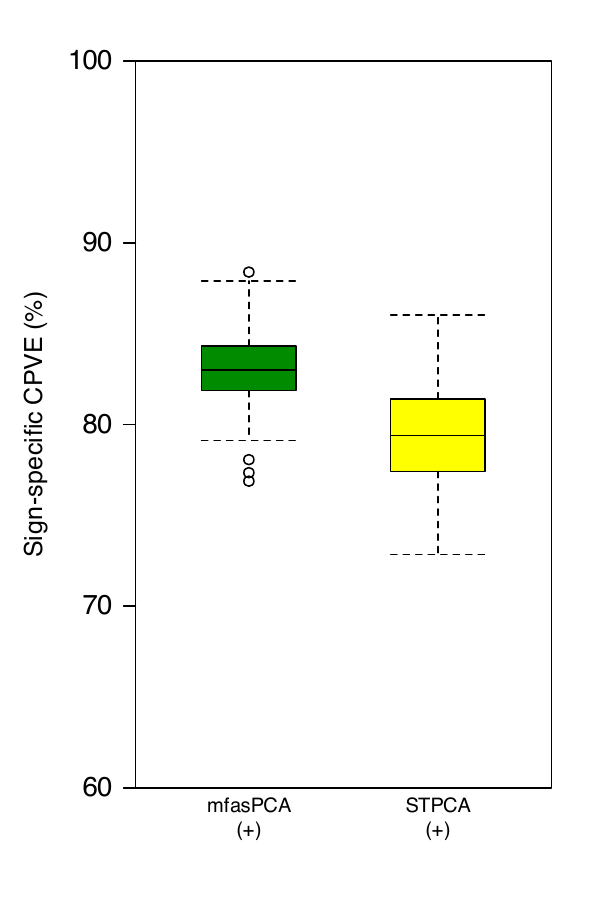}
        \end{subfigure} &
        \begin{subfigure}[t]{4cm}
            \includegraphics[width=\textwidth,height=6cm]{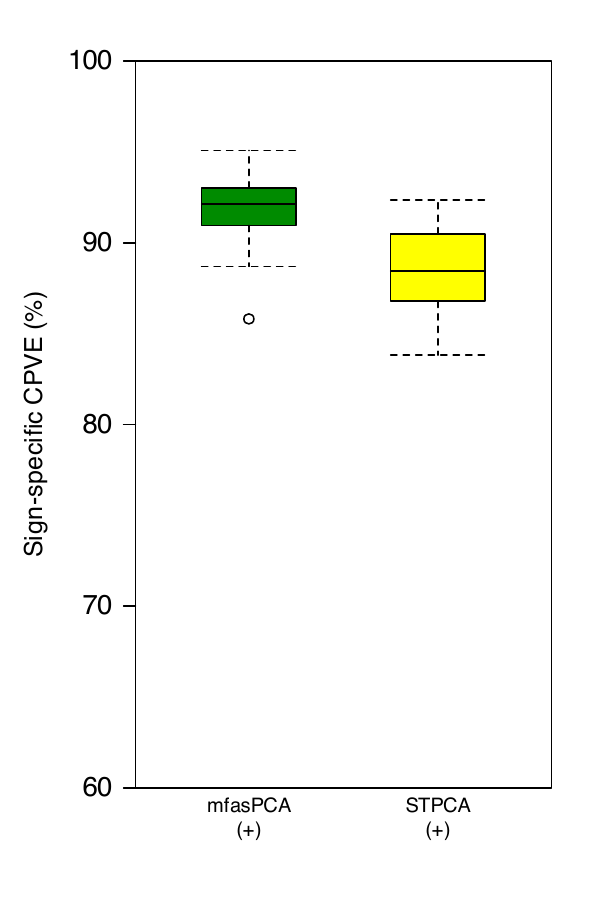}
        \end{subfigure} \\
        \begin{subfigure}[t]{4cm}
            \includegraphics[width=\textwidth,height=6cm]{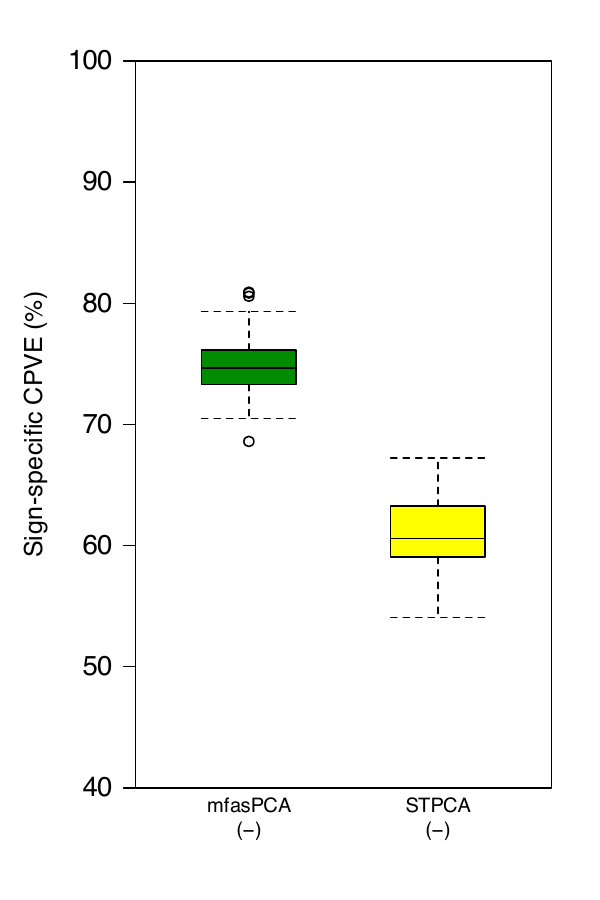}
        \end{subfigure} &
        \begin{subfigure}[t]{4cm}
            \includegraphics[width=\textwidth,height=6cm]{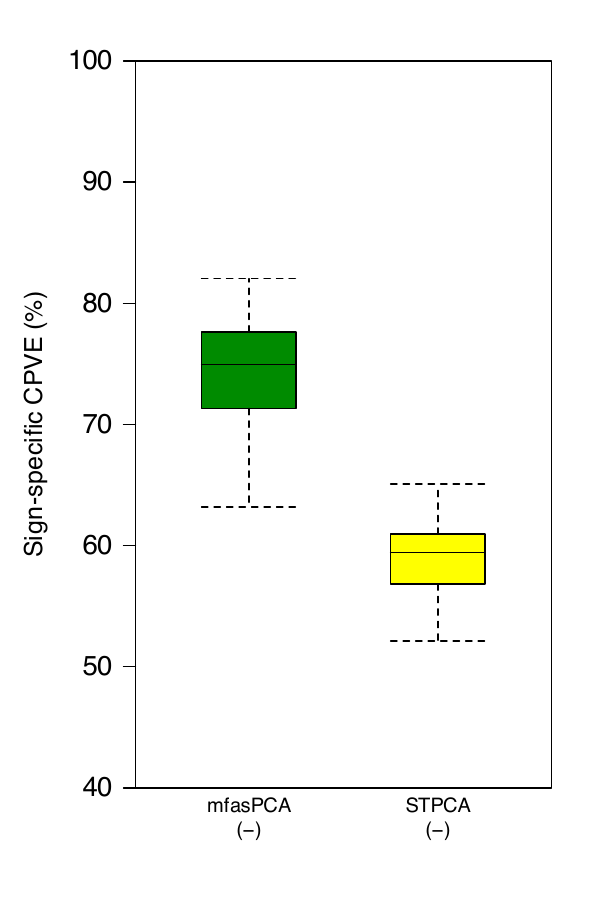}
        \end{subfigure} &
        \begin{subfigure}[t]{4cm}
            \includegraphics[width=\textwidth,height=6cm]{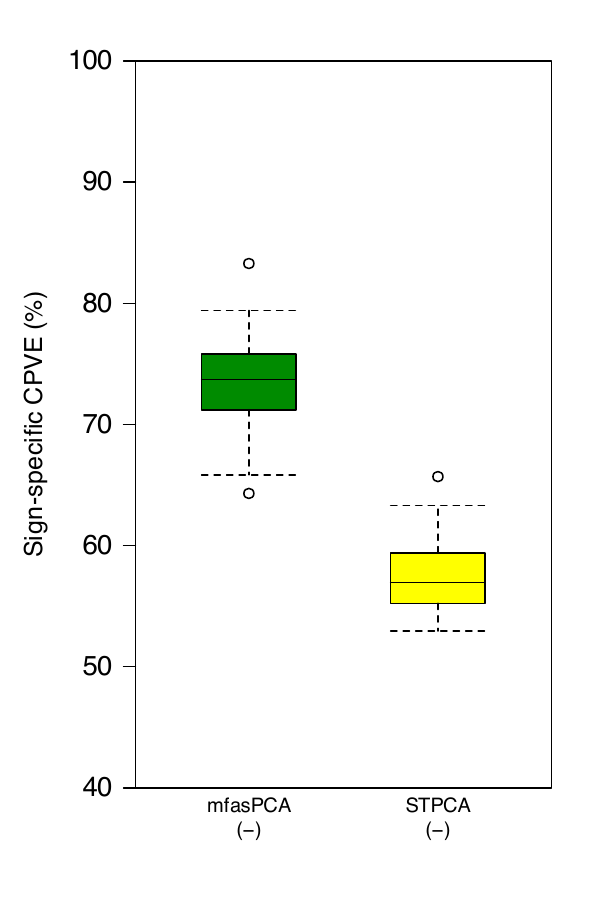}
        \end{subfigure} &
        \begin{subfigure}[t]{4cm}
            \includegraphics[width=\textwidth,height=6cm]{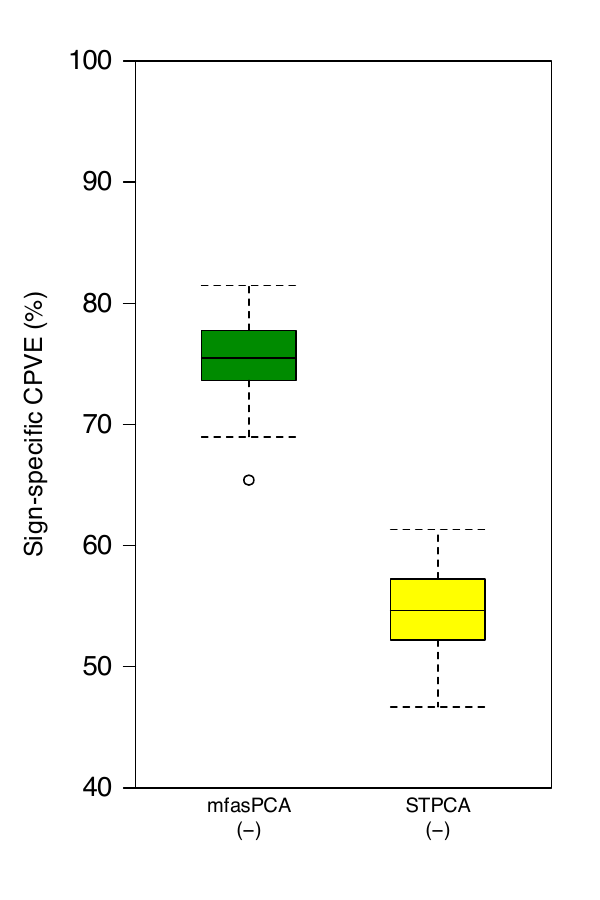}
        \end{subfigure}
    \end{tabular}
    \caption[Radial-distance weights ($r=120$ km)]{Sign-specific CPVE for mfasPCA and STPCA across 50 simulated datasets with \textit{radial-distance weights ($r=120$ km)}. \textbf{Top row:} top five positive components. \textbf{Bottom row:} top five negative components}
    \label{fig:w3-radial-120km}
\end{figure}

\begin{figure}[!htbp]
    \centering
    \begin{tabular}{cccc}
        \multicolumn{1}{c}{(a) $\rho=0.3$} & \multicolumn{1}{c}{(b) $\rho=0.5$} & \multicolumn{1}{c}{(c) $\rho=0.7$} & \multicolumn{1}{c}{(d) $\rho=0.9$} \\
        \begin{subfigure}[t]{4cm}
            \includegraphics[width=\textwidth,height=6cm]{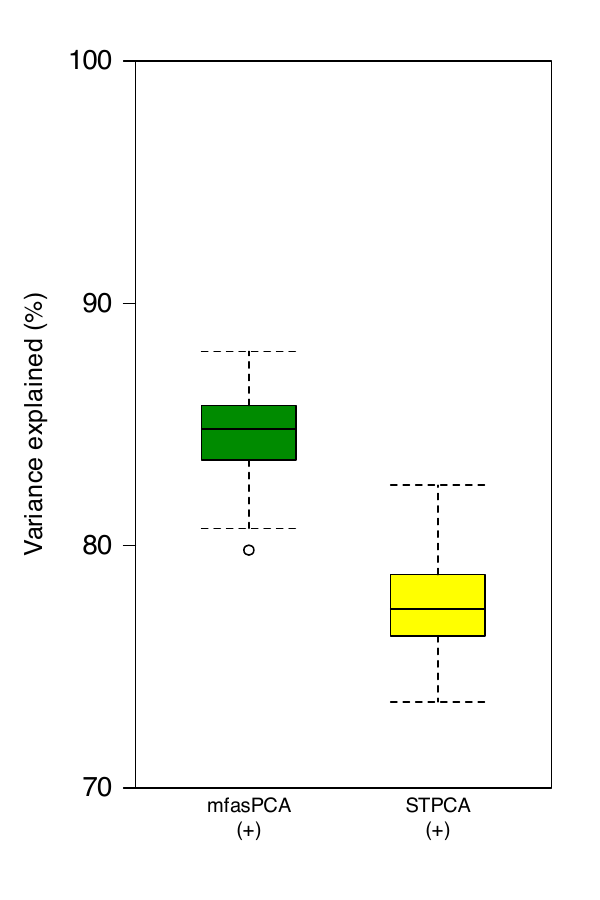}
        \end{subfigure} &
        \begin{subfigure}[t]{4cm}
            \includegraphics[width=\textwidth,height=6cm]{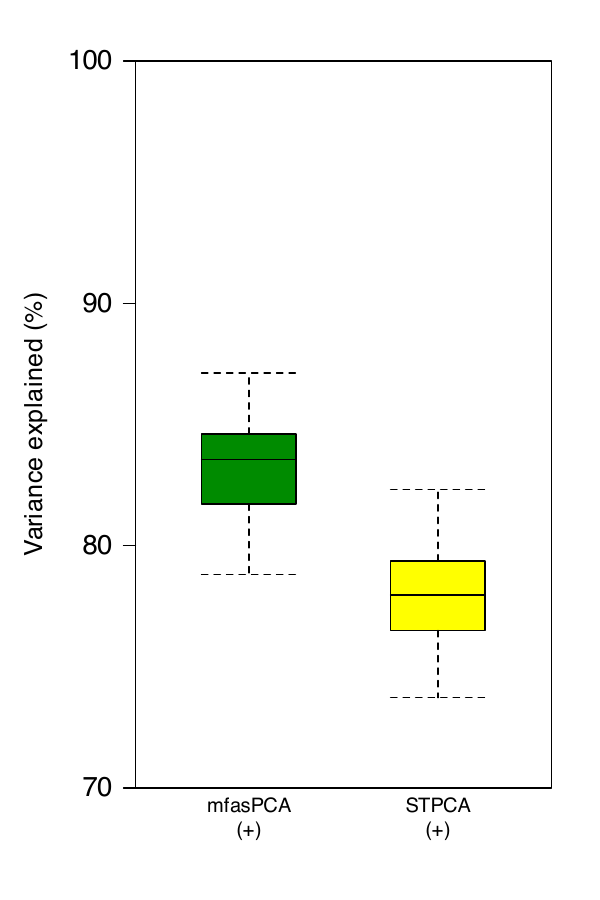}
        \end{subfigure} &
        \begin{subfigure}[t]{4cm}
            \includegraphics[width=\textwidth,height=6cm]{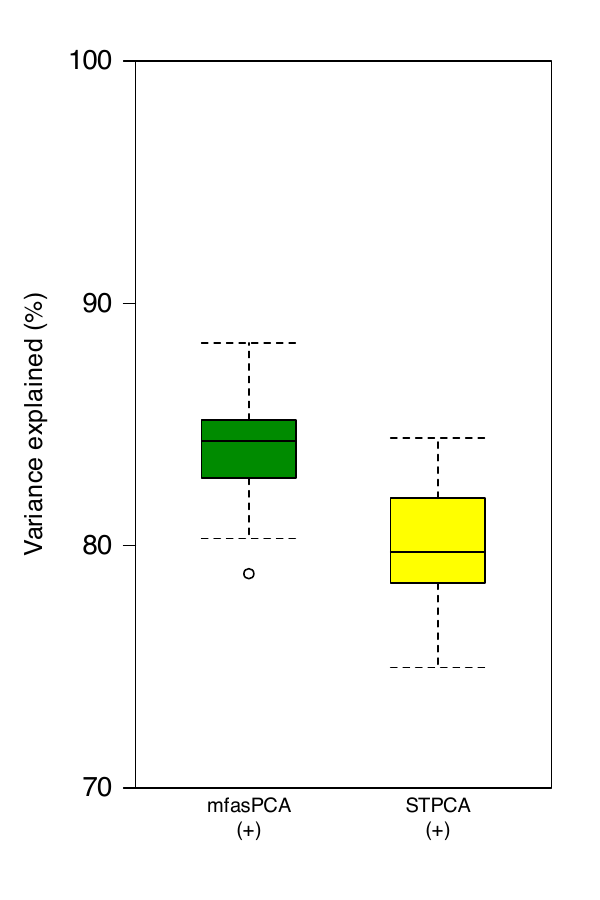}
        \end{subfigure} &
        \begin{subfigure}[t]{4cm}
            \includegraphics[width=\textwidth,height=6cm]{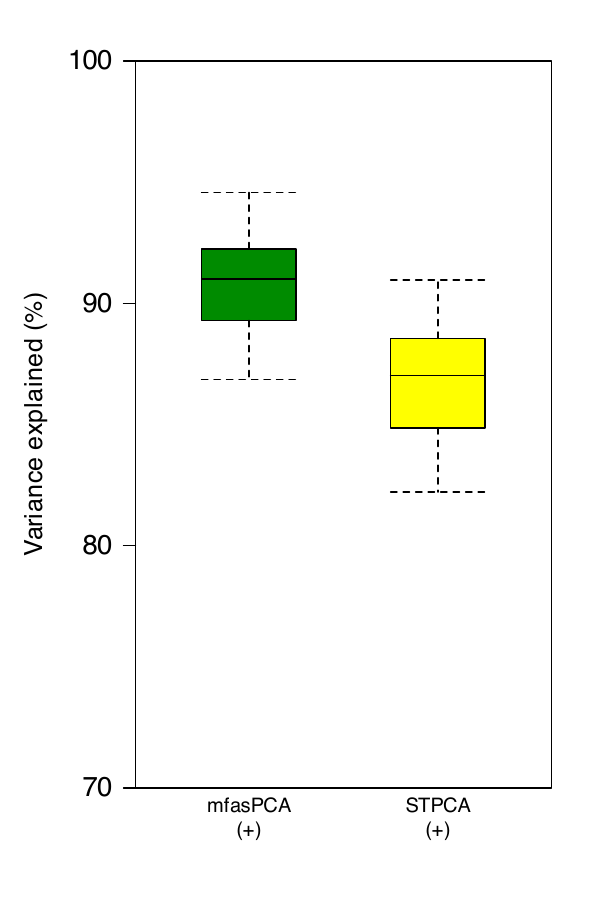}
        \end{subfigure} \\
        \begin{subfigure}[t]{4cm}
            \includegraphics[width=\textwidth,height=6cm]{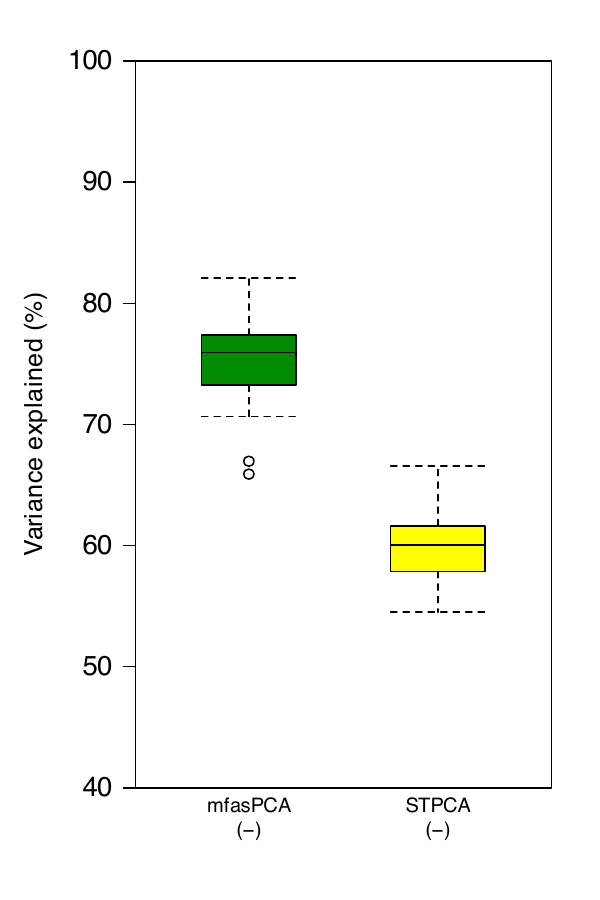}
        \end{subfigure} &
        \begin{subfigure}[t]{4cm}
            \includegraphics[width=\textwidth,height=6cm]{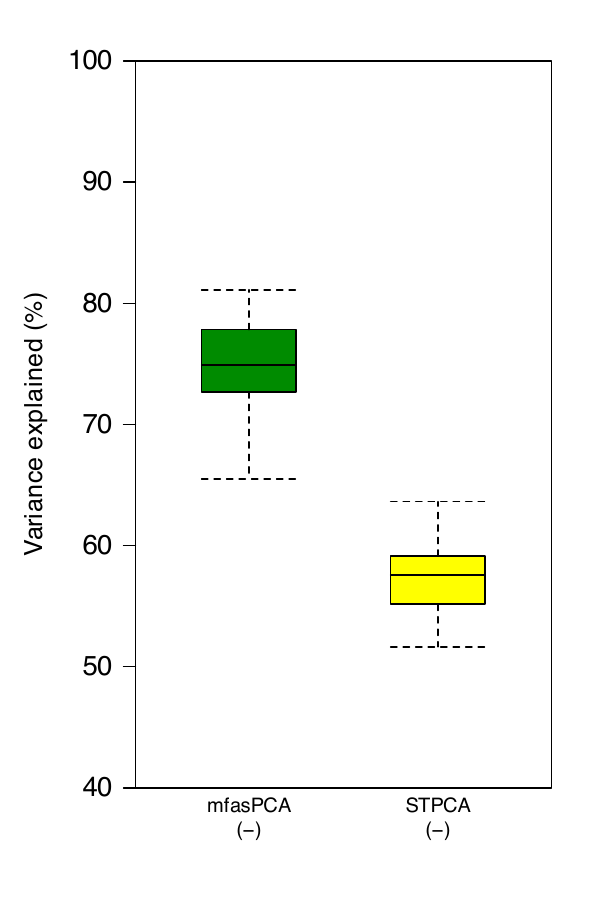}
        \end{subfigure} &
        \begin{subfigure}[t]{4cm}
            \includegraphics[width=\textwidth,height=6cm]{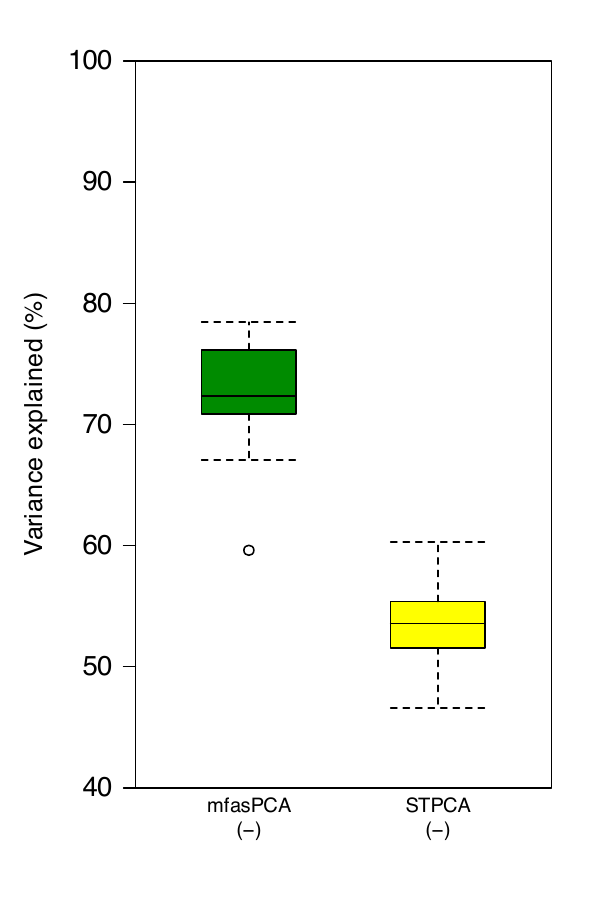}
        \end{subfigure} &
        \begin{subfigure}[t]{4cm}
            \includegraphics[width=\textwidth,height=6cm]{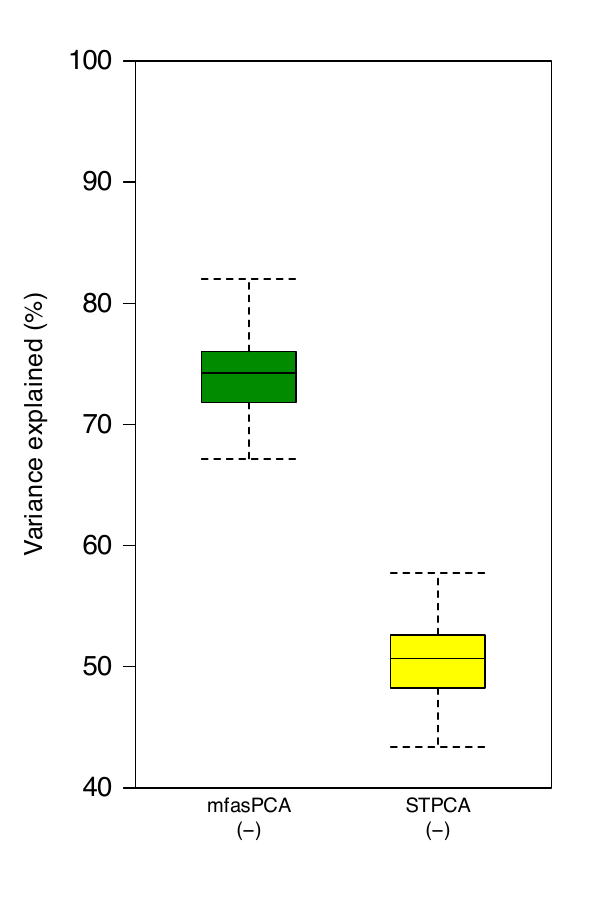}
        \end{subfigure}
    \end{tabular}
    \caption[Shared-boundary weights]{Sign-specific CPVE for mfasPCA and STPCA across 50 simulated datasets with \textit{shared-boundary weights}. \textbf{Top row:} top six positive components. \textbf{Bottom row:} top five negative components}
    \label{fig:w8-shared-boundary}
\end{figure}


\clearpage

\subsection{Negative spatial dependence ($\rho \in \{-0.3,-0.5,-0.7,-0.9\}$) under shared-boundary weights}

\begin{figure}[!htbp]
    \centering
    \begin{tabular}{cccc}
        \multicolumn{1}{c}{(a) $\rho=-0.3$} & \multicolumn{1}{c}{(b) $\rho=-0.5$} & \multicolumn{1}{c}{(c) $\rho=-0.7$} & \multicolumn{1}{c}{(d) $\rho=-0.9$} \\
        \begin{subfigure}[t]{4cm}
            \includegraphics[width=\textwidth,height=6cm]{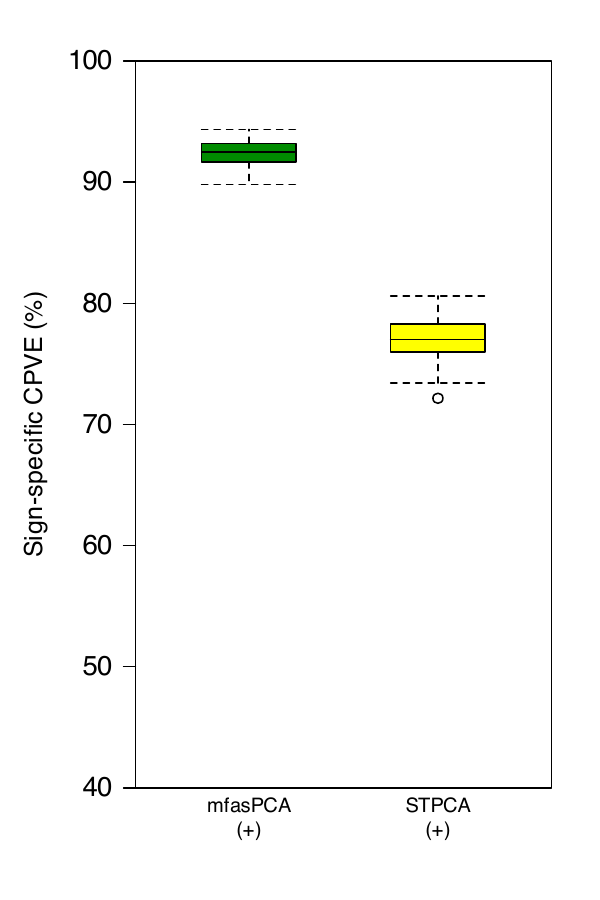}
        \end{subfigure} &
        \begin{subfigure}[t]{4cm}
            \includegraphics[width=\textwidth,height=6cm]{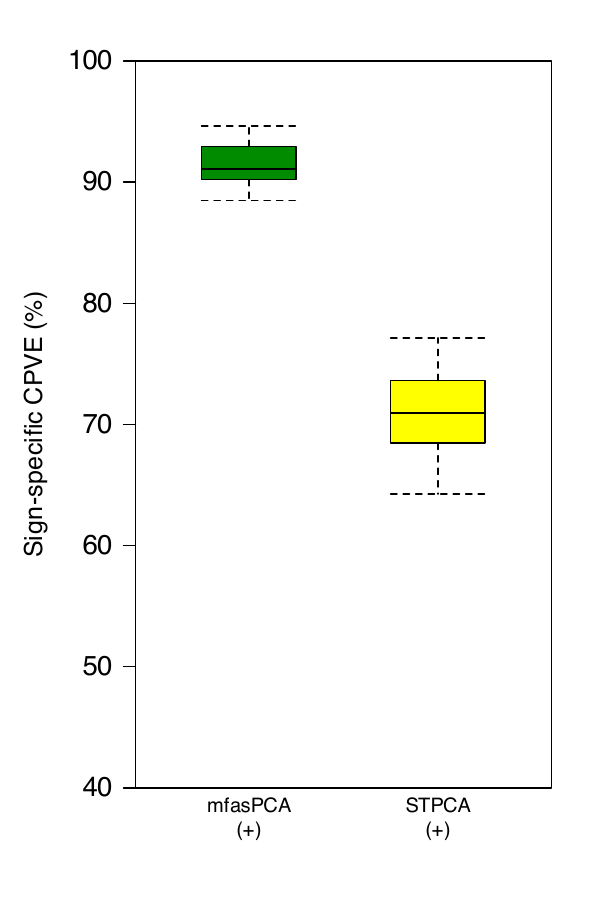}
        \end{subfigure} &
        \begin{subfigure}[t]{4cm}
            \includegraphics[width=\textwidth,height=6cm]{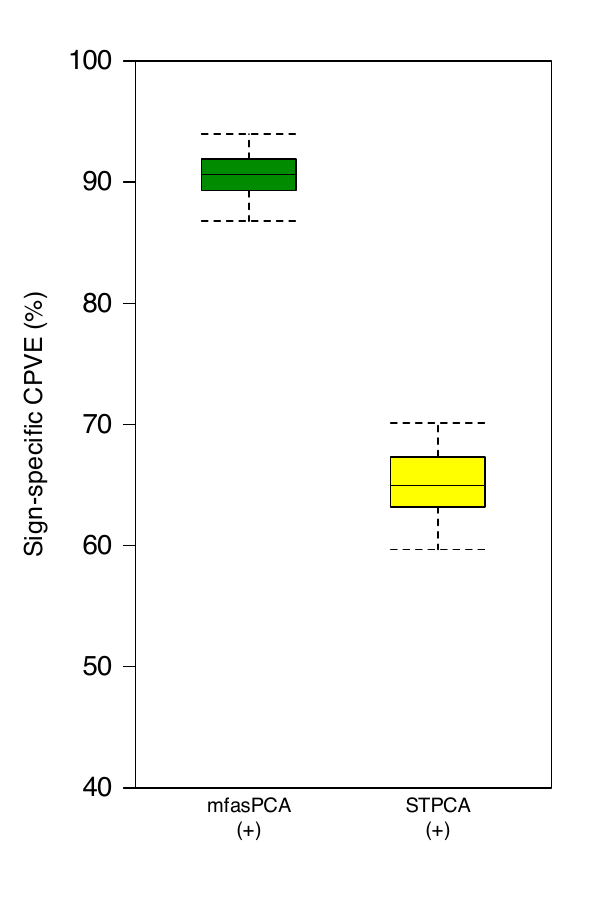}
        \end{subfigure} &
        \begin{subfigure}[t]{4cm}
            \includegraphics[width=\textwidth,height=6cm]{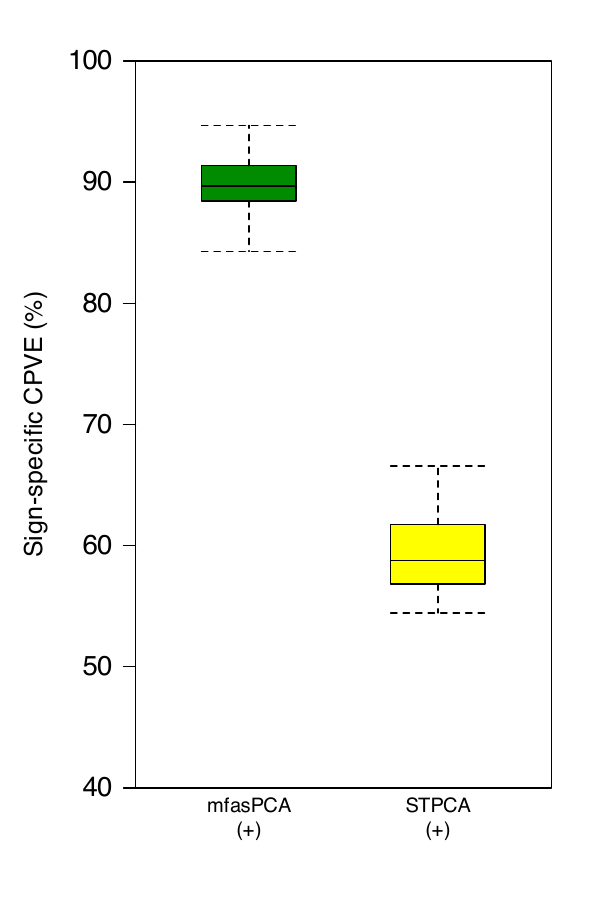}
        \end{subfigure} \\
        \begin{subfigure}[t]{4cm}
            \includegraphics[width=\textwidth,height=6cm]{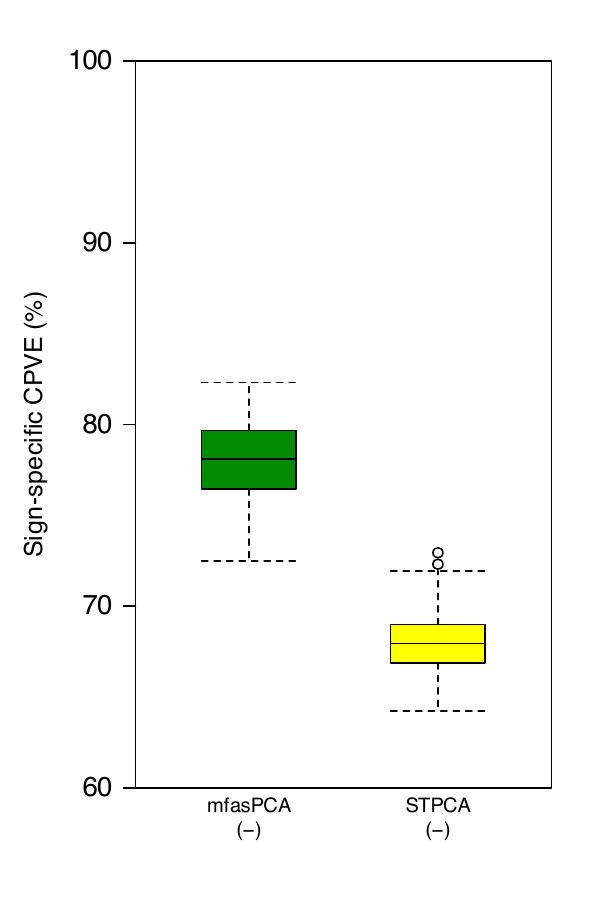}
        \end{subfigure} &
        \begin{subfigure}[t]{4cm}
            \includegraphics[width=\textwidth,height=6cm]{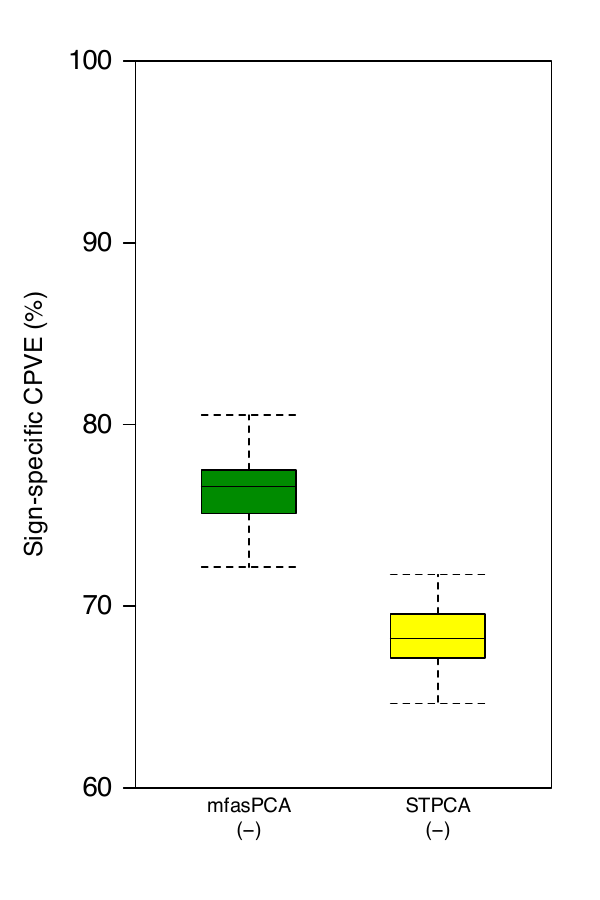}
        \end{subfigure} &
        \begin{subfigure}[t]{4cm}
            \includegraphics[width=\textwidth,height=6cm]{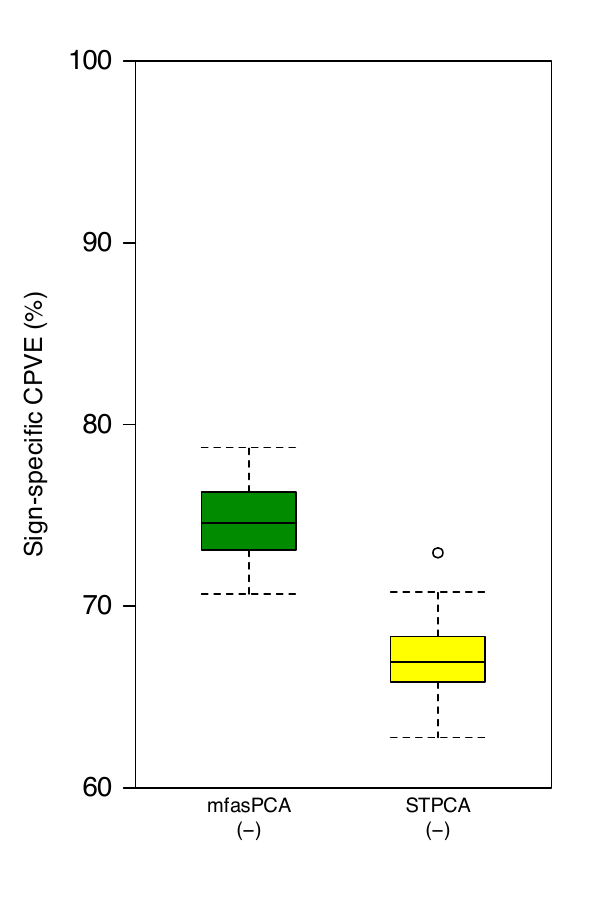}
        \end{subfigure} &
        \begin{subfigure}[t]{4cm}
            \includegraphics[width=\textwidth,height=6cm]{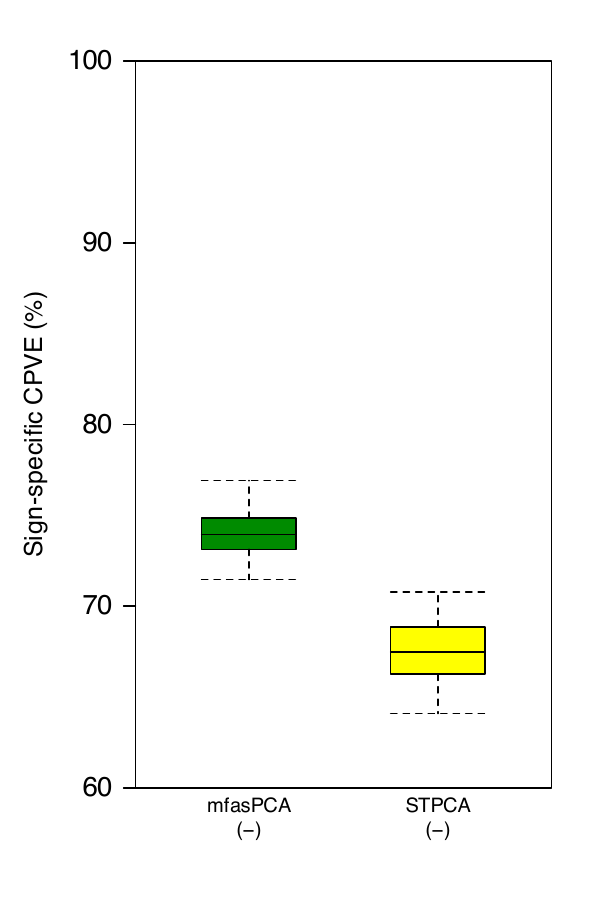}
        \end{subfigure}
    \end{tabular}
    \caption[Shared-boundary weights (negative rho)]{CPVE for mfasPCA and STPCA across 50 simulated datasets with \textit{shared-boundary weights}. \textbf{Top row:} top eight positive components. \textbf{Bottom row:} top six negative components}
    \label{fig:w8-shared-boundary-neg}
\end{figure}


\clearpage

\subsection{Polish data: Extended CPVE and permutation-test results}

\begin{table}[!htbp]
    \centering
    \footnotesize                       
    \setlength{\tabcolsep}{3pt}         
    \renewcommand{\arraystretch}{0.9}   
    \captionsetup{font=footnotesize}    
    \caption{Cumulative proportion of variance explained (\%) by the first two positive and negative components for mfasPCA and STPCA across nine spatial weight matrices for the Polish data}
    \label{tab:cpve_across_weights}
    \begin{threeparttable}
        \begin{tabular}{clrrrr}
            \toprule
            No. & Weight &
            \multicolumn{1}{c}{\begin{tabular}{@{}c@{}}CPVE (+)\\ mfasPCA\end{tabular}} &
            \multicolumn{1}{c}{\begin{tabular}{@{}c@{}}CPVE (+)\\ STPCA\end{tabular}} &
            \multicolumn{1}{c}{\begin{tabular}{@{}c@{}}CPVE (-)\\ mfasPCA\end{tabular}} &
            \multicolumn{1}{c}{\begin{tabular}{@{}c@{}}CPVE (-)\\ STPCA\end{tabular}} \\
            \midrule
            1 & kNN (standard, k = 4)              & 99.16 & 95.25 & 57.84 & 76.10 \\
            2 & kNN (symmetric, k = 4)             & 99.25 & 96.42 & 60.40 & 76.60 \\
            3 & Radial distance                    & 99.95 & 99.22 & 62.66 & 68.86 \\
            4 & Power distance (alpha = 2)         & 99.68 & 97.25 & 70.98 & 82.78 \\
            5 & Exponential distance (alpha = 2)   & 99.77 & 98.71 & 58.91 & 76.61 \\
            6 & Double-power distance (k = 3)      & 99.71 & 98.64 & 55.97 & 74.19 \\
            7 & Rook contiguity                    & 98.89 & 86.68 & 57.38 & 81.88 \\
            8 & Shared-boundary                    & 98.59 & 85.35 & 63.00 & 83.84 \\
            9 & Combined distance--boundary        & 98.66 & 86.81 & 63.98 & 84.25 \\
            \bottomrule
        \end{tabular}
    \end{threeparttable}
\end{table}

\begin{table}[!htbp]
    \centering
    \footnotesize
    \setlength{\tabcolsep}{3pt}
    \caption{Omnibus eigenvalue-based permutation tests and per-eigen Holm tests for mfasPCA across nine spatial weight matrices}
    \label{tab:randtests_mfasPCA}
    \begin{threeparttable}
        \begin{tabular}{clccc}
            \toprule
            No. & Weight & Omnibus p(+) & Omnibus p(--) & Sig PCs (per-eigen) \\
            \midrule
            1 & kNN (standard, k = 4)            & 0.288          & 0.674          & P:2 \\
            2 & kNN (symmetric, k = 4)           & 0.205          & 0.677          & P:2 \\
            3 & Radial distance                  & 0.160          & 0.829          & --  \\
            4 & Power distance (alpha = 2)       & 0.093          & 0.392          & P:2 \\
            5 & Exponential distance (alpha = 2) & 0.102          & 0.800          & P:2 \\
            6 & Double-power distance (k = 3)    & 0.085          & 0.920          & P:2 \\
            7 & Rook contiguity                  & 0.300          & 0.627          & P:2 \\
            8 & Shared-boundary                  & 0.022$^{*}$    & 0.618          & P:2 \\
            9 & Combined distance--boundary      & 0.022$^{*}$    & 0.579          & P:2 \\
            \bottomrule
        \end{tabular}
        \begin{tablenotes}
            \footnotesize
            \item[] Omnibus p-values are reported for positive (+) and negative (--) subspaces, with significance indicated by stars. The final column lists the within-sign components (P for positive, N for negative) that are significant at $\alpha = 0.05$ after Holm adjustment; `--' indicates no significant components. $^{*}p < 0.05$, $^{**}p < 0.01$.
        \end{tablenotes}
    \end{threeparttable}
\end{table}

\begin{table}[!htbp]
    \centering
    \footnotesize
    \setlength{\tabcolsep}{3pt}
    \caption{Omnibus eigenvalue-based permutation tests and per-eigen Holm tests for STPCA across nine spatial weight matrices}
    \label{tab:randtests_stpca}
    \begin{threeparttable}
        \begin{tabular}{clcc>{\raggedright\arraybackslash}p{3.3cm}}
        \toprule
        No. & Weight & Omnibus p(+) & Omnibus p($-$) & Sig PCs (per-eigen) \\
        \midrule
        1 & kNN (standard, k = 4)            & 0.130          & 0.150          & -- \\
        2 & kNN (symmetric, k = 4)           & 0.129          & 0.198          & -- \\
        3 & Radial distance                  & 0.123          & 0.721          & -- \\
        4 & Power distance (alpha = 2)       & 0.046$^{*}$    & 0.129          & -- \\
        5 & Exponential distance (alpha = 2) & 0.052          & 0.129          & N:24,25,26,27 \\
        6 & Double-power distance (k = 3)    & 0.082          & 0.290          & N:30 \\
        7 & Rook contiguity                  & 0.114          & 0.008$^{**}$   & N:17,18,19 \\
        8 & Shared-boundary                  & 0.126          & 0.013$^{*}$    & -- \\
        9 & Combined distance--boundary      & 0.135          & 0.016$^{*}$    & N:15,16,17,18,19,20,21,\\
          &                                  &                &                & 22,23,24,28 \\
        \bottomrule
        \end{tabular}
        \begin{tablenotes}
            \footnotesize
            \item Omnibus p-values are reported for positive (+) and negative ($-$) subspaces, with significance indicated by stars. The final column lists the within-sign components (P for positive, N for negative) that are significant at $\alpha = 0.05$ after Holm adjustment; `--' indicates no significant components. $^{*} p < 0.05$, $^{**} p < 0.01$.
        \end{tablenotes}
        \vspace{1em}
        The per-eigen column reports all components that remain significant after Holm adjustment within each sign-specific family; some significant STPCA components occur at higher ranks.
    \end{threeparttable}
\end{table}

\begin{table}[!htbp]
    \centering
    \small
    \setlength{\tabcolsep}{4pt}
    \caption{Sequential cumulative-sum permutation tests for mfasPCA across nine spatial weight matrices. Listed are only components with significant spatial signal (Holm-adjusted sequential $p$-value $\le 0.05$)}
    \label{tab:sequential_mfasPCA}
    \begin{threeparttable}
        \begin{tabular}{cllccc}
            \toprule
            No. & Weight & Sign & PC & Holm seq.\ $p$ & Significant \\
            \midrule
            1 & kNN (standard, k = 4)            & Positive & 2 & 0.0210 & *  \\
            2 & kNN (symmetric, k = 4)           & Positive & 2 & 0.0354 & *  \\
            4 & Power distance (alpha = 2)       & Positive & 2 & 0.0060 & ** \\
            5 & Exponential distance (alpha = 2) & Positive & 2 & 0.0055 & ** \\
            6 & Double-power distance (k = 3)    & Positive & 2 & 0.0265 & *  \\
            7 & Rook contiguity                  & Positive & 2 & 0.0189 & *  \\
            8 & Shared-boundary                  & Positive & 2 & 0.0168 & *  \\
            9 & Combined distance--boundary      & Positive & 2 & 0.0189 & *  \\
            \bottomrule
        \end{tabular}
        \begin{tablenotes}
            \footnotesize
            \item $^{*} p < 0.05$, $^{**} p < 0.01$.
        \end{tablenotes}
    \end{threeparttable}
\end{table}

\begin{table}[!htbp]
    \centering
    \small
    \setlength{\tabcolsep}{4pt}
    \caption{Sequential cumulative-sum permutation tests for STPCA across nine spatial weight matrices. Listed are only components with significant spatial signal (Holm-adjusted sequential $p$-value $\le 0.05$)}
    \label{tab:sequential_STPCA}
    \begin{threeparttable}
        \begin{tabular}{cllccc}
            \toprule
            No. & Weight & Sign & PC & Holm seq.\ $p$ & Significant \\
            \midrule
            5 & Exponential distance (alpha = 2) & Negative & 12 & 0.0405 & *  \\
            5 & Exponential distance (alpha = 2) & Negative & 13 & 0.0442 & *  \\
            5 & Exponential distance (alpha = 2) & Negative & 14 & 0.0150 & *  \\
            5 & Exponential distance (alpha = 2) & Negative & 15 & 0.0174 & *  \\
            5 & Exponential distance (alpha = 2) & Negative & 16 & 0.0336 & *  \\
            9 & Combined distance--boundary      & Negative & 11 & 0.0300 & *  \\
            9 & Combined distance--boundary      & Negative & 12 & 0.0110 & *  \\
            9 & Combined distance--boundary      & Negative & 13 & 0.0054 & ** \\
            9 & Combined distance--boundary      & Negative & 14 & 0.0054 & ** \\
            9 & Combined distance--boundary      & Negative & 15 & 0.0054 & ** \\
            9 & Combined distance--boundary      & Negative & 16 & 0.0032 & ** \\
            9 & Combined distance--boundary      & Negative & 17 & 0.0032 & ** \\
            9 & Combined distance--boundary      & Negative & 18 & 0.0032 & ** \\
            9 & Combined distance--boundary      & Negative & 19 & 0.0032 & ** \\
            9 & Combined distance--boundary      & Negative & 20 & 0.0032 & ** \\
            9 & Combined distance--boundary      & Negative & 21 & 0.0054 & ** \\
            9 & Combined distance--boundary      & Negative & 22 & 0.0054 & ** \\
            9 & Combined distance--boundary      & Negative & 23 & 0.0252 & *  \\
            9 & Combined distance--boundary      & Negative & 24 & 0.0342 & *  \\
            \bottomrule
        \end{tabular}
        \begin{tablenotes}
            \footnotesize
            \item $^{*} p < 0.05$, $^{**} p < 0.01$.
        \end{tablenotes}
    \end{threeparttable}
\end{table}

\FloatBarrier


\vspace*{0.5em}
\bibliographystyle{sn-basic}
\bibliography{library.bib}